\documentclass[twocolumn]{aastex63}

\usepackage{color,comment,multirow,booktabs,upgreek}
\usepackage{multirow,enumitem}
\usepackage{CJK}
\graphicspath{{./}{images/}}

\newcommand{\hi}{\ifmmode{\rm HI}\else{H\/{\sc i}}\fi} 
\newcommand{\glon}{\ifmmode{\ell}\else{$\ell$}\fi} 
\newcommand{\glat}{\ifmmode{b}\else{$b$}\fi} 
\newcommand{\vlsr}{\ifmmode{V_\mathrm{LSR}}\else{$V_\mathrm{LSR}$}\fi} 
\newcommand{\vwind}{\ifmmode{V_\mathrm{w}}\else{$V_\mathrm{w}$}\fi} 
\newcommand{\de}{\ifmmode{^\circ}\else{$^\circ$}\fi} 
\newcommand {\kms}{\ifmmode{\rm km \, s^{-1}}\else{$\rm km \, s^{-1}$}\fi}

\newcommand {\moyr}{\,{\rm M_\odot\,\rm yr}^{-1}}

\shorttitle{CRs in Star-forming Galactic Disks}
\shortauthors{Armillotta et al.}

\begin{document}
\begin{CJK*}{UTF8}{gbsn}

\title{Cosmic-Ray Transport in Simulations of Star-forming Galactic Disks}%

\correspondingauthor{Lucia Armillotta}
\email{lucia.armillotta@princeton.edu}

\author[0000-0002-5708-1927]{Lucia Armillotta}
\affiliation{Department of Astrophysical Sciences, Princeton University, Princeton, NJ 08544, USA}

\author[0000-0002-0509-9113]{Eve C. Ostriker}
\affiliation{Department of Astrophysical Sciences, Princeton University, Princeton, NJ 08544, USA}

\author[0000-0002-2624-3399]{Yan-Fei Jiang(姜燕飞)}
\affiliation{Center for Computational Astrophysics, Flatiron Institute,
New York, NY 10010, USA}

\begin{abstract}
Cosmic ray transport on galactic scales depends on the detailed properties of the magnetized, multiphase interstellar medium (ISM).
In this work, we post-process a high-resolution TIGRESS magnetohydrodynamic  simulation modeling a local galactic disk patch 
with a two-moment fluid algorithm for cosmic ray transport. We consider a variety of prescriptions for the cosmic rays, from a simple purely diffusive formalism with 
constant scattering coefficient, to a physically-motivated model in which the scattering coefficient is set by critical balance between streaming-driven Alfv\'en wave excitation and damping mediated by local gas properties.  
We separately focus on cosmic rays with kinetic energies of $\sim 1$ GeV (high-energy) and $\sim 30$~MeV (low-energy), respectively important for ISM dynamics and chemistry. We find that simultaneously accounting for advection, streaming, and diffusion of cosmic rays is crucial for properly modeling their transport. Advection dominates in the high-velocity, low-density, hot phase, while diffusion and streaming are more important in higher density, cooler phases. Our physically-motivated model shows that there is no single diffusivity for cosmic-ray transport: the scattering coefficient varies by four or more orders of magnitude,  maximal at density $n_\mathrm{H} \sim 0.01\,  \mathrm{cm}^{-3}$. Ion-neutral damping of Alfv\'en waves results in strong diffusion and nearly uniform cosmic ray pressure within most of the mass of the ISM.   
However, cosmic rays are trapped near the disk midplane by the higher scattering rate in the surrounding lower-density, higher-ionization gas. 
The transport of high-energy cosmic rays differs from that of low-energy cosmic rays, with less effective diffusion and greater energy losses for the latter.
\end{abstract}


\vspace*{1cm}

\section{Introduction}
\label{Introduction}

Cosmic rays (CRs) are charged particles moving with relativistic speeds, observed over more than ten orders of magnitude in energy with a (broken) power-law distribution. Mainly generated within disk galaxies through shock acceleration in supernova remnants \citep[e.g.][]{Bell78, Blandford&Ostriker78, Schlickeiser89}, CRs easily spread throughout the interstellar medium (ISM) thanks to their quasi-collisionless nature. In the Milky Way's disk, the energy density of CRs, dominated by protons with kinetic energies of a few GeV \citep[see reviews by][]{Strong+07,Grenier+15}, is approximately in equipartition with the thermal, turbulent and magnetic energy densities \citep[e.g.][]{Boulares&Cox90, Beck01}. This suggests that CRs can significantly contribute to the dynamics of the ISM, potentially aiding in the internal support against gravity and/or  helping to drive galactic winds. Additionally, CR ionization is very important in the dense gas that is shielded to UV, providing heating, driving chemical reactions, and maintaining the coupling to magnetic fields \citep[e.g.][]{Padovani+20}. CRs therefore play several important roles in the evolution of galaxies. 

The interaction between CRs and the surrounding gas is mostly mediated by the ambient magnetic field. Being charged particles, CRs gyrate around and stream along magnetic field lines, while scattering off of magnetic fluctuations on spatial scales of order the CR gyroradius. Scattering reduces the mean free path and effective propagation speed of CRs, thus allowing them to couple with the background thermal gas. 

There are two main scenarios for the origin of magnetic fluctuations that scatter CRs, 
namely ``self-confinement'' and ``extrinsic turbulence.'' In the former scenario, the fluctuations are Alfv\'{e}n waves amplified by resonant streaming instabilities of CRs that develop when the bulk flow speed of the CR distribution exceeds the Alfv\'en speed in the background plasma \citep{Kulsrud&Pearce69,Wentzel74}. Scattering by resonant Alfv\'{e}n waves isotropizes the CRs in the reference frame of the wave, tending to reducing streaming to the local Alfv\'{e}n speed \citep[e.g.][]{Kulsrud05, Bai2019}. However, damping mechanisms, including ion-neutral damping \citep{Kulsrud&Pearce69}, nonlinear Landau damping  \citep{Kulsrud05}, linear Landau damping \citep{Wiener+18} and turbulent damping \citep{Farmer&Goldreich04, Lazarian16, Holguin+19}, 
limit Alfv\'en wave amplification and therefore the CR scattering rate. 
In the extrinsic turbulence picture, the magnetic fluctuations are driven by mechanisms independent of CRs, such as turbulent cascades or other energy injection sources \citep[e.g.][]{Chandran00, Yan&Lazarian02}. The same damping mechanisms mentioned above would also dissipate the magnetic energy of extrinsically-driven MHD waves, thus reducing the rate of CR scattering \citep [e.g.][]{Xu&Lazarian17}.

In both scenarios, 
the net CR flux is down the pressure gradient, and the magnetic field mediates transfer of momentum from the CR distribution to the background gas. In addition to momentum, in the self-confinement regime damping of Alfv\'{e}n waves transfers energy to the surrounding gas at nearly the same rate waves are exited by CRs. In the extrinsic-turbulence scenario, provided that the MHD waves have no preferred direction of propagation, CRs do not stream along with the waves. As a consequence, there is no transfer of energy from the CR distribution to the waves and, due to wave damping, from the waves to the gas. Instead, energy can flow from the waves to the CRs through second-order Fermi acceleration \citep[see reviews by][for a detailed overview of the two scenarios]{Zweibel13, Zweibel17}.

Since frequent wave-particle scattering can make the CR mean free path very short compared to other length scales of interest, in most astrophysical studies of ISM dynamics it is appropriate to treat CRs as a fluid. The transport of the CR fluid can be described in terms of diffusion relative to the hydromagnetic wave frame and advection along with the background magnetic field by thermal gas. In the self-confinement picture, the wave frame moves at the Alfv\'en speed, so this streaming has to be included together with diffusion and advection in the fluid treatment. 

Estimates for the Milky-Way disk suggests that self-confinement via resonant streaming instability is the dominant effect mediating transport for CRs with kinetic energies lower than a few tens of GeV {\citep[e.g.][]{Zweibel13, Zweibel17, Evoli+18}}. For CRs with higher energies, the growth rate of streaming instability rapidly decreases with increasing CR 
energy while the background turbulence has higher amplitude, so that scattering by extrinsic turbulence becomes more and more important \citep[][see also \autoref{sigma} and \autoref{spectrum}]{Skilling71,Blasi2012}. Since the majority of the total energy density in CRs is held in particles with kinetic energies of a few GeV, while ionization is provided by CRs at even lower energy, the self-confinement CR transport framework is most relevant to understanding the effects of CRs on the background thermal gas.
As we discuss in \autoref{sigma}, for our calculations (focusing on GeV and lower energy) we shall consider self-generated waves rather than 
external turbulence for scattering.  
We do not investigate the CR acceleration mechanism itself.

As interaction with CRs represents a significant source of energy and momentum for the surrounding gas, understanding how they impact the ISM dynamics on galactic scales has been central in recent studies of galaxy evolution. Both analytic models \citep[e.g.][]{Breitschwerdt+91, Everett+08, Dorfi&Breitschwerdt12, Mao&Ostriker18} and magnetohydrodynamical (MHD) simulations of isolated galaxies or cosmological zoom-ins  \citep[e.g.][]{Hanasz2013,Pakmor+16,Ruszkowski+17,Hopkins2021,Werhahn+21} and portions of ISM \citep[e.g.][]{Girichidis+16, Simpson+16, Farber+18, Girichidis+18} have demonstrated that CRs may play an important role in driving galactic outflows, regulating the level of star formation in disks, and shaping the multiphase gas distribution in the circumgalactic medium. However, the degree to which CRs affect these phenomena is strongly sensitive to the way different CR transport mechanisms, i.e.~diffusion, streaming and advection, are treated in the model \citep[e.g.][]{Ruszkowski+17, Chan+19}. 

The uncertainty regarding a fluid prescription for
CR transport is mainly due to the complicated microphysical processes at play and to the consequent difficulty of connecting the microscales comparable to the CR gyroradius, where scattering takes place, to the macroscales of the galaxies. Historically, most studies of CR propagation on galactic scales have focused on our Galaxy and have made use of direct measurements of CR energy density and  abundances of nuclei to constrain the details of the transport process, generally treated via an energy dependent diffusive formalism (e.g.~\citealt{Cummings+16, Guo+16, Johannesson+16, Korsmeier&Cuoco16}, see also review by \citealt{Amato&Blasi18} and references therein). This approach is very effective in representing the observable consequences of CR propagation to reproduce most of the available data 
in great detail. However, the prescriptions for the underlying gas distribution are generally highly simplified, assume spatially-constant CR diffusivity that ignores the multiphase structure of the gas, and often neglect bulk transport via advection and streaming.
These assumptions are certainly inaccurate \citep[e.g.][]{Krumholz+20, Crocker+20, Hopkins2021}. 
Clearly, 
treating the different mechanisms involved in the CR transport as a function of the background gas properties is required for a more physical characterization
of CR propagation on galactic scales and coupling with the surrounding plasma. At the same time, numerical studies of CR-ISM interactions are most meaningful if the ISM treatment accurately represents the physics of the multiphase, magnetized gas (including self-consistent treatment of star formation and feedback) at sufficiently high spatial resolution.  

Beyond ISM dynamics, understanding how CRs propagate within galaxies is also crucial to investigate their effect on the chemistry of the gas.
While CRs with relatively high kinetic energies (a few GeV) interact with the background gas mostly through collisionless processes, CRs with kinetic energies lower than 100~MeV are an important source of collisional ionization and heating of the ISM. 
While their small contribution to the total CR energy density makes low-energy CRs irrelevant to galactic-scale gas dynamics,
they deeply impact the thermal, chemical, and dynamical evolution of the densest regions of the ISM, which are otherwise shielded from ionizing photons \citep[see reviews by][]{Grenier+15, Padovani+20}. In particular, by heating and ionizing the background gas, CRs affect its temperature and  couple it to the magnetic field, respectively. Both these effects are crucial to the internal dynamics of dense molecular clouds, including self-gravitating fragmentation, and as a consequence to the rate and character of star formation.

The goal of this paper is to investigate the propagation of CRs in a galactic environment (mass-containing disk + low-density corona) with conditions typical of the Sun's environment in the Milky Way. For this purpose, we extract a set of snapshots from the TIGRESS\footnote{ Three-phase Interstellar medium in Galaxies Resolving Evolution with Star formation and Supernova feedback} MHD simulation modeling a patch of galactic disk representative of our solar neighborhood \citep{Kim&Ostriker17,Kim&Ostriker18}. For each snapshot, we compute the propagation of CRs depending on the underlying distribution of thermal gas density, velocity, and magnetic field. The advantage of the TIGRESS simulations is that star cluster formation and feedback from supernovae are modeled in a self-consistent manner. This provides
a realistic representation of the multiphase ISM and of the distribution of supernovae -- assumed to be the only source of CRs in our models -- within it. The original 
TIGRESS simulations do not 
include CRs, so in this work we calculate the transport of CRs by post-processing the selected simulation snapshots. The back-reaction of thermal gas and magnetic field to the CR pressure is therefore not directly investigated in this paper.

In this work, we shall consider a variety of models to compute the transport of CRs, from simple models with either constant diffusion or  streaming only, to a more detailed model in which the rate of CR scattering varies with the properties of the background gas in line with the predictions of the self-confinement scenario. These models are separately applied to high-energy ($\sim 1$~GeV) and low-energy ($\sim 30$~MeV) CR protons since their propagation evolves in different ways.  These two energies are chosen as representative of the portion of the CR distribution that is most important for dynamics and for chemistry, respectively. While the former are almost collisionless, the latter undergo more significant kinetic energy losses due to their effective Coulomb interactions with the dense ISM.  Moreover, the growth rate of Alfv\'{e}n waves depends on the CR energy, implying different drift velocities for CRs with different energies. 

The layout of the paper is as follows. In \autoref{Method}, we briefly describe the TIGRESS framework and provide the details of the CR transport models used to infer the distribution of CRs in the solar neighborhood environment modeled by TIGRESS. In \autoref{sec:GeV_const} and \autoref{Gev-Variablesigma}, we analyze the distribution of high-energy CRs
predicted by transport models assuming spatially-constant and variable scattering coefficients, respectively.
In \autoref{Low-EnergyCRs}, we present our results for the distribution of low-energy CRs assuming variable scattering coefficient only. In \autoref{Discussion}, we discuss our work in relation to observational findings and other recent computational work. Finally, in \autoref{Conclusions}, we summarize our main results.

\section{Methods}
\label{Method}

\subsection{MHD simulation}
\label{MHD simulation}

The MHD simulation post-processed in this work is performed with the TIGRESS framework \citep{Kim&Ostriker17}, in which local patches of galactic disks are self-consistently modeled with resolved star formation and supernova feedback. Here, we briefly summarize the relevant features of the simulation, and refer to \citet{Kim&Ostriker17} for a more detailed description.

The TIGRESS framework is built on the grid-based MHD code \textit{Athena} \citep{Stone+08}. The ideal MHD equations are solved in a shearing-periodic box \citep{Stone&Gardiner10} representing a kiloparsec-sized patch of a differentially-rotating galactic disk. This treatment guarantees uniformly high spatial resolution, which is crucial for a realistic representation of the multiphase ISM. For the  study of CR propagation, this is particularly important since transport is quite different in different thermal phases of the gas.  The low-density, hot ISM achieves very high velocity in winds that are escaping from the disk;  the moderate-density, moderate-velocity warm gas fills most of the mid-plane volume and makes up the majority of the ISM mass -- and also participates in  extraplanar fountain flows; and the high-density gas hosts star-forming regions.  The ionization state (which determines wave damping rates) and Alfv\'en speeds (which can limit streaming speeds) are also quite different in the different phases.   

Additional physics in TIGRESS includes self-gravity from gas and young stars, a fixed external gravitational potential representing the stellar disk and the dark matter halo, optically thin cooling, and grain photoelectric heating. Sink particles are implemented to follow the formation of and gas accretion onto
star clusters in regions where gravitational collapse occurs  \citep[see also][for an update in the treatment of sink particle accretion]{Kim+20}. Each sink/star particle is treated as a star cluster with coeval stellar population that fully samples the Kroupa initial mass function \citep{Kroupa01}. Young massive stars (star particle age $t_\mathrm{sp} \lesssim 40$~Myr) provide feedback to the  ISM in the form of far-ultraviolet (FUV) radiation and supernova explosions. The instantaneous FUV luminosity and supernova rate for each star cluster are determined from the \textsc{STARBURST99} population synthesis model \citep{Leitherer+99}.

While TIGRESS simulations for several different galactic environments have been completed \citep{Kim+20}, in this work we analyze the simulation modeling the solar neighborhood environment. This simulation is based on the same model parameters for which resolution studies were presented in \citet{Kim&Ostriker17}, and for which the fountain and wind flows were analyzed in \citet{Kim&Ostriker18} and \citet{Vijayan+20}. This model adopts galactocentric distance $R_\mathrm{0} = 8$~kpc, angular velocity of local galactic rotation $\Omega = 28 \, \kms$~kpc$^{-1}$, shear parameter $d \,\mathrm{ln} \,\Omega / d\, \mathrm{ln}\, R = -1$, and initial gas surface density $\Sigma = 13 \,\moyr$. The simulation we analyze has box size $L_\mathrm{x} = L_\mathrm{y} = 1024$~pc and $L_\mathrm{z} = 7168$~pc with a uniform spatial resolution $\Delta x = 8$~pc. 
While other versions of this model have been run at resolution down to $\Delta x=1$~pc, we choose the present simulation for computational efficiency.  \citet{Kim&Ostriker17, Kim&Ostriker18} demonstrated that a spatial resolution of 8~pc is sufficient to achieve robust convergence of several ISM and outflow properties, and in \autoref{AppendixB} we verify that a resolution of 8~pc guarantees convergence of CR properties as well. We find that models with resolution $\Delta x = 16 $~pc are still converged, while models with lower resolution ($\Delta x \geq 32 $~pc) are not converged in the distribution of CR pressure and are characterized by large temporal fluctuations.

\begin{figure}
\centering
\includegraphics[width=0.48\textwidth]{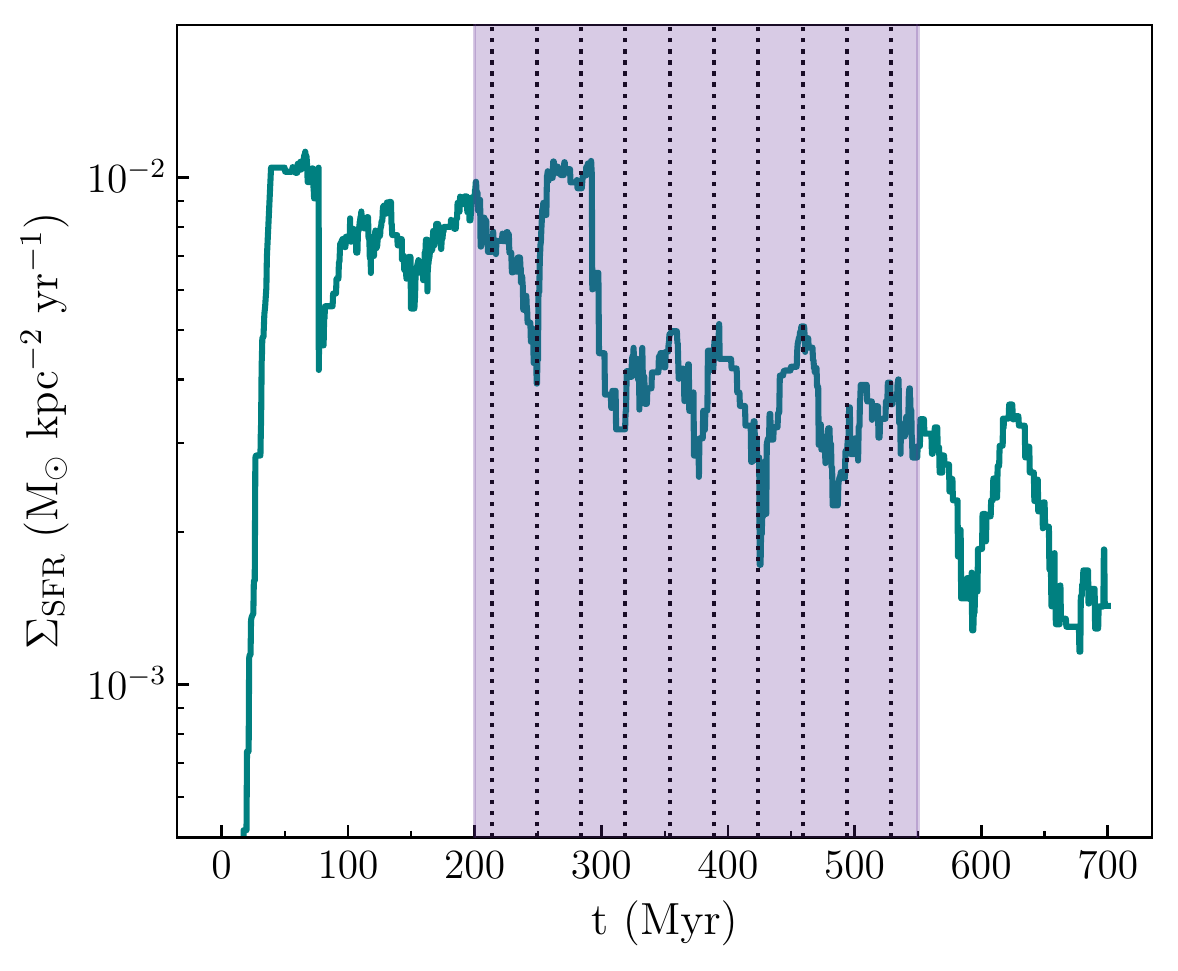}
\caption{Star formation rate per unit area $\Sigma_\mathrm{SFR}$ as a function of time for clusters younger than 40~Myr in the TIGRESS simulation modeling the solar neighborhood environment. The shaded region denotes the time range investigated in this paper. The vertical dotted lines indicate the times of the snapshots post-processed with the CR transport code.}
\label{SFR}
\end{figure}

As discussed in \citet{Kim&Ostriker17}, the TIGRESS simulations (and similar simulations from other groups such as  \citealt{Gatto2017}) are subject to transient effects at early times.
After $t\approx100$~Myr, the system has reached a self-regulated state: feedback from young massive stars drives turbulent motions and heats the ISM, thus providing the turbulent, thermal, and magnetic support needed to offset the vertical weight of the gas. Only a small fraction of the gas collapses to create the star clusters that supply the energy to maintain the ISM equilibrium. Some of the gas that is heated and accelerated by supernova explosions breaks out of the galactic plane into the coronal region, driving multiphase outflows consisting of hot winds and warm fountains. For the present work, we investigate the time range  $200-550$~Myr, covering many star-formation/feedback cycles and outflow/inflow events \citep[see \autoref{SFR} in this paper and Figure 3 in][]{Vijayan+20}. We select and post-process 10 snapshots at equal intervals within this time range (vertical dotted lines in \autoref{SFR}).

\begin{figure*}
\centering
\includegraphics[width=\textwidth]{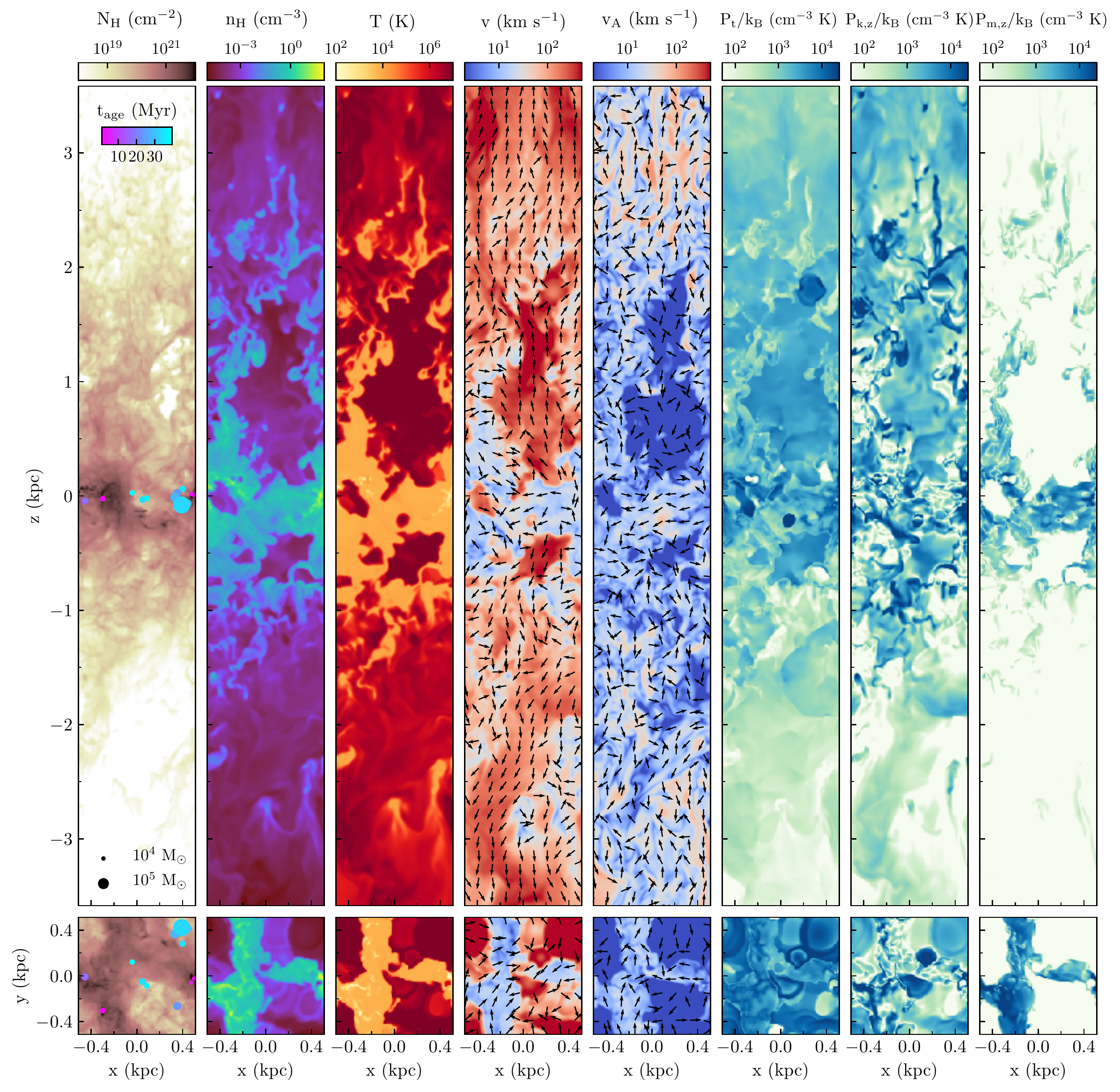}
\caption{Sample snapshot ($t = 286$~Myr) from the TIGRESS simulation modeling the solar environment. The far left panel shows the hydrogen number density projected along the $y$- (\textit{top panel}) and $z$-direction (\textit{bottom panel}). The projected positions of young ($t_\mathrm{sp} \leqslant 40$~Myr) star particles are shown as colored circles, with size and color indicating their mass and mass-weighted age, respectively. Continuing to the right, the panels show the slices through the center of the simulation box of hydrogen number density $n_\mathrm{H}$, gas temperature $T$, gas speed $v$, Alfv\'{e}n speed $v_\mathrm{A}$, thermal pressure $P_\mathrm{t}$, kinetic vertical pressure $P_\mathrm{k,z}$ and magnetic vertical stress $P_\mathrm{m,z}$. The arrows overlaid on the gas velocity and Alfv\'{e}n speed slices indicate the projected directions of the gas velocity and Alfv\'{e}n speed, respectively. The thermal pressure, kinetic pressure and magnetic stress are divided by the Boltzmann constant $k_\mathrm{B} = 1.38 \times 10^{-16}$~erg~K$^{-1}$.}
\label{MHDsim}
\end{figure*}

\autoref{MHDsim} displays the distribution on the grid of several quantities from a sample MHD simulation snapshot at $t=286$~Myr, when a strong outflow driven by supernova feedback is present. The upper (lower) set of panels shows $x-z$ ($x-y$) projections along $\hat y$ ($\hat z$) or slices at $y=0$ ($z=0$). From left to right, the upper/lower panels show: the hydrogen column density $N_\mathrm{H}$ overlaid with the star particle positions, hydrogen number density $n_\mathrm{H}=\rho/(1.4 m_p)$, gas temperature $T$, gas speed $v$ and direction, Alfv\'{e}n speed $v_\mathrm{A} = B / \sqrt{4 \pi \rho}$ and direction, thermal pressure $P_\mathrm{t}$, vertical kinetic pressure $P_\mathrm{k,z} = \rho v_\mathrm{z}^2$, and vertical magnetic stress $P_\mathrm{m,z} = (B_\mathrm{x}^2 + B_\mathrm{y}^2 - B_\mathrm{z}^2)/8 \pi$.  Here, $\rho$ is the gas mass density, $B$ is the magnetic field magnitude, $v_\mathrm{z}$ is the gas velocity in the vertical direction,  and $B_\mathrm{x}$, $B_\mathrm{y}$, $B_\mathrm{z}$ are the magnetic field components along the $x$-, $y$- and $z$-directions, respectively. 

Thermal pressure, vertical kinetic pressure and vertical magnetic stress provide support against the vertical weight of the gas. The arrows in the gas velocity and Alfv\'{e}n speed slices indicate the projected direction of the gas velocity and Alfv\'{e}n speed, respectively. We note that, while $v$ and $v_\mathrm{A}$ are comparable in the warm/cold ($T\lesssim10^4$~K) and moderate/high-density ($n_\mathrm{H} \gtrsim 0.1$~cm$^{-3}$) phase of the gas ($v \simeq v_\mathrm{A} \simeq 10\,\kms$), 
the gas velocity dominates
in the hot ($T > 10^6$~K) and rarefied ($n_\mathrm{H} \lesssim 10^{-3}$~cm$^{-3}$) phase ($v \gg 100\,\kms$ and $v_\mathrm{A} \lesssim 1 \,\kms$). Moreover, while the gas-velocity streamlines are outflowing for the hot gas in the extra-planar region ($\vert z \vert \gtrsim 300$~pc), 
the motions are turbulent within the warm/cold gas.  The magnetic field lines are primarily horizontal near the mid-plane, while aligning more (but not entirely) with the outflow velocities in the extraplanar region. 

\begin{figure*}[t]
\centering
\includegraphics[width=0.9\textwidth]{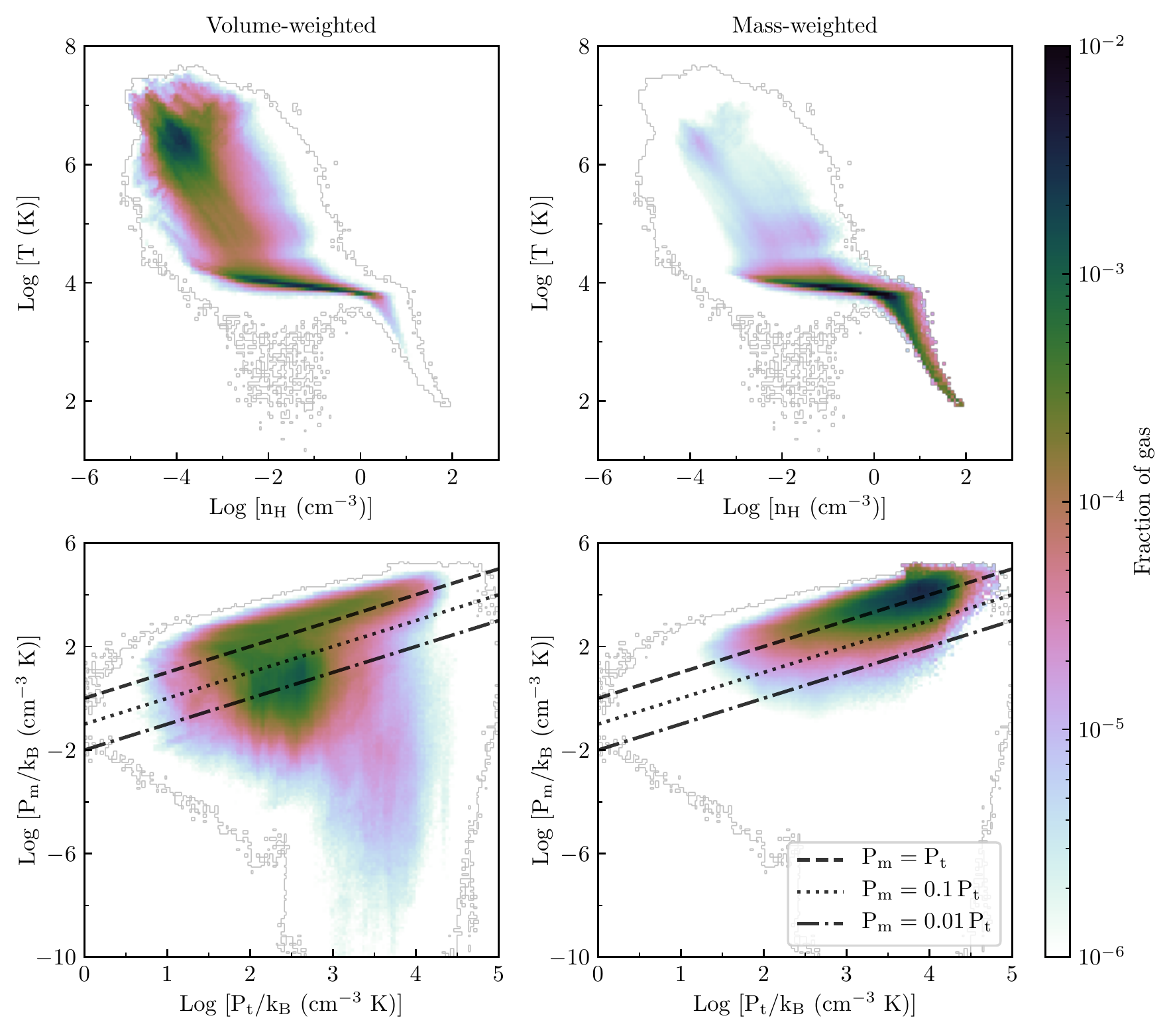}
\caption{Time-averaged gas
distribution in the density $n_\mathrm{H}$ and temperature $T$ phase plane (\textit{top panels}) and in the thermal pressure $P_\mathrm{t}/k_\mathrm{B}$ and magnetic pressure $P_\mathrm{m}/k_\mathrm{B}$ phase plane (\textit{bottom panels}). The distributions are computed as a two-dimensional probability density functions showing the volume (\textit{left panels}) and mass (\textit{right panels}) of gas within each logarithmic bin, normalised by the bin area. The dashed black lines in the bottom panels denote the thermal - magnetic pressure equipartition curve, while the dotted and dot-dashed lines denote the relations $P_\mathrm{m} = 0.1 \,P_\mathrm{t}$ and $P_\mathrm{m} = 0.01 \, P_\mathrm{t}$, respectively.}
\label{fig:PDF}
\end{figure*}

\autoref{fig:PDF} shows the volume-weighted and mass-weighted temperature-density and magnetic pressure-thermal pressure phase diagrams averaged over the 10 selected snapshots. The magnetic pressure is defined as $P_\mathrm{m} = (B_\mathrm{x}^2 + B_\mathrm{y}^2 + B_\mathrm{z}^2)/8 \pi$. The dashed line in the magnetic pressure-thermal pressure diagram denotes equipartition, i.e. plasma $\beta=1$, while the dotted and dot-dashed lines indicate the relations plasma $\beta=10$ and 100, respectively.
The temperature-density diagrams indicate that the hot and warm gas components dominate in terms of volume -- with the former widely distributed in the extra-planar region and the latter located mostly in the galactic disk and in a few clouds/filaments at higher latitudes (see \autoref{MHDsim}). The warm phase dominates the mass, with some contribution from the denser cold phase. The magnetic-thermal pressure diagram shows that, in terms of mass, most of the gas is characterized by rough equality between thermal and magnetic pressure. As clearly visible in \autoref{MHDsim}, thermal and magnetic pressure are nearly in equipartition in the denser portions of  ISM. At higher latitudes, the magnetic pressure decreases much faster than the thermal pressure, especially in regions occupied by rarefied, hot gas. This explains why, in terms of volume, a significant fraction of gas has magnetic pressure well below the equipartition curve, while having moderate  thermal pressure ($P_\mathrm{m}\sim 0.1 - 0.01\, P_\mathrm{t}$ and $\log(P_\mathrm{t}/k_B) \sim 2-3$). The locus with extremely low $P_\mathrm{m}$ and moderately high $P_\mathrm{t}$ in the bottom-left panel represents the interior of superbubbles.

\subsection{Post-processing with cosmic ray}

Each snapshot selected from the MHD simulation is post-processed with the algorithm for CR transport implemented in the \textit{Athena}++ code \citep {Stone+20} by \citet{Jiang&Oh18}. CRs are treated as a relativistic fluid, whose energy and momentum evolution (in the absence of external sources and collisional losses) is described by the following two moment equations:
\begin{equation}
\frac{\partial e_\mathrm{c}}{\partial t} + \mathbf{\nabla} \cdot \mathbf{F_\mathrm{c}} = - (\mathbf{v} + \mathbf{v_\mathrm{s}}) \cdot 
\tensor{\mathrm{\sigma}}_\mathrm{tot} \cdot   [  \mathbf{F_\mathrm{c}} - \mathbf{v} \cdot (\tensor{{\mathbf{P}}}_\mathrm{c} + e_\mathrm{c} \tensor{\mathbf{I}}) ] 
\;,
\label{CRenergy}
\end{equation}
\begin{equation}
\frac{1}{v_\mathrm{m}^2} \frac{\partial \mathbf{F_\mathrm{c}}}{\partial t} + \mathbf{\nabla} \cdot \tensor{\mathbf{P}}_\mathrm{c} = - \tensor{\mathrm{\sigma}}_\mathrm{tot} \cdot [  \mathbf{F_\mathrm{c}} - \mathbf{v} \cdot (\tensor{{\mathbf{P}}}_\mathrm{c} + e_\mathrm{c} \tensor{\mathbf{I}}) ] \;,
\label{CRflux}
\end{equation}
where $e_\mathrm{c}$ and $\mathbf{F_\mathrm{c}}$ are the energy density and energy flux, respectively. We take the CR pressure tensor  
as approximately isotropic in the streaming frame, i.e.  $\tensor{\mathbf{P}}_\mathrm{c} \equiv P_\mathrm{c}\tensor{\mathbf{I}}$,  
with $P_\mathrm{c} = (\gamma_\mathrm{c} -1) \,e_\mathrm{c} = e_\mathrm{c}/3$, where $\gamma_\mathrm{c} = 4/3$ is the adiabatic index of the relativistic fluid, and $\tensor{\mathbf{I}}$ is the identity tensor. With these assumptions, the second term in the square brackets of \autoref{CRenergy} and \autoref{CRflux} becomes $(4/3)\mathbf{v}e_\mathrm{c}$.
These transport equations are supplemented by additional source  and sink terms, to represent 
injection of CR energy from supernovae and collisional losses (see \autoref{Injection} and \autoref{Losses}). 

The CR streaming velocity, 
\begin{equation}
\mathbf{v_\mathrm{s}} = - \mathbf{v_\mathrm{A,i}}
\, \frac{\mathbf{B} \cdot (\nabla \cdot \tensor{\mathbf{P}}_\mathrm{c})}{\vert \mathbf{B} \cdot (\nabla \cdot \tensor{\mathbf{P}}_\mathrm{c})\vert}
= -\mathbf{v_\mathrm{A,i}} 
\frac{\hat{B} \cdot \nabla P_\mathrm{c}}{\vert \hat{B} \cdot \nabla P_\mathrm{c}\vert}
\;,
\label{vs}
\end{equation}
is defined to have the same magnitude as the local Alfv\'{e}n speed in the ions, $v_{\rm A,i} \equiv \mathbf{B}/\sqrt{4\pi\rho_i}$, oriented along the local magnetic field and pointing down the CR pressure gradient. We note that the ion density, $\rho_i$, is the same as $\rho$ for gas that is high enough temperature to be fully ionized (so that $|\mathbf{v}_\mathrm{s}| = v_A$), but is low compared to $\rho$ in the warm/cool gas (so that $|\mathbf{v}_\mathrm{s}| = v_\mathrm{A,i} \gg v_A$); see \autoref{IonFrac}.  

The speed $v_\mathrm{m}$ represents the maximum velocity CRs can propagate in the simulation. In principle, this should be equal to the speed of light. However,  \citet{Jiang&Oh18} demonstrated that the simulation outcomes are not sensitive to the exact value of $v_\mathrm{m}$ as long as $v_\mathrm{m}$ is much larger than any other speed in the simulation; this ``reduced speed of light'' approximation is discussed in the context of two-moment radiation methods in \citet{SkinnerOstriker2013}. Here, we adopt $v_\mathrm{m}=10^4\, \kms$, and, since all our simulations reach a steady state (see below), our results are insensitive to this choice.

The diagonal tensor  $\tensor{\mathbf{\sigma}}_\mathrm{tot}$ encodes the response to particle-wave interactions that cannot be resolved at macroscopic scales in the ISM. Along the direction of the magnetic field, the total coefficient,  
\begin{equation}\label{eq:sigmapar}
    \sigma_{\rm tot,\parallel}^{-1}= \sigma_\parallel^{-1} + \frac{v_\mathrm{A,i}}{|\hat B \cdot \nabla P_\mathrm{c} |} (P_\mathrm{c} + e_\mathrm{c}) \, ,
\end{equation}
allows for both scattering and streaming, while in the directions perpendicular to the magnetic field there is only scattering, 
\begin{equation}\label{eq:sigmaperp}
    \sigma_{\rm tot,\perp}= \sigma_{\perp}\, .
\end{equation}
For the relativistic case, $\sigma_\parallel=\nu_\parallel/c^2$ and $\sigma_\perp=\nu_\perp/c^2$ for $\nu_\parallel$ the scattering rate parallel to $\hat B$ due to Alfv\'en waves that are resonant with the CR gyro-motion 
(see \autoref{sigma}), and $\nu_\perp$ an  effective perpendicular scattering rate.

In \autoref{CRenergy} and \autoref{CRflux}, the left-hand side (LHS) describes the transport of CRs in the simulation frame, while the right-hand side (RHS) represents source and sink terms for the CR energy density or flux. 
In \autoref{CRenergy}, the term $- \mathbf{v} \cdot [ \tensor{\mathrm{\sigma}}_\mathrm{tot} \cdot   ( \mathbf{F_\mathrm{c}} - 4/3 \mathbf{v} e_\mathrm{c}) ]$ 
describes the direct CR pressure work done on or by the gas; in steady state this reduces to $\mathbf{v}\cdot\nabla P_\mathrm{c}$ (and can be either positive or negative). The term $- \mathbf{v_\mathrm{s}} \cdot [ \tensor{\mathrm{\sigma}}_\mathrm{tot} \cdot   ( \mathbf{F_\mathrm{c}} - 4/3 \mathbf{v} e_\mathrm{c}) ]$ represents
the rate of energy transferred to the gas via wave damping; 
in steady state this becomes $ \mathbf{v}_\mathrm{s}\cdot\nabla P_\mathrm{c}$. The term proportional to $\mathbf{v}_\mathrm{s}$ is always negative because CRs always stream down the CR pressure gradient. The RHS terms of \autoref{CRflux} are written as the product of the particle-wave interaction coefficient and the flux evaluated in the rest frame of the fluid. This term asymptotes to zero in the absence of CR scattering  (yielding $\sigma \rightarrow 0$), either because wave damping is extremely strong or because there is no wave growth. 

In steady state, \autoref{CRflux} reduces to the canonical expression for $\mathbf{F_\mathrm{c}}$,
\begin{equation}
\mathbf{F_\mathrm{c}} =   \frac{4}{3}\,e_\mathrm{c}\, (\mathbf{v} + \mathbf{v_\mathrm{s}})  
- \tensor{\mathbf{\sigma}}^{-1} \cdot \nabla P_\mathrm{c}\, ,
\label{SteadyFlux}
\end{equation}
obtained by combining \autoref{CRflux} -- \autoref{eq:sigmaperp}. In this limit, the CR flux can be decomposed into three components: the advective flux $\mathbf{F_\mathrm{a}} = 4/3\,  e_\mathrm{c} \mathbf{v} $, the streaming flux $\mathbf{F_\mathrm{s}} =  4/3\, e_\mathrm{c} \mathbf{v_\mathrm{s}} $, and the diffusive flux $\mathbf{F_\mathrm{d}} =
- \tensor{\mathbf{\sigma}}^{-1} \cdot \nabla P_\mathrm{c}$. 
In the following sections, we analyze the contribution of each of these components to the total flux once the overall CR distribution has reached a steady state, i.e. when $\partial e_\mathrm{c, tot}/ \partial {t} \simeq \partial \mathbf{F_\mathrm{c, tot}}/ \partial {t} \simeq 0$, where $e_\mathrm{c, tot}$ and $\mathbf{F_\mathrm{c, tot}}$ are the energy and flux density integrated over the entire simulation box. In particular, we compare the three components of the CR propagation speed, i.e. gas-advection velocity $v$, Alfv\'{e}n speed $v_\mathrm{A,i}$, and diffusive speed relative to the waves,  defined as 
\begin{equation}\label{eq:vd_def}
  v_\mathrm{d} = \frac{3}{4} \, \frac{\vert \mathbf{F_\mathrm{d}}\vert}{e_\mathrm{c}} = - \frac{1}{4} \tensor{\mathbf{\sigma}}^{-1} \cdot \frac{\nabla e_\mathrm{c}}{e_\mathrm{c}}  \, \, .   
\end{equation}

Below, we explain how we compute some of the terms appearing in \autoref{CRenergy} and \autoref{CRflux}  as well as additional explicit source and sink terms. In particular, in \autoref{Injection} and \autoref{Losses} we describe how injection of CRs from supernova explosion and collisional losses are included in the code through their respective source and sink terms,  
while in \autoref{sigma} we present the different approaches used to calculate the scattering coefficients $\sigma_\parallel$ and $\sigma_\perp$. In \autoref{spectrum} and \autoref{IonFrac}, we show how some quantities relevant for the calculation of the scattering rate are computed. Finally, in \autoref{Summary}, we summarize the models of CR transport explored in this work.

\subsubsection{Cosmic ray injection}
\label{Injection}

For a star cluster particle of mass $m_\mathrm{sp}$ and mass-weighted age $t_\mathrm{sp}$, we calculate the rate of injected CR energy as $\dot{E}_\mathrm{c, sp} = \epsilon_\mathrm{c} \, E_\mathrm{SN} \,\dot{N}_\mathrm{SN}$, where $\epsilon_\mathrm{c}$ is the fraction of supernova energy that goes into production of CRs, $E_\mathrm{SN} = 10^{51}$~erg is the energy released by an individual supernova event, and $\dot{N}_\mathrm{SN} = m_\mathrm{sp} \,\xi_\mathrm{SN} (t_\mathrm{sp})$ is the number of supernovae per unit time. $\xi_\mathrm{SN}$, defined as the number of supernovae per unit time per star cluster mass measured at a given time $t_\mathrm{sp}$, is determined from the \textsc{STARBURST99} code \citep[see][]{Kim&Ostriker17}. 

The injection of CR energy from supernovae enters in the RHS of \autoref{CRenergy} through a source term $Q$. We assume that the injected energy is distributed around each star cluster particle following a Gaussian profile, and, in each cell, we calculate the injected CR energy density per unit time as 
\begin{equation}
Q = \frac{1}{2 \pi \sqrt{2 \pi} \, \sigma_\mathrm{inj}^3 }\,\sum_{\mathrm{sp} =1}^{N_\mathrm{sp}} \dot{E}_\mathrm{c, sp} \cdot \mathrm{exp} (-r_\mathrm{sp}^2/2 \sigma_\mathrm{inj}^2)\;,
\end{equation}
where the sum is taken over all the star cluster particles in the simulation box. $r_\mathrm{sp}$ is the distance between the cell center and the star particle, while $\sigma_\mathrm{inj}$ is the standard deviation of the distribution. We explore different values of $\sigma_\mathrm{inj}$, from $2 \Delta x$ to $10 \Delta x$, and we find that the final CR distribution is almost independent of this choice. 

In most of the CR transport models analyzed in this work, we assume that 10\% of the supernova energy is converted into CR energy 
\citep[$\epsilon_\mathrm{c} = 0.1$, e.g.][]{Morlino&Caprioli12,Ackermann+13}. 
We point out that $e_\mathrm{c}$ linearly scales with $\epsilon_\mathrm{c}$ ($\partial e_\mathrm{c}/ \partial t \propto Q$, where $Q$ does not depend on $e_\mathrm{c}$). 
Therefore, our reported results for CR energy density or pressure could be renormalized to a different fraction of the SN energy injection rate simply by multiplying by $\epsilon_\mathrm{c}/0.1$  
(exceptions are presented in \autoref{Low-EnergyCRs}). 

In the case that the sum of other RHS terms in \autoref{CRenergy} is negligible compared to the injected CR energy density, in steady-state the average flux along the z-direction, $\langle F_\mathrm{c,z} \rangle$, can be written as $0.5 \epsilon_\mathrm{c}  E_\mathrm{SN} \langle \Sigma_\mathrm{SFR} \rangle / m_\star$, where $m_\star = 95.5 M_\odot$ is the total mass of new stars per supernova and $\Sigma_\mathrm{SFR}$ is the star formation rate density. In the TIGRESS simulation analyzed in this paper, the average value of $\Sigma_\mathrm{SFR}$ is $\simeq 4 \times 10^{-3}$~$\moyr$~kpc$^{-2}$ \citep{Kim&Ostriker17}, which, given our assumption $\epsilon_\mathrm{c} = 0.1$, corresponds to $\langle F_\mathrm{c,z} \rangle = 2\times10^{45}$~erg~yr$^{-1}$~kpc$^{-2}$. We note, however, that the average flux can be reduced/increased relative to this by up to a factor $3$ due to the energy transferred to/from the gas (terms on the RHS of \autoref{CRenergy}).

\subsubsection{Energy losses}
\label{Losses}

CRs lose their energy due to collisional interactions with the surrounding gas.  As CR energy losses are proportional to the gas density, the dense ISM is the place where losses are expected to be more significant. Ionization of atomic and molecular hydrogen is the main mechanism responsible for energy losses of CRs with kinetic energies $E_\mathrm{k} \equiv E - m_\mathrm{p}c^2 \lesssim 100$~MeV, with $E$ the total relativistic energy, while losses due to pion production via elastic collisions with ambient atoms are dominant for CRs with kinetic energies $E_\mathrm{k}  \gtrsim 1$~GeV.

Due to collisions with the ambient gas, individual CRs lose energy at a rate
\begin{equation}
\frac{dE}{dt} = - v_\mathrm{p} L(E)  n_\mathrm{H}  
\equiv - \Lambda_\mathrm{coll}(E)  E n_\mathrm{H}
\;,
\label{eq:lossrate}
\end{equation}
where $L(E)$ is the energy loss function, defined as the product of the energy lost per ionization event and the cross section of the collisional interaction \citep[see review by][]{Padovani+20}, and $v_\mathrm{p}$ is the proton velocity,
\begin{equation}
v_\mathrm{p} = c \, \sqrt{1-\left( \frac{m_\mathrm{p} c^2}{E}\right)^2}\,,
\label{eq:proton_velocity}
\end{equation} 
with $m_\mathrm{p}$ the proton mass. 
Considering a population of CRs with different energies, the energy lost per unit time per unit volume, $\Gamma_\mathrm{loss}$, would therefore be 
\begin{equation}
\Gamma_\mathrm{loss} = - n_\mathrm{H} \int 
\Lambda_\mathrm{coll}(E) E
n_\mathrm{c}(E_\mathrm{k})\, d E_\mathrm{k}\;,
\label{Gammaloss}
\end{equation}
where $n_\mathrm{c}(E_\mathrm{k})$ is the number of CRs per unit volume and unit kinetic energy and the integral is evaluated over the entire CR energy spectrum. 

In practice, \autoref{Gammaloss} might be evaluated as a discrete sum over a finite number of energy bins. However, for the calculations performed in this work, we use the so-called `single bin' approximation, i.e. we assume that all CRs are characterized by a single energy $E$.
\autoref{Gammaloss} then becomes
\begin{equation}
\Gamma_\mathrm{loss} 
= - \Lambda_\mathrm{coll} (E)\, n_\mathrm{H} e_\mathrm{c} \;.
\label{Gammaloss2}
\end{equation}
As explained in \autoref{Introduction}, we want to analyze the transport of both CRs with kinetic energies of about 1~GeV, which dominate the CR energy budget and are therefore dynamically important for the surrounding gas, and CRs with kinetic energies of about 30~MeV, which play a fundamental role in the process of gas ionization and heating \citep[e.g.][]{Draine11}.
For this reason, we perform two different sets of simulations: in one set we adopt $\Lambda_\mathrm{coll} = 4 \times 10^{-16}$ cm$^{3}$~s$^{-1}$, representative of CRs with kinetic energies of about 1~GeV, while in the other we adopt $\Lambda_\mathrm{coll}  = 9 \times 10^{-16}$ cm$^{3}$~s$^{-1}$, representative of CRs with kinetic energies of about 30~MeV. The value of the proton loss function at a given energy is extracted from the gray line in Figure~2 of \citet{Padovani+20}, representing the loss function for a medium of pure atomic hydrogen, and multiplied by a factor 1.21, to account for elements heavier than hydrogen. In the following, we will refer to CR protons with $E_\mathrm{k} \simeq 1$~GeV as high-energy CRs and to CR protons with $E_\mathrm{k} \simeq 30$~MeV as low-energy CRs.

Since collisional losses affect not only the energy density of CRs, but also their flux, we update both the RHS of \autoref{CRenergy} and the RHS of \autoref{CRflux} adding the term $\Gamma_\mathrm{E_c, loss} =- \Lambda_\mathrm{coll} (E) n_\mathrm{H} e_\mathrm{c}$ and $\Gamma_\mathrm{F_c, loss}  = - \Lambda_\mathrm{coll} (E) n_\mathrm{H} \mathbf{F}_\mathrm{c}/v_\mathrm{p}^2$, respectively. 

\subsubsection{Scattering coefficient}
\label{sigma}

In \autoref{Introduction}, we have seen that there are two main processes responsible for CR scattering, namely `self-confinement' and `extrinsic turbulence'. In the first scenario, CRs are scattered by  Alfv\`{e}n waves that the CRs themselves excite, while in the second scenario CRs are scattered by the background turbulent magnetic field. The self-confinement mechanism dominates the scattering for CRs with kinetic energies lower than 100 GeV \citep{Zweibel13, Zweibel17}, and it is, therefore, relevant for the range of energies we are interested to study in this paper.

In the CR transport algorithm adopted here, the degree of scattering is parametrized by the scattering coefficients $\sigma_\parallel$ and $\sigma_\perp$ in the CR flux equation (see \autoref{CRflux}). The most common approach that has been adopted in MHD (and HD) simulations is to assume constant values for the scattering coefficients based on empirical estimates in the Milky Way. These estimates are inferred using CR propagation models based on 
analytic 
prescriptions for the gas distribution and/or assuming spatially-constant isotropic diffusion (see \autoref{Introduction} and references therein).
While these models are able to match many observed CR properties, they often neglect a number of factors that may be key for a full understanding of the physics behind the transport of CRs on galactic scales, 
especially the role of advection and local variations of the background gas properties (e.g.~magnetic field structure, gas density, ionization fraction).

In this work, we follow two different general approaches. First, in  \autoref{sec:GeV_const},   we perform simulations with a spatially-constant values for the scattering coefficients. 
While $\sigma_\parallel$ represents the gyro-resonant scattering rate along the local magnetic field direction, $\sigma_\perp$ can be understood as scattering along unresolved fluctuations of the mean magnetic field.
We explore a range of values for $\sigma_\parallel$ going from $10^{-27}$~cm$^{-2}$s to $10^{-30}$~cm$^{-2}$s, where $\sigma_\parallel \sim 10^{-28}-10^{-29}$~cm$^{-2}$ is the scattering coefficient usually adopted for CR protons of a few GeV in simulations of Milky Way-like environments. The range of $\sigma_\parallel$ and $\sigma_\perp$ explored in this work is listed in \autoref{SummaryTable} (see \autoref{Summary}). 
Second, in \autoref{Gev-Variablesigma}, we derive the scattering coefficient $\sigma_\parallel$ in a self-consistent manner based on the predictions of the quasi-linear theory for the growth of Alfv\`{e}n gyro-resonant waves and assuming balance between the rate of wave growth and the rate of wave damping \citep{Kulsrud&Pearce69}. CRs interact with Alfv\`{e}n waves that they themselves drive via resonant streaming instability. 

Given a distribution of CRs that is isotropic in a frame moving at drift speed $v_D$ with respect to the gas velocity along the magnetic field, from \citet{Kulsrud05} the growth rate of resonant Alfv\`{e}n waves in a fully ionized plasma is 
\begin{equation}
\Gamma_\mathrm{stream} (p_1) = \frac{\pi}{4}\frac{\Omega_0 m_p}{\rho}\left( \frac{v_\mathrm{D}}{v_\mathrm{A}} -1   \right)n_1,
\label{GrowthRate1}
\end{equation}
where 
\begin{equation}
n_1 \equiv 4 \pi p_1 \int_{p_1}^\infty p F(p) dp\;.
\label{eq:n1}
\end{equation}
Here, $\Omega_0 = e \vert \bf{B} \vert / (m_\mathrm{p} c)$ is the cyclotron frequency for $e$ the electron charge, $c$ the speed of light, $m_\mathrm{p}$ the proton mass,  and $F(p)$  the CR distribution function in momentum space in the streaming frame (see \autoref{spectrum} for a description of how $F(p)$ is computed in the code).  The momentum $p_1=m_p \Omega_0/k$ is the resonant value for wavenumber $k$.  The momentum $p_1$ corresponds to the component along the magnetic field, i.e. $p_1 = \mathbf{p}\cdot \hat B$ for relativistic momentum $p = [(E/c)^2 - (m_pc)^2]^{1/2}$ and $E= E_k + mc^2$ the total relativistic energy.  In general, the growth rate depends on particle energy since the spectrum enters in $n_1$.  
In \autoref{AppendixA1}, we show how $n_1$ relates to the CR number density $n_\mathrm{c}$ and energy density $e_\mathrm{c}$ for our parameterization of the CR distribution as a broken power law (see \autoref{spectrum}).  
For a pure power law distribution, $\Gamma_{\mathrm{stream}}(p_1) \sim \Omega_0 n_\mathrm{c}(p>p_1)/n_H$ with an order-unity coefficient, i.e. the growth rate at $p_1$ scales with the total number density of CRs with momentum  exceeding $p_1$.

We can also relate the  CR drift velocity to the fluxes as 
$v_\mathrm{D}= (3/4) \,(F_\mathrm{c,\parallel}-F_\mathrm{a,\parallel})/ e_\mathrm{c}$, which in steady state (see \autoref{SteadyFlux}) becomes $v_\mathrm{D}=(3/4) \,(F_\mathrm{s,\parallel}+F_\mathrm{d,\parallel})/ e_\mathrm{c} = v_\mathrm{A} + \vert \mathbf{\hat{B}} \cdot \nabla P_\mathrm{c}\vert\,/ (4 P_\mathrm{c} \sigma_\parallel)$, with $F_\mathrm{c,\parallel}$, $F_\mathrm{a,\parallel}$,  $F_\mathrm{s,\parallel}$ and $F_\mathrm{d,\parallel}$ the components of the total, advective, streaming, and diffusive flux along the magnetic field direction, and $\mathbf{{\hat{B}}}$ the magnetic field direction.  Substituting in for $v_\mathrm{D}/v_\mathrm{A}$ in \autoref{GrowthRate1}, the growth rate can be rewritten as
\begin{equation}
\Gamma_\mathrm{stream} (p_1) = \frac{\pi^2}{4} \frac{\Omega_0  m_\mathrm{p} v_\mathrm{A}}{B^2} \frac{\vert \mathbf{{\hat{B}}} \cdot \nabla  P_\mathrm{c}\vert}{ \sigma_\parallel P_\mathrm{c}}\, n_\mathrm{1} \;.
\label{GrowthRate2}
\end{equation}

The growth of Alfv\`{e}n waves is hampered by damping mechanisms that causes those waves to dissipate. Here, we consider two main damping mechanisms, ion-neutral damping and nonlinear Landau damping. 

The ion-neutral damping arises from friction between ions and neutrals in partially ionized gas. In this regime, Alfv\'en waves propagate only in the ions (nearly decoupled from neutrals) at the scales where wave-particle interaction takes place, since the collision frequency is typically much lower than the frequency of resonant waves. Alfv\'en waves in the ions are damped by collisions with neutrals at a rate \citep{Kulsrud&Pearce69}
\begin{equation}
\Gamma_\mathrm{damp,in} = \frac{1}{2} \frac{n_\mathrm{n} m_\mathrm{n}}{m_\mathrm{n}+m_\mathrm{i}} \langle \sigma v \rangle_\mathrm{in}\;,
\label{DampingRateIN}
\end{equation}
where $n_\mathrm{n}$ is the neutral number density, $m_\mathrm{n}$ is the mean mass of neutrals, $m_\mathrm{i}$ is the mean mass of ions (see \autoref{IonFrac} for the definition of neutral and ion mass and density) and $\langle \sigma v\rangle_\mathrm{in}$ is the rate coefficient for  ion-neutral collisions
\citep[$\sim 3\times10^{-9}$~cm$^3$~s$^{-1}$,][Table 2.1]{Draine11}. 

\autoref{GrowthRate1} is derived under the assumption that the background plasma is fully ionized.   In the decoupled regime, the resonant Alfv\`{e}n waves propagate at the ion Alfv\`{e}n speed $v_\mathrm{A,i} = B / \sqrt{4 \pi \rho_\mathrm{i}}$ -- with $\rho_\mathrm{i}$ the ion mass density -- rather than at the Alfv\`{e}n speed  $v_\mathrm{A} = B / \sqrt{4 \pi \rho}$, which applies either for $\rho \approx \rho_i$ (nearly fully ionized plasma) or for wavelengths at which the neutrals and ions are well coupled \citep[see][]{Plotnikov2021}. 
In \autoref{GrowthRate2}, this can be accounted for with the substitution $v_\mathrm{A} \rightarrow v_\mathrm{A,i}$ to obtain $\Gamma_\mathrm{stream,i}$.  

In the simplest version of the self-confinement scenario \citep{Kulsrud&Pearce69,Kulsrud&Cesarsky1971}, it is assumed that wave growth and damping balance.  Setting
 $\Gamma_\mathrm{stream,i} = \Gamma_\mathrm{damp,in}$, the parallel scattering coefficient becomes
\begin{equation}
\sigma_\mathrm{\parallel,in} (p_1) = \frac {\pi}{8} \, \frac{\vert \mathbf{\hat{B}} \cdot \nabla  P_\mathrm{c}\vert}{v_\mathrm{A,i} P_\mathrm{c}}  \frac{\Omega_0}{ n_\mathrm{n}\langle \sigma v \rangle_\mathrm{in}} \, \frac{m_\mathrm{p} (m_\mathrm{n} + m_\mathrm{i})}{ m_\mathrm{i}  m_\mathrm{n}}  \frac{n_1}{n_\mathrm{i}} \;.
\label{IN}
\end{equation}

The nonlinear Landau damping occurs when thermal ions have a Landau resonance with the beat wave formed by the interaction of two resonant Alfv\`{e}n waves. The rate of nonlinear Landau damping is \citep{Kulsrud05}
\begin{equation}
\Gamma_\mathrm{damp,nll} =0.3 \, \Omega \, \frac{v_\mathrm{t,i}}{c} \left(\frac{\delta B}{B}\right)^2 \;,
\label{DampingRateNLL}
\end{equation}
where $\Omega = \Omega_0 / \gamma (p_1)$ is the relativistic cyclotron frequency, with $\gamma$ the Lorentz factor of CRs with momentum $p_1$, $v_\mathrm{t,i}$ is the ion thermal velocity (which we set equal to the gas sound speed), and ${\delta B}/{B}$ is the magnetic field fluctuation at the resonant scale. The quasi-linear theory predicts that the  scattering rate is $\nu_s \sim \Omega ({\delta B}/{B})^2$, while the scattering coefficient is $\sigma_\parallel  \sim \nu_s/v_\mathrm{p}^2 \sim \Omega ({\delta B}/{B})^2 /v_\mathrm{p}^2$ so that $\Gamma_\mathrm{damp,nll} =0.3 (v_\mathrm{t,i} v_\mathrm{p}^2/c)\sigma_\mathrm{\parallel}$. Again assuming $\Gamma_\mathrm{stream} = \Gamma_\mathrm{damp,nll}$ for self-confinement, the parallel scattering coefficient becomes
\begin{equation}
\sigma_\mathrm{\parallel,nll} (p_1) = \sqrt{\frac{\pi}{16} \, \frac{\vert \mathbf{{\hat{B}}} \cdot \nabla P_\mathrm{c}\vert}{v_\mathrm{A,i} P_\mathrm{c}} \frac{\Omega_0 c}{ 0.3 v_\mathrm{t,i} v_\mathrm{p}^2} \frac{m_\mathrm{p}}{m_\mathrm{i}} \frac{n_\mathrm{1}}{n_\mathrm{i}}}\;
\label{NLL}
\end{equation}
for nonlinear Landau damping.\footnote{Strictly speaking, the wave energy growth rate is $2 \Gamma_\mathrm{stream,i}$,  while the theoretical scattering rate coefficient is $(\pi/8) (\delta B/B)^2 \Omega$; taken together this would introduce a factor $0.8$ inside the square root of \autoref{NLL}.}
In the code, the local scattering coefficient is set by the damping mechanism that contributes the most to the Alfv\`{e}n wave dissipation, i.e. $\sigma_\parallel$ is equal to the minimum between the results of \autoref{IN} and \autoref{NLL}. In \autoref{Gev-Variablesigma} and \autoref{Low-EnergyCRs}, we see that the ion-neutral damping mechanism dominates in the cooler and denser portions of the ISM, while the nonlinear Landau damping mechanism dominates in the hot and ionized phase of the gas. 

\subsubsection{CR spectrum}
\label{spectrum}
In this section, we explain how we compute the distribution function of CR protons in momentum space, $F(p)$, relevant for the calculation of $n_\mathrm{1}$ in \autoref{IN} and \autoref{NLL}. $F(p)$ is related to the number of CRs per unit volume and unit energy $n_\mathrm{c}(E_\mathrm{k})$ as
\begin{equation}
F(p) = \frac{n_\mathrm{c}(E_\mathrm{k})}{4 \pi p^2} \frac{dE_\mathrm{k}}{dp}\;.
\end{equation}
In turn, $n_\mathrm{c}(E_\mathrm{k})$ can be written as a function of the CR energy-flux spectrum $j(E_\mathrm{k})$ as $n_\mathrm{c}(E_\mathrm{k}) = j(E_\mathrm{k})/v_\mathrm{p}$. Here, we adopt the spectrum of CR protons proposed by \citet{Padovani+18} for the solar-neighborhood,
\begin{equation}
j(E_\mathrm{k}) = C  \frac{E_\mathrm{k}^{\delta}}{(E_\mathrm{k}+E_\mathrm{t})^{2.7+\delta}}~\mathrm{eV}^{-1}~\mathrm{cm}^{-2}~\mathrm{s}^{-1}\;,
\label{CRspectrum}
\end{equation}
where the adopted value for $E_\mathrm{t}$ is 650 MeV. The high-energy slope of this function, $-2.7$, is well determined \citep[e.g.][]{Aguilar+14, Aguilar+15}, while the low-energy slope $\delta$ is uncertain. A simple extrapolation of the Voyager~1 data down energies of 1~MeV predicts $\delta \approx 0.1$ \citep{Cummings+16}. However, a slope $\delta \approx 0.1$ fails to reproduce the CR ionisation rate measured in local diffuse clouds ($n\approx100$~cm$^{-3}$, $T\approx 100$~K) from $\rm{H_3^+}$ emission \citep[e.g.][]{Indriolo&McCall12}. \citet{Padovani+18} found that the low-energy slope required to reproduce the observed CR ionisation rate at the edges of molecular clouds must rise towards low energy, with best fit $\delta = -0.8$ . 
The authors however noticed that the average Galactic value of $\delta$ is likely to lie between $-0.8$ and 0.1. 
In fact, $\delta$ is expected to increase (spectral flattening) within clouds as low-energy CRs preferentially lose energy ionizing  and heating the ambient gas (see \autoref{Losses}). 

In this work, we adopt two different approaches for the calculation of $j(E_\mathrm{k})$ (\autoref{CRspectrum}) depending on whether we model the propagation of high-energy or low-energy CRs. In simulations of high-energy CRs, we adopt a spatially-constant value of $\delta$. We explore three values of the low-energy slope: $\delta = -0.35$ (default simulation), $\delta = 0.1$ and $\delta = -0.8$ (the results of these two cases are discussed in \autoref{AppendixA3}). The normalization factor $C$ is evaluated in each cell depending on the local value of the CR energy density. Since CRs with kinetic energies of about 1~GeV dominate the total-energy budget of CRs with kinetic energy above $E_\mathrm{t}$, we can assume $e_\mathrm{c} \simeq \int_{E_t}^\infty E n_\mathrm{c}(E_\mathrm{k}) dE_\mathrm{k} $ for the high-energy CRs. In any given cell, $C$ can then be calculated as
\begin{equation}
C = {e_\mathrm{c}(\rm{GeV})} \left( {\int_{E_t}^\infty \frac{E E_\mathrm{k}^{\delta}}{v_\mathrm{p} \,(E_\mathrm{k}+E_\mathrm{t})^{2.7+\delta}} \,dE_\mathrm{k}}\right)^{-1}~\frac{\mathrm{eV}^{1.7}}{\mathrm{cm^2}~\mathrm{s}}\;
\label{SpecNorm}
\end{equation}
where $e_\mathrm{c} (\rm{GeV})$ is from the high-energy CRs. The value of $C$ is then used in normalizing the spectrum which is input to the scattering rate (\autoref{sigma}) as well as the CR ionization rate (\autoref{IonFrac}) calculations.

In simulations of low-energy CRs, we instead calculate the local value of $\delta$ based on the local energy density of both low-energy and high-energy CRs. For the low-energy CRs, $e_\mathrm{c} = E n_\mathrm{c} (E_\mathrm{k}) dE_\mathrm{k}$ represents the energy density of CRs with kinetic energy between $E_\mathrm{k} - dE_\mathrm{k}/2$ and $E_\mathrm{k} + dE_\mathrm{k}/2$, where we adopt $E_\mathrm{k}=30$~MeV and an energy width bin $dE_\mathrm{k}$ equal to 1~MeV.  
We then calculate the low-energy slope of the CR spectrum as
\begin{equation}
\delta = \frac{\mathrm{log}\left(e_\mathrm{c}(\mathrm{MeV})/C) + \log(v_\mathrm{p} \,(E_\mathrm{k}+E_\mathrm{t})^{2.7}\right) - \mathrm{log}(E dE_\mathrm{k})}
{\mathrm{log}\,(E_\mathrm{k}) - \mathrm{log}\,(E_\mathrm{k} + E_\mathrm{t})} \;,
\label{delta_spectrum}
\end{equation}
where now $e_\mathrm{c}(\rm{MeV})$, $v_\mathrm{p}$, $E_\mathrm{k}$, and $E$ refer to the low-energy CRs, while the value of $C$ is taken from the corresponding default simulation of high-energy CRs. This is possible because the kinetic energy is mainly contained in the higher-energy portion of the spectrum in \autoref{CRspectrum}, so for a given  total CR energy input rate (taken as 10\% of the SN energy) the normalization constant $C$  is nearly independent of $\delta$ 
for the range we consider (see also \autoref{AppendixA3}, where we show that the pressure of high-energy CRs is almost independent of the adopted $\delta$).
Since $C$ is proportional to $e_\mathrm{c}({\rm GeV})$, 
$\delta$ from \autoref{delta_spectrum} depends on the relative energy
deposited in high- and low-energy CRs, but not on the absolute level.  
For the high-energy CRs, we assume that a fraction
$\epsilon_\mathrm{c} (\rm{GeV}) = 0.1$ of 
the SN energy input rate is deposited at  $E_\mathrm{k} \gtrsim E_\mathrm{t}$. For the low-energy CRs, we must make an assumption about the CR injection spectrum in order to calculate the corresponding
energy deposition fraction $\epsilon_\mathrm{c} (\rm{MeV})$.  We explore three different values of the low-energy slope of the injection spectrum: $\delta_\mathrm{inj} = 0.1$, $\delta_\mathrm{inj} = -0.35$, and $\delta_\mathrm{inj} = -0.8$.
For these values of $\delta$, the fractions of CRs with $E_\mathrm{k}=30\pm 1/2$~MeV are $0.005$, $0.02$ and $0.07$, corresponding to $\epsilon_\mathrm{c}  = 5\times10^{-4}$, $2\times10^{-3}$ and $7\times10^{-3}$, respectively.

\subsubsection{CR ionization rate and ionization fraction}
\label{IonFrac} 

In this section, we explain how the ion and neutral densities are calculated in \autoref{IN} and \autoref{NLL}. The ion number density is calculated as $n_\mathrm{i} = x_\mathrm{i} n_\mathrm{H}$, where the hydrogen number density $n_\mathrm{H}$ is an output of the MHD simulation and $x_\mathrm{i}$ is the ion fraction. For gas at $T>2 \times 10^4$~K, the ion faction is calculated from the values tabulated by \citet{Sutherland&Dopita93}, while, for gas at $T \leqslant 2\times 10^4$~K, the ion fraction is calculated as \citep{Draine11}
\begin{equation}
x_\mathrm{i} = x_\mathrm{M} + \frac{\sqrt{ \left( \beta + \chi + x_\mathrm{M}\right)^2 +4\beta} - \left( \beta + \chi + x_\mathrm{M}\right)}{2}\;,
\label{EqIonFrac}
\end{equation}
where $x_\mathrm{M} = 1.68\times10^{-4}$ is adopted for the ion fraction of species with ionization potential $<13.6$~eV (the largest contributor from the metals is $C^+$), while the second term on the RHS is the fraction of ionized hydrogen $x_\mathrm{H^+}$.~In \autoref{EqIonFrac}, $\beta$ is defined as $\zeta_\mathrm{H}/(\alpha_\mathrm{rr} n_\mathrm{H})$, where $\zeta_\mathrm{H}$ is the CR ionization rate per hydrogen atom and $\alpha_\mathrm{rr}=1.42\times10^{-12}$~cm$^{3}$~s$^{-1}$ is adopted for the rate coefficient for radiative recombination of ionized hydrogen, while $\chi$ is defined as $\alpha_\mathrm{gr}/\alpha_\mathrm{rr}$, where $\alpha_\mathrm{gr} = 2.83\times10^{-14}$~cm$^{3}$~s$^{-1}$ is adopted for the grain-assisted recombination rate coefficient.  Note that we have chosen this value to be representative of the cold  neutral medium ($T \simeq 100$~K, $n_\mathrm{H} \simeq 10-100$~cm$^{-3}$), rather than the warm neutral medium ($T \simeq 10^4$~K, $n_\mathrm{H} \simeq 0.1-1$~cm$^{-3}$), where $\alpha_\mathrm{gr}$ is actually smaller. The reason is that $x_\mathrm{i} \approx \beta^{1/2}$ at the typical densities of the warm medium ($4 \beta \gg \beta + \chi +x_\mathrm{M}$) and changing the value of $\alpha_\mathrm{gr}$ marginally affects the value of $x_\mathrm{i}$. 
For warm gas (most of the neutrals), 
the ion fraction can be approximated as $x_i=0.008 (\zeta_\mathrm{H}/10^{-16}~{\rm s}^{-1})^{1/2}~(n_\mathrm{H}/1~{\rm cm}^{-3})^{-1/2}$. Given the CR ionization rate per hydrogen atom of $\sim 3\times 10^{-16}~{\rm s}^{-1}$ measured in local diffuse clouds, the ion number density at the average densities of the local ISM ($n_\mathrm{H} \simeq 0.1-1$~cm$^{-3}$) is $\sim 0.02$~cm$^{-3}$.

The CR ionization rate per atomic hydrogen $\zeta_\mathrm{H}$ accounts for ionization due to CR nuclei and secondary electrons produced by primary ionization events. It can be approximated as $\zeta_\mathrm{H} = 1.5 \, \zeta_\mathrm{c}$, where $\zeta_\mathrm{c}$ is the ionization rate per atomic hydrogen due to nuclei only (primary ionization rate), and it is calculated as 
\begin{equation}
\zeta_\mathrm{c} = \int_{E_\mathrm{k,min}}^{E_\mathrm{k,max}} \frac{v_\mathrm{p} \,n_\mathrm{c}(E_\mathrm{k}) L_\mathrm{ion}(E_\mathrm{k})}{\epsilon} dE_\mathrm{k}
\label{CRionrate}
\end{equation}
\citep{Padovani+20}. In \autoref{CRionrate}, $n_\mathrm{c}(E_\mathrm{k})$ is computed as explained in \autoref{spectrum}, $\epsilon \approx 50$~eV is the average energy lost by each proton per ionization event and $L_\mathrm{ion}(E_\mathrm{k})$ is the proton loss function due to hydrogen ionization. We adopt the power-law approximation proposed by \citet{Silsbee+19},
\begin{equation}
L_\mathrm{ion}(E_\mathrm{k}) = L_\mathrm{0} \left(\frac{E_\mathrm{k}}{E_0}\right)^{-0.82}
\label{lossfunction}
\end{equation} 
where $L_0 = 1.27\times10^{-15}$~eV~cm$^2$ and $E_0=1$~MeV. \autoref{lossfunction} holds over the range of kinetic energies between $10^5$ and $10^9$~eV, where CR losses due to ionization of atomic and molecular hydrogen are relevant (see also \autoref{Losses}). 
The minimum kinetic energy for CRs, $E_\mathrm{k,min}$, is unknown since Voyager~1 does not probe energies below 1 MeV. We therefore assume, following \citet{Padovani+18}, that the lower limit of the integral in \autoref{CRionrate} is $E_\mathrm{k,min} = 10^5$~eV. The upper limit is $E_\mathrm{k,max} = 10^9$~eV as Coulomb losses are negligible above that density. In \autoref{AppendixA2}, we show how the value of $\zeta_\mathrm{c}$ depends on the low-energy slope of the spectrum, on the CR pressure through the normalization factor C (\autoref{SpecNorm}), and on the choice of $E_\mathrm{k,min}$.

From $n_\mathrm{i}$, we compute the ion mass density -- relevant for the calculation of the ion Alfv\'{e}n speed -- as $\rho_{i} = \mu_\mathrm{i} m_\mathrm{p} n_\mathrm{i}$, where $\mu_\mathrm{i}$ is the ion mean molecular weight. For gas at $T > 2 \times 10^4$~K, we adopt $\mu_\mathrm{i} \approx 2 \, \mu$, where $\mu$ is the total mean molecular weight tabulated by \citet{Sutherland&Dopita93} as a function of temperature. For gas at $T \leqslant 2 \times 10^4$~K, we calculate the ion mean molecular weight as $\mu_\mathrm{i} = (x_\mathrm{H^+} m_\mathrm{p} + x_\mathrm{M} m_\mathrm{M})/(x_\mathrm{i}m_\mathrm{p})$, with $m_\mathrm{M} \approx 12\, m_\mathrm{p}$ the mean ion mass of species with ionization potential larger than 13.6~eV. 

Finally, in \autoref{IN}, we calculate the neutral mass density as $n_\mathrm{n} m_\mathrm{n} = \rho - \rho_\mathrm{i}$ and the mean ion mass as $m_\mathrm{i} = \mu_\mathrm{i} m_\mathrm{p}$. Moreover, we assume that the mean neutral mass is $m_\mathrm{n} \approx 2 \, m_\mathrm{p}$ for gas at  $T < 100$~K, where hydrogen is predominantly in molecular form, and $m_\mathrm{n} \approx m_\mathrm{p}$ for gas at  $100 \leqslant T \leqslant 2 \times 10^4$~K, where hydrogen is predominantly in atomic form. 

\subsubsection{Summary of CR transport models}
\label{Summary}

The algorithm for CR propagation presented in the previous sections is applied to the 10 snapshots selected from the TIGRESS simulation modeling the solar neighborhood environment (see \autoref{MHD simulation}). The energy and flux densities of CRs are evolved through space and time according to \autoref{CRenergy} and \autoref{CRflux}, while the background MHD quantities are frozen in time. We stop and analyze the simulations once the overall distributions of CR energy density has reached a steady state, i.e. $(e_\mathrm{c,tot} (t) - e_\mathrm{c,tot} (t - 0.1 \,\mathrm{Myr}))/e_\mathrm{c,tot} (t) < 10^{-6}$, with $e_\mathrm{c,tot} = \int_\mathrm{Vol} e_\mathrm{c} dx^3$.

Our goal is to explore the predictions of different models of CR propagation, and we therefore consider several different models in which the parameters are treated differently. The models explored in this work are listed in \autoref{SummaryTable}.
We separately investigate the propagation of high-energy ($E_\mathrm{k} \sim 1$~GeV) and low-energy ($E_\mathrm{k} = 30$~MeV) CRs adopting two different values of $\Lambda_\mathrm{coll}$, the rate coefficient for collisional losses (see \autoref{Gammaloss2}). 

First, in \autoref{sec:GeV_const}, we consider high-energy CRs with spatially-constant scattering coefficients. We consider propagation models with (1) only diffusion ($v_\mathrm{s} = 0$, $v = 0$), (2) only streaming ($\sigma_\parallel = \sigma_\perp \gg 1$, $v = 0$), (3) both diffusion and streaming but no  advection ($v = 0$), and (4) diffusion, streaming and advection. For the latter two cases we explore different combinations of spatially-constant $\sigma_\parallel$ and $\sigma_\perp$. Note that we set the streaming speed to to the magnitude of the ideal Alfv\'{e}n speed, $v_A=B/(4\pi \rho)^{1/2}$ for $\rho$ the total gas density, and, in models without advection, we neglect the effect of collisional losses setting $\Lambda_\mathrm{coll} = 0$.

Second, in \autoref{Gev-Variablesigma} we consider physically-motivated models  (including diffusion, streaming, and advection) in which $\sigma_\parallel$ 
varies based on the local CR pressure and gas properties (see \autoref{sigma} for details). In these models, 
the streaming velocity is set to $v_\mathrm{A,i}$. 
Calculating the scattering coefficient in a self-consistent manner requires making an assumption for the low-energy slope of the CR energy spectrum $\delta$ (see \autoref{CRspectrum}), since $\sigma_\mathrm{\parallel}$ depends on the ionization fraction $x_i$, and $x_i$ in warm/cold gas depends on the ionization rate produced by low-energy CRs. Here, we consider three different values of $\delta$. Also, we model the propagation of CRs either in the absence (we set $\sigma_\perp \gg 1$) or in the presence of diffusion perpendicular to the magnetic field direction. For the latter case, we consider either isotropic ($\sigma_\perp = \sigma_\parallel$) or anisotropic diffusion (with $\sigma_\perp = 10\,\sigma_\parallel$).

For low-energy CRs, in \autoref{Low-EnergyCRs} we investigate propagation models with variable scattering coefficient only. All models include streaming, advection, and diffusion parallel to the magnetic field direction. We explore the effect of three different assumptions for the low-energy slope of the CR injection spectrum $\delta_\mathrm{inj}$, which entails different fractions of supernova energy going into production of low-energy CRs. 

\begin{table}
\caption{List of CR transport models}
\setlength{\tabcolsep}{4pt}
\centering
\begin{tabular}{ccccccc}
\multicolumn{7}{c}{\textbf{High-energy CRs}} \\
\multicolumn{7}{c}{($E_\mathrm{k} \simeq 1$~GeV, $\Lambda_\mathrm{coll} = 4 \times 10^{-16}$ cm$^{3}$~s$^{-1}$)}\\
\noalign{\vspace{5pt}}\hline\hline\noalign{\vspace{3pt}}
\multicolumn{7}{l}{\textbf{1}. Diffusion only, $\sigma_\parallel = 10^{-28}$~cm$^{-2}$~s, $\sigma_\perp = 10 \, \sigma_\parallel$, $\Lambda_\mathrm{coll} = 0$}\\
\noalign{\vspace{3pt}}\hline\noalign{\vspace{3pt}}
\multicolumn{7}{l}{\textbf{2}. Streaming only, $\vert v_\mathrm{s} \vert = \vert v_\mathrm{A} \vert$, $\Lambda_\mathrm{coll} = 0$}\\
\noalign{\vspace{3pt}}\hline\noalign{\vspace{3pt}}
\multicolumn{7}{l}{\textbf{3}. Diffusion and streaming, $\vert v_\mathrm{s} \vert = \vert v_\mathrm{A} \vert$, $\Lambda_\mathrm{coll} = 0$}\\
\noalign{\vspace{1pt}}\noalign{\vspace{1pt}}
$\sigma_\parallel$~(cm$^{-2}$~s)& $10^{-27}$ & $10^{-28}$& $10^{-28}$& $10^{-28}$& $10^{-29}$& $10^{-30}$\\
$\sigma_\perp$~(cm$^{-2}$~s) & $10^{-26}$ & $10^{-27}$& $10^{-28}$ & $10^{-29}$& $10^{-28}$& $10^{-29}$\\
\noalign{\vspace{3pt}}\hline\noalign{\vspace{3pt}}
\multicolumn{7}{l}{\textbf{4}. Diffusion, streaming and advection, $\vert v_\mathrm{s} \vert = \vert v_\mathrm{A} \vert$}\\
\noalign{\vspace{1pt}}\noalign{\vspace{1pt}}
$\sigma_\parallel$~(cm$^{-2}$~s)& $10^{-27}$ & $10^{-28}$& $10^{-29}$& $10^{-28}$& & \\
$\sigma_\perp$~(cm$^{-2}$~s) & $10^{-26}$ & $10^{-27}$& $10^{-28}$& $-$& &\\
\noalign{\vspace{3pt}}\hline\noalign{\vspace{3pt}}
\multicolumn{7}{l}{\textbf{5}. Self-consistent model, variable $\sigma_\parallel$, $\vert v_\mathrm{s} \vert = \vert v_\mathrm{A,i} \vert$}\\
\noalign{\vspace{1pt}}\noalign{\vspace{1pt}}
$\delta$& $-0.35$& $-0.35$& $-0.35$& $0.1$& $-0.8$ &$-0.8$ \\
$\sigma_\perp$& $-$& $10 \sigma_\parallel$& $\sigma_\parallel$ & $-$&$-$& $10 \sigma_\parallel$\\

\noalign{\vspace{3pt}}\hline\hline\noalign{\vspace{15pt}}
\multicolumn{7}{c}{\textbf{Low-energy CRs}}\\
\multicolumn{7}{c}{($E_\mathrm{k} = 30$~MeV, $\Lambda_\mathrm{coll} = 9 \times 10^{-16}$ cm$^{3}$~s$^{-1}$)}\\
\noalign{\vspace{5pt}}\hline\hline\noalign{\vspace{3pt}}
\multicolumn{7}{l}{\textbf{1}. Self-consistent model, variable $\sigma_\parallel$, $\vert v_\mathrm{s} \vert = \vert v_\mathrm{A,i} \vert$}\\
\noalign{\vspace{2pt}}\noalign{\vspace{1pt}}
$\delta_\mathrm{inj}$& $-0.8$& $-0.35$ & $-1.0$& & &\\
\noalign{\vspace{5pt}}\hline\hline\noalign{\vspace{3pt}}
\end{tabular} 
\label{SummaryTable}
\end{table}

\section{High-energy cosmic rays:\\ models with spatially-constant scattering coefficient}\label{sec:GeV_const}

In this section, we consider CR transport models in which the scattering rate coefficient is set to a spatially-constant value. This is helpful for gauging the effects of different values of $\sigma$, and also useful for making contact to the many works in the literature that have adopted spatially-constant $\sigma$.


\subsection{Models without advection}
\label{NoAdv}

\begin{figure*}
\centering
\includegraphics[width=0.8\textwidth]{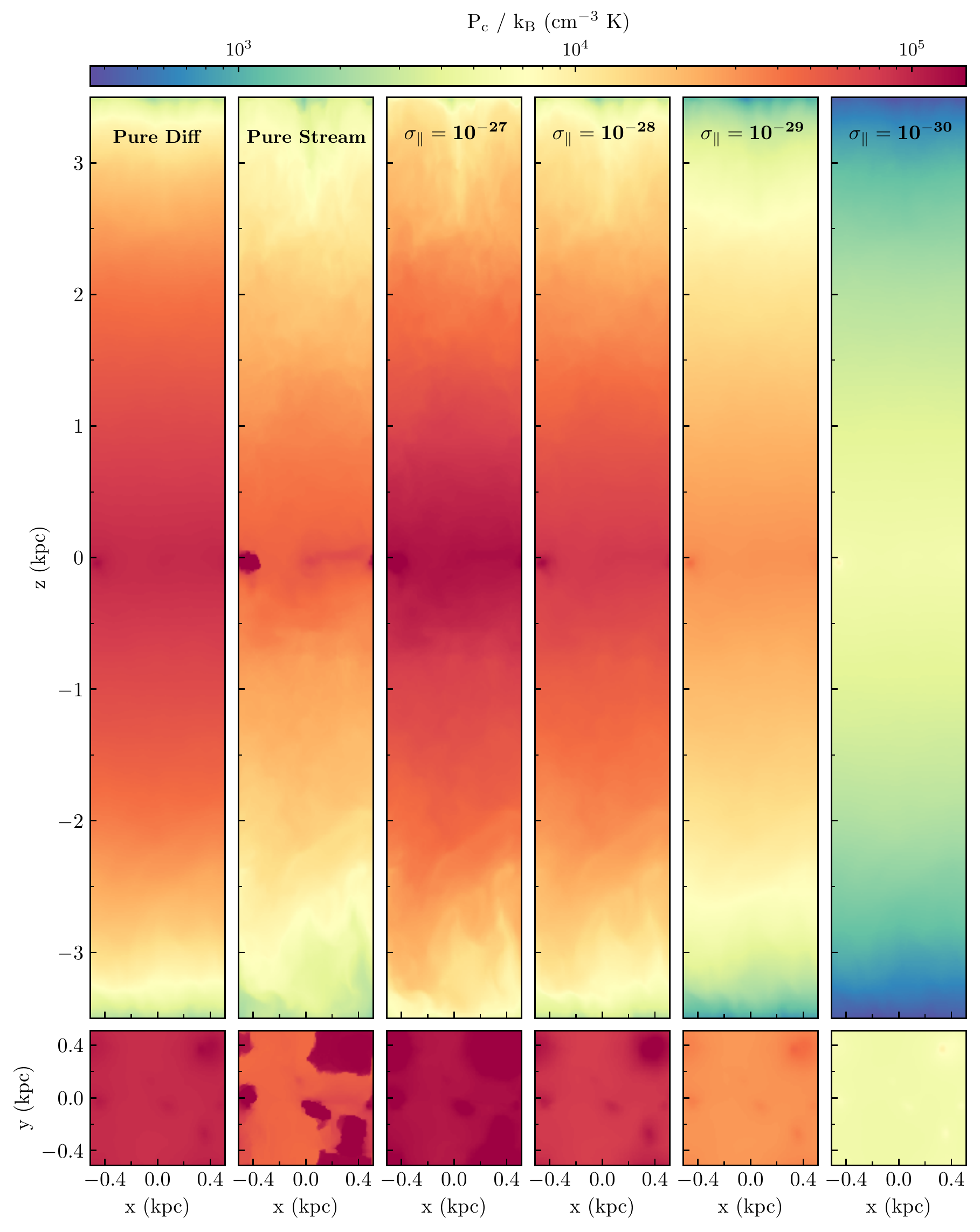}
\caption{Distribution on the grid of the CR pressure for different models of CR transport neglecting advection, showing $y=0$ slices of  $P_\mathrm{c}$. The first two panels on the left are models with pure CR diffusion ($\sigma_\parallel = 10^{-28}$~cm$^{-2}$~s) and pure CR streaming, respectively. The remaining panels are for models with both CR diffusion and streaming. These  adopt different values of $\sigma_\parallel$ (from left to right: $10^{-27}$~cm$^{-2}$~s, $10^{-28}$~cm$^{-2}$~s, $10^{-29}$~cm$^{-2}$~s and $10^{-30}$~cm$^{-2}$~s) and $\sigma_\perp$ ($\sigma_\perp = 10 \,\sigma_\parallel$). The $t = 286$~Myr TIGRESS snapshot is used (see \autoref{MHDsim}).
}
\label{fig:noadv_snaps}
\end{figure*}

We start with the analysis of CR transport models neglecting advection. These models have been applied to a single TIGRESS snapshot ($t=286$~Myr, \autoref{MHDsim}) only, rather than to the full set of 10 snapshots. \autoref{fig:noadv_snaps} shows the distribution on the grid of CR pressure predicted by the different models. The first two panels on the left refer to the models assuming pure diffusion and pure streaming, respectively. In the model with pure diffusion, $\sigma_\parallel$ is chosen to be $10^{-28}$~cm$^{-2}$~s. The other models include both diffusion and streaming and are performed with different values of  $\sigma_\parallel$, from $10^{-27}$~cm$^{-2}$ to $10^{-30}$~cm$^{-2}$. An immediate conclusion from \autoref{fig:noadv_snaps} is that in the absence of advection, regardless of the CR propagation model,  the distribution of CR pressure is very smooth across the grid compared to the distribution of the magneto-hydrodynamical quantities shown in \autoref{MHDsim}. The model with pure streaming and, to a lesser extent, the models with relatively high scattering coefficient predict a higher CR pressure in  proximity to CR injection sites (see distribution of young star clusters in \autoref{MHDsim}). Streaming of CRs is quite ineffective within expanding supernova bubbles, where the magnetic field is chaotic and the Alfv\'{e}n speed is extremely low ($\ll 1\, \kms$).  For the same reason, a steady state is not reached before 1~Gyr in the simulation accounting for CR streaming only. Diffusion is clearly crucial for spreading CRs beyond their injection sites. Also evident from  \autoref{fig:noadv_snaps}, and consistent with expectations, is that the CR pressure decreases at higher $\sigma_\parallel$ since diffusion becomes more and more effective ($F_\mathrm{d} \propto 1/\sigma $).

\begin{figure}
\includegraphics[width=0.48\textwidth]{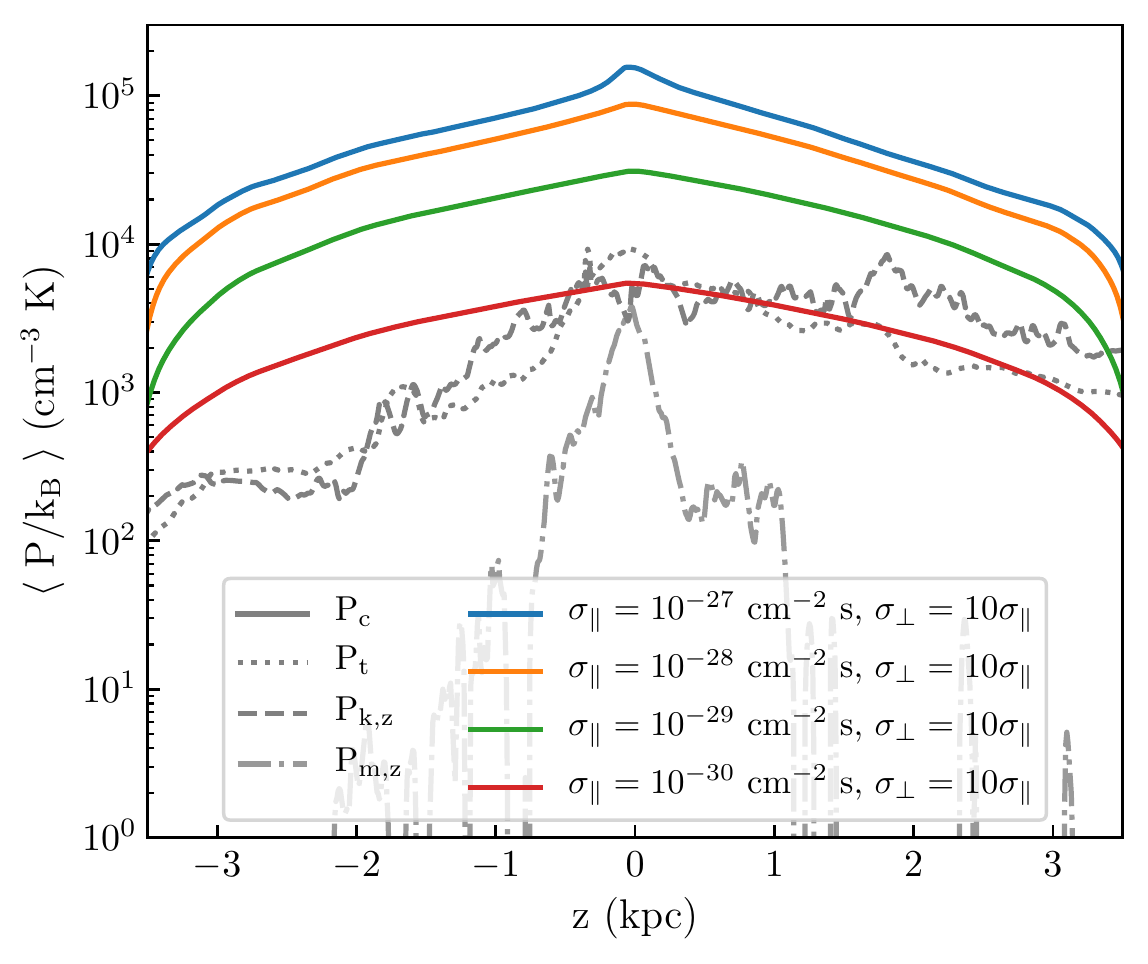}
\caption{Horizontally-averaged CR pressure $P_\mathrm{c}$ as a function of $z$ for different models of CR transport including both diffusion and streaming (solid lines). Each color corresponds to a given value of $\sigma_\parallel$, from $10^{-27}$~cm$^{-2}$~s (blue line) to $10^{-30}$~cm$^{-2}$~s (red line). All these models assume $\sigma_\perp = 10 \,\sigma_\parallel$. The gray lines show the horizontally-averaged profiles of thermal pressure $P_\mathrm{t}$ (dotted line), vertical kinetic pressure $P_\mathrm{k,z}$ (dashed line) and vertical magnetic stress $P_\mathrm{m,z}$ (dot-dashed line). These profiles are obtained by post-processing the TIGRESS snapshot at $t = 286$~Myr.}
\label{fig:noadv_profiles}
\end{figure}

In \autoref{fig:noadv_profiles}, we show the horizontally-averaged vertical profiles of $P_\mathrm{c}$ predicted by the four models with both streaming and diffusion. As noted above, the value of $P_\mathrm{c}$ at a given $z$ is lower for smaller $\sigma_\parallel$. In the mid-plane, $P_\mathrm{c}$ becomes comparable with the other relevant pressures if we assume $\sigma_\parallel = 10^{-30}$~cm$^{-2}$~s. We point out that this value is lower than the range $\sigma \sim 10^{-29}-10^{-28}$~cm$^{-2}$~s predicted by traditional studies of CR propagation in our Galaxy that neglect advection and do not employ detailed magnetic field structure (see \autoref{Comparison with other works} for a discussion).
The comparison with the horizontally-averaged profiles of thermal, kinetic and magnetic pressure (dotted, dashed and dot-dashed gray lines, respectively) confirms that the distribution of CR pressure is extremely uniform compared to that of the other pressures, even in cases where streaming is the dominant mechanism of CR transport (i.e. $\sigma_\parallel > 10^{-29}$~cm$^{-2}$~s; see \autoref{Streaming vs diffusion}). As pointed out in \autoref{MHD simulation}, in much of the volume magnetic field lines are mostly tangled. With random changes of the magnetic field orientation, streaming transport resembles diffusion on scales larger than the coherence length of the field line, and contributes to produce a uniform distribution of CRs across space. We note, however, that there is a greater degree of large-scale field alignment near the mid-plane -- where the preferentially horizontal field helps confine CRs --, and at high latitude regions -- where the enhanced vertical alignment does help transport CRs out of the disk. 

\subsubsection{Streaming vs diffusive transport}
\label{Streaming vs diffusion}

\begin{figure*}
\centering
\includegraphics[width=0.9\textwidth]{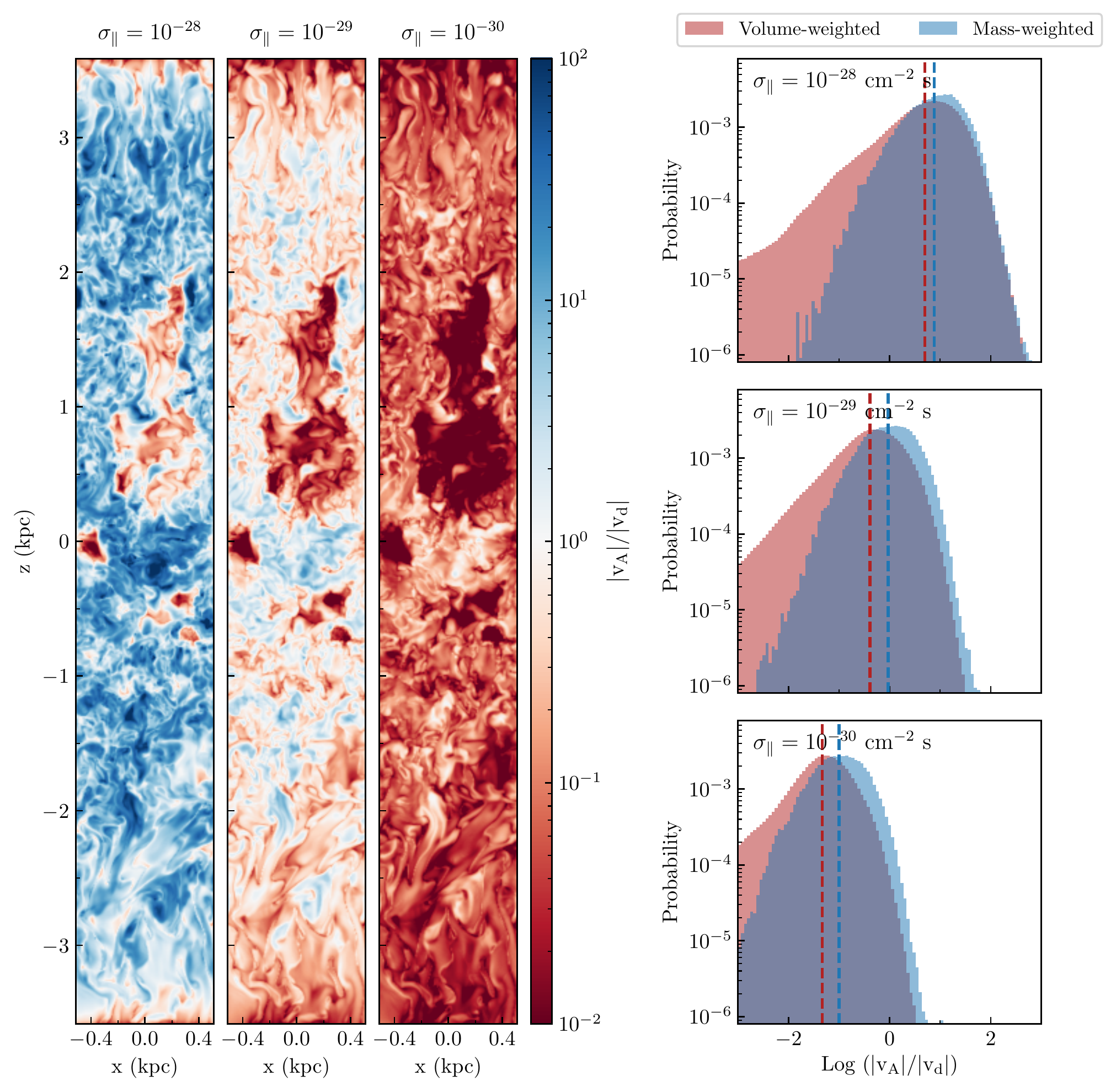}
\caption{Analysis of the relative contribution of streaming and diffusion to the overall CR propagation in the absence of advection. \textit{Left side}: distribution on the grid of the ratio between Alfv\'{e}n speed $\vert v_\mathrm{A} \vert$ and diffusive speed $\vert v_\mathrm{d} \vert$ in models with $\sigma_\parallel = 10^{-28}$~cm$^{-2}$~s (\textit{left panel}), $\sigma_\parallel = 10^{-29}$~cm$^{-2}$~s (\textit{middle panel}) and $\sigma_\parallel = 10^{-30}$~cm$^{-2}$~s (\textit{right panel}). \textit{Right side}: volume-weighted (red histograms) and mass-weighted (blue histograms) probability distributions of $\vert v_\mathrm{A} \vert/\vert v_\mathrm{d} \vert$ for $\sigma_\parallel = 10^{-28}$~cm$^{-2}$~s (\textit{top panel}), $\sigma_\parallel = 10^{-29}$~cm$^{-2}$~s (\textit{middle panel}) and $\sigma_\parallel = 10^{-30}$~cm$^{-2}$~s (\textit{bottom panel}).  The red and blue dashed lines indicate the median values of the volume-weighted and mass-weighted distributions, respectively. 
}
\label{fig:FsvsFd}
\end{figure*}

We investigate the relative importance of streaming and diffusive transport,  evaluating the ratio of Alfv\'{e}n speed and diffusive speed (\autoref{eq:vd_def}) across the simulation box. The left panel of \autoref{fig:FsvsFd} shows the distribution on the grid of $\vert v_\mathrm{A} \vert/\vert v_\mathrm{d} \vert$ in models with different choices of $\sigma_\parallel$, from $10^{-28}$~cm$^{-2}$~s to $10^{-30}$~cm$^{-2}$~s. Streaming transport largely dominates in the model with $\sigma_\parallel = 10^{-28}$~cm$^{-2}$~s, except for a few regions characterized by low Alfv\'{e}n speeds ($v_\mathrm{A} \ll 1 \,\kms$, see \autoref{MHDsim}). A visual comparison between the Alfv\'{e}n speed snapshot and the density and temperature snapshots in \autoref{MHDsim} shows that low Alfv\'{e}n speeds occur within expanding supernova bubbles and at the base of the hot winds generated by their blow-out. The ratio $\vert v_\mathrm{A} \vert/\vert v_\mathrm{d} \vert$ is closer to unity in the model with $\sigma_\parallel = 10^{-29}$~cm$^{-2}$~s, indicating an equivalent contribution of streaming and diffusion, except for the regions with $v_\mathrm{A} \ll 1 \,\kms$, where diffusion is more important. Instead, diffusive transport is largely dominant in the model with $\sigma_\parallel = 10^{-30}$~cm$^{-2}$~s.

The right panel of \autoref{fig:FsvsFd} shows the volume-weighted (red histograms) and mass-weighted (blue histograms) probability distributions of $\vert v_\mathrm{A} \vert/\vert v_\mathrm{d} \vert$ across the simulation domain for the three different choices of $\sigma_\parallel$. In all models, the mass-weighted distributions present more pronounced peaks and less extended tails towards low values of $\vert v_\mathrm{A} \vert/\vert v_\mathrm{d} \vert$ compared to the volume-weighted distributions. This is because the regions at higher density, which contribute the most to the mass budget (see \autoref{fig:PDF}), are characterized by larger Alfv\'{e}n speeds ($v_\mathrm{A} \gtrsim 10 \, \kms$ for $n_\mathrm{H} > 0.1$~cm$^{-3}$, see \autoref{MHDsim}) and, therefore, significant CR streaming. The difference between the volume-weighted and mass-weighted distribution is reflected in slightly different median values, with the volume-weighted median systematically lower than the mass-weighted median. Regardless of the weight chosen to analyze the distribution of $\vert v_\mathrm{A} \vert/\vert v_\mathrm{d} \vert$, the evidence discussed in the previous paragraph is confirmed: streaming is the dominant transport mechanism in the model assuming $\sigma_\parallel = 10^{-28}$~cm$^{-2}$~s, while diffusion is the dominant mechanism in the model assuming $\sigma_\parallel = 10^{-30}$~cm$^{-2}$~s. 

\subsubsection{Diffusion perpendicular to the magnetic field lines}
\label{Perpendicular Diffusion}

\begin{figure*}
\includegraphics[width=\textwidth]{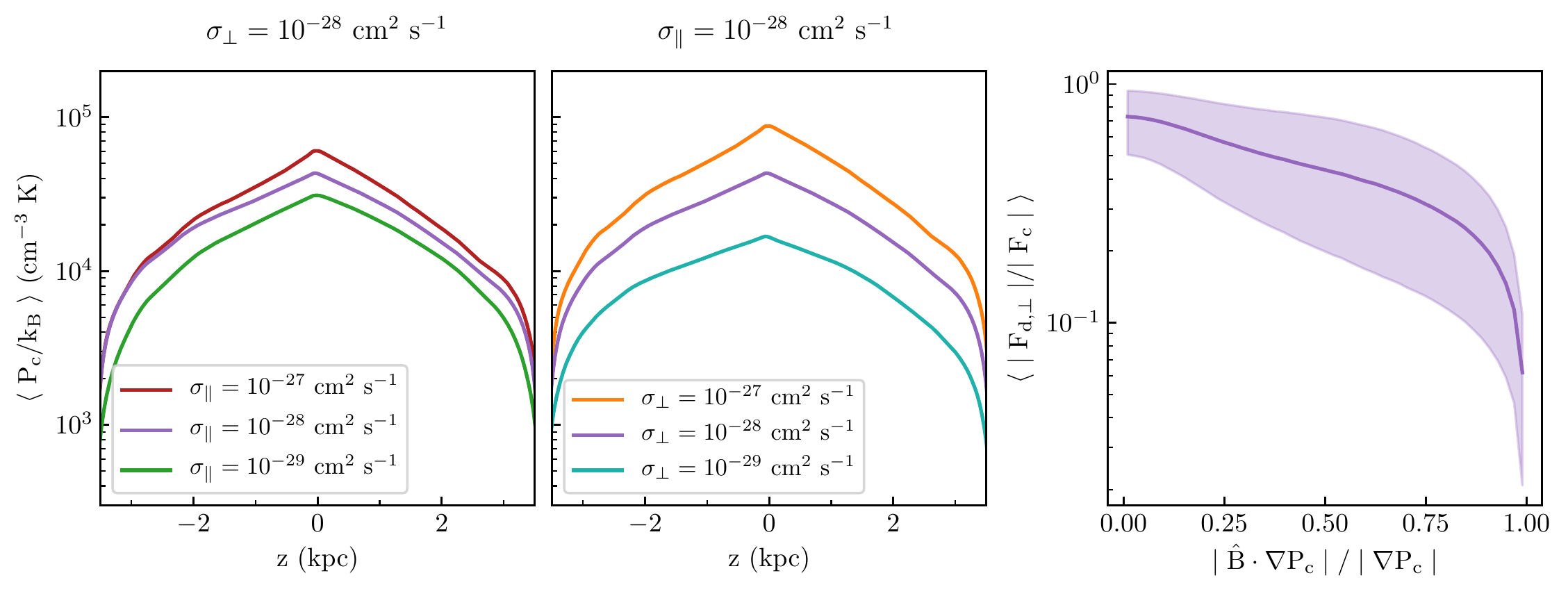}
\caption{Analysis of the relative effects of diffusion parallel and perpendicular to the magnetic field direction in models neglecting CR advection. \textit{Left panel}: horizontally-averaged CR pressure as a function of $z$ for models with same $\sigma_\perp = 10^{-28}$~cm$^{-2}$~s and different $\sigma_\parallel$, from $10^{-27}$~cm$^{-2}$~s (red line) to $10^{-29}$~cm$^{-2}$~s (green line). \textit{Middle panel}: same as in the left panel, but for models with same $\sigma_\parallel = 10^{-28}$~cm$^{-2}$ and $\sigma_\perp$ ranging from $10^{-27}$~cm$^{-2}$~s (orange line) to $10^{-29}$~cm$^{-2}$~s (cyan line). \textit{Right panel}: average ratio of $\vert F_\mathrm{d,\perp} \vert$ to $\vert F_\mathrm{c} \vert$ as a function of the cosine of the angle between the magnetic field direction and the CR  pressure gradient direction, for the 
model with $\sigma_\parallel \equiv \sigma_\perp = 10^{-28}$~cm$^{-2}$~s (purple line in the three panels). The shaded area covers the 16th and 84th percentiles of the distribution.
}
\label{fig:diffpar-diffperp}
\end{figure*}

So far, we have focused on the effect on CR transport produced by different choices of $\sigma_\parallel$. Since all the analyzed models assume $\sigma_\perp = 10 \, \sigma_\parallel$, an increase/decrease of $\sigma_\parallel$  has always implied an increase/decrease of $\sigma_\perp$ by the same factor. In this section, we investigate the extent to which diffusion perpendicular to the magnetic field contributes to the overall CR propagation by comparing the results of models with different ratios of $\sigma_\parallel$ and $\sigma_\perp$. The left panel of \autoref{fig:diffpar-diffperp} displays the average vertical profile of CR pressure for models with same $\sigma_\perp = 10^{-28}$~cm$^{-2}$~s and different $\sigma_\parallel$, ranging from $10^{-27}$~cm$^{-2}$~s to $10^{-29}$~cm$^{-2}$~s. In contrast, the middle panel shows the average vertical profile of CR pressure for models with same $\sigma_\parallel = 10^{-28}$~cm$^{-2}$~s and different $\sigma_\perp$. As expected, CR pressure $P_\mathrm{c}$ increases with $\sigma_\parallel$ when $\sigma_\perp$ is constant, while $P_\mathrm{c}$ increases with $\sigma_\perp$ when $\sigma_\parallel$ is constant, but the sensitivity to changes is not the same.    
In both panels the purple lines represent the same model with $\sigma_\perp = \sigma_\parallel = 10^{-28}$~cm$^{-2}$~s  with either an increase/decrease of $\sigma_\parallel$ (red/green line on left) or increase/decrease of $\sigma_\perp$ (orange/cyan line on right).  Evidently, varying $\sigma_\perp$ rather than $\sigma_\parallel$ entails a greater change in CR pressure.
For example, in the mid-plane, $P_\mathrm{c}$ decreases by a factor $\sim 3$ when $\sigma_\perp$ decreases from $10^{-28}$~cm$^{-2}$~s to $10^{-29}$~cm$^{-2}$~s, while it decreases by a factor $\sim 1.3$ when $\sigma_\parallel$ decreases from $10^{-28}$~cm$^{-2}$~s to $10^{-29}$~cm$^{-2}$~s.

In the right panel of \autoref{fig:diffpar-diffperp}, we analyze the ratio of the diffusive flux perpendicular to the magnetic field, $F_\mathrm{d,\perp}$, to the total CR flux, as a function of $\cos\theta=\vert \hat{B} \cdot \nabla P_\mathrm{c}\vert/\vert \nabla P_\mathrm{c}\vert $.
The analysis is performed for the model adopting $\sigma_\perp = \sigma_\parallel = 10^{-28}$~cm$^{-2}$~s. The average ratio $\vert F_\mathrm{d,\perp} \vert $/$\vert F_\mathrm{c} \vert$ increases when the magnetic pressure gradient is not aligned with the magnetic field,
and becomes larger than 0.5 for 
$\cos\theta\lesssim 0.25 $. This behavior indicates that diffusion perpendicular to the magnetic field direction is the main propagation mechanism in regions where the magnetic field is nearly perpendicular to the CR pressure gradient. Diffusion perpendicular to the magnetic field direction is therefore crucial for the propagation of CRs that would be otherwise confined, either by a tangled magnetic field (at high altitude) or by a mostly-horizontal magnetic field (near the midplane). This result explains the significant variation of CR pressure led by variations of $\sigma_\perp$.

\subsection{Models including advection}
\label{Adv}

\begin{figure}
\centering
\includegraphics[width=0.45\textwidth]{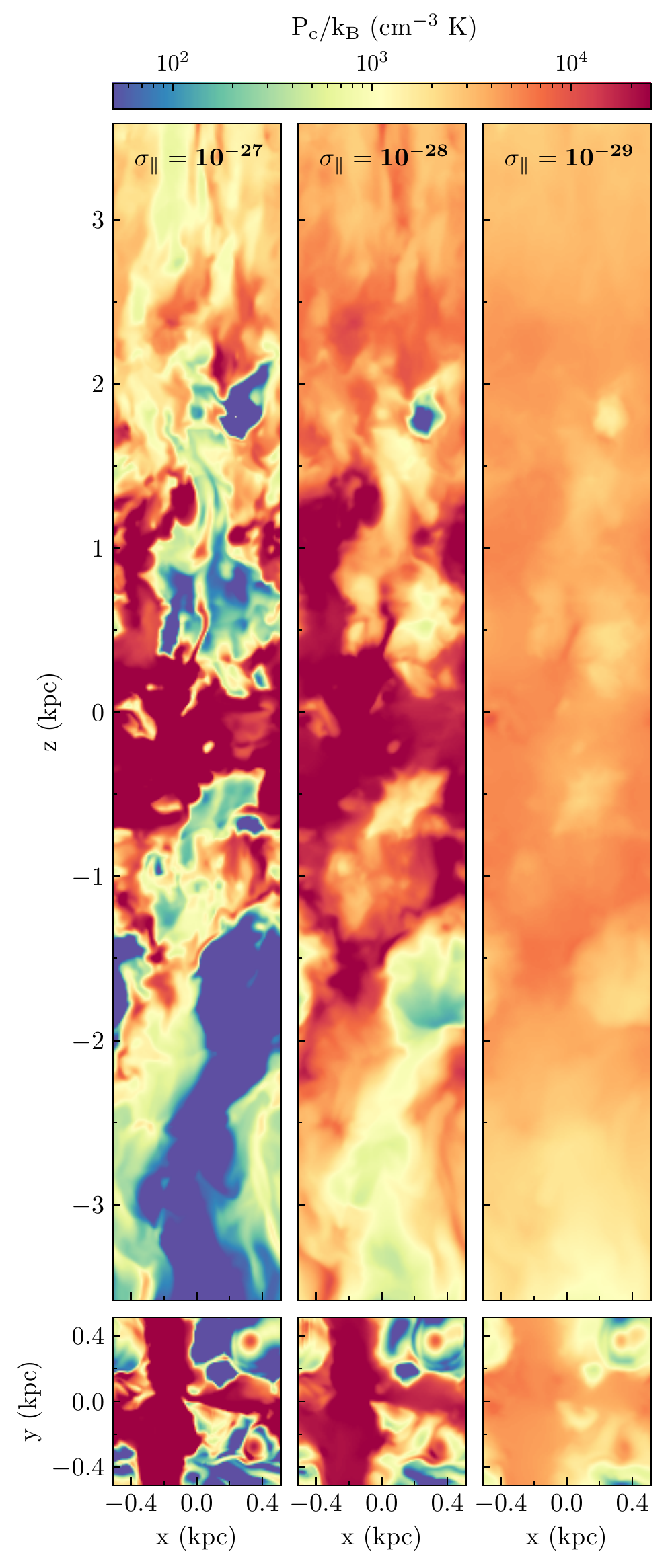}
\caption{Distribution on the grid of CR pressure for different models of CR transport including advection, showing $y=0$ slices of $P_\mathrm{c}$.  While all these models assume spatially-constant scattering, they adopt different values of the scattering coefficient: $\sigma_\parallel = 10^{-27}$~cm$^{-2}$~s (\textit{left panel}), $\sigma_\parallel = 10^{-28}$~cm$^{-2}$~s (\textit{middle panel}) and $\sigma_\parallel = 10^{-29}$~cm$^{-2}$~s (\textit{right panel}).  As in \autoref{fig:noadv_snaps}, the $t = 286$~Myr TIGRESS snapshot is used (see \autoref{MHDsim}). 
}
\label{fig:Gev_snaps}
\end{figure}

In this section, we present the predictions of CR propagation models with spatially-constant scattering including advective transport, in addition to streaming and diffusion. 
\autoref{fig:Gev_snaps} shows the distribution on the grid of CR pressure for three different choices of $\sigma_\parallel$, from $10^{-27}$~cm$^{-2}$~s to $10^{-29}$~cm$^{-2}$~s, for a single MHD snapshot at $t=286$~Myr. Except for the case with low scattering coefficient ($\sigma_\parallel = 10^{-29}$~cm$^{-2}$~s), where the high diffusivity produces a relatively uniform CR pressure across the grid, the distribution of CRs closely follows the gas distribution (see \autoref{MHDsim}). CRs accumulate in regions with
high density and low temperature, where the relatively-low gas velocities ($v<50\,\kms$)  do not foster their removal. By contrast, CRs in regions with hot and fast-moving winds ($v \gg 100\,\kms$) are rapidly advected away from the mid-plane. 
\autoref{MHDsim} shows that the velocity streamlines of the hot winds 
channel gas 
out of the disk, allowing CRs coupled to the hot phase gas to escape through these ``chimneys.''  The 
importance of advective transport 
is particularly evident in the model with high scattering coefficient ($\sigma_\parallel = 10^{-27}$~cm$^{-2}$~s), where CR diffusion is negligible (see \autoref{Streaming vs diffusion} and \autoref{Streaming vs Diffusion vs Advection}). 
The correlation between CRs and the  density/temperature distribution in the left panel of \autoref{fig:Gev_snaps}  contrasts strongly with the very smooth CR pressure profile in the third panel of \autoref{fig:noadv_snaps}, and more generally the smooth CR distributions in all the models without advection.

\begin{figure}
\centering
\includegraphics[width=0.48\textwidth]{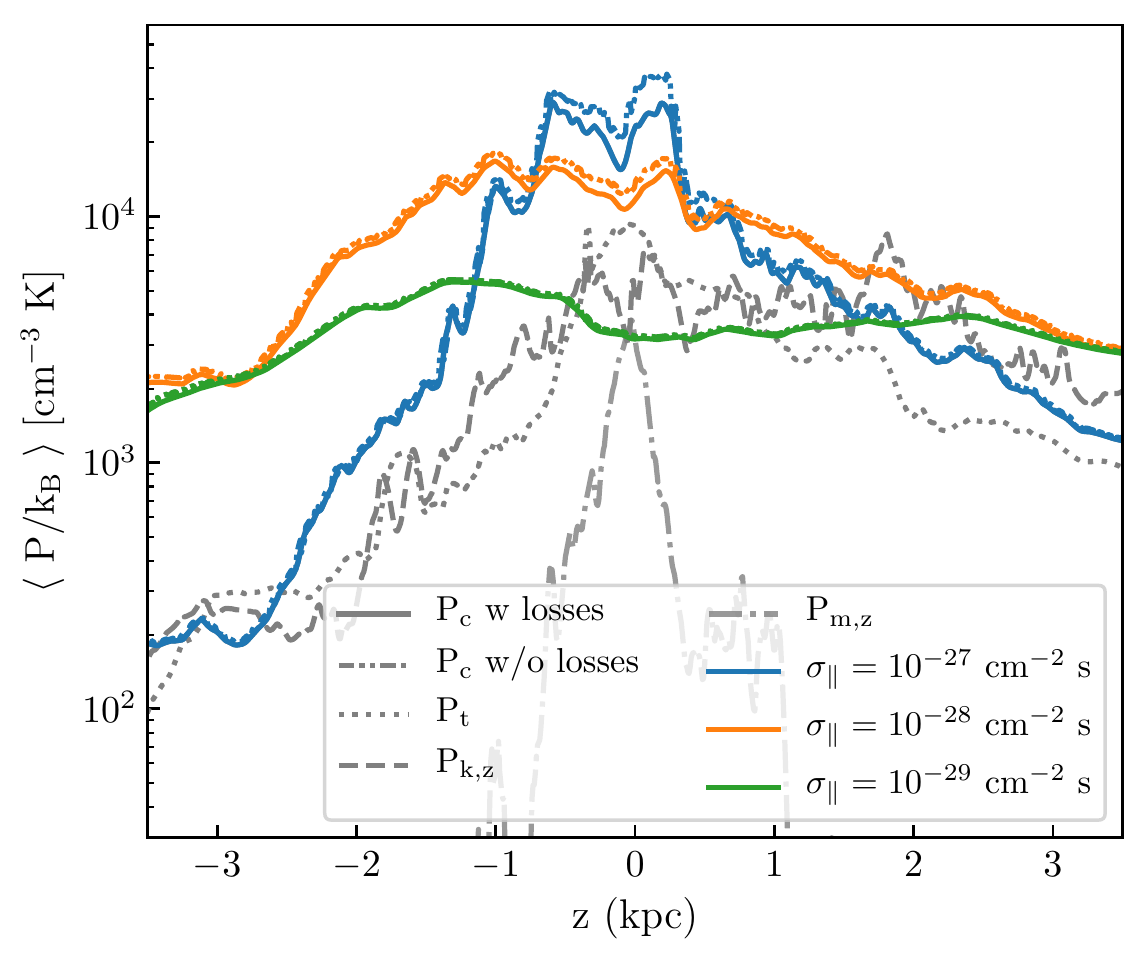}
\caption{Horizontally-averaged CR pressure $P_\mathrm{c}$ as a function of $z$ for different models of CR transport including diffusion, streaming, and advection, for the same $t = 286$~Myr snapshot as  \autoref{fig:noadv_profiles}. Different colors correspond to different 
$\sigma_\parallel$: $10^{-27}$~cm$^{-2}$~s (blue), $10^{-28}$~cm$^{-2}$~s (orange), and $10^{-29}$~cm$^{-2}$~s (green).  The dot-dot-dashed lines show models neglecting CR collisional losses, while the solid lines are for models that include losses. 
The gray lines show the horizontally-averaged vertical profiles of thermal pressure $P_\mathrm{t}$ (dotted), vertical kinetic pressure $P_\mathrm{k,z}$ (dashed) and vertical magnetic stress $P_\mathrm{m,z}$ (dot-dashed) from the MHD snapshot. 
}
\label{fig:Gev_profiles}
\end{figure}

\autoref{fig:Gev_profiles} shows  horizontally-averaged vertical profiles of $P_\mathrm{c}$ as in \autoref{fig:noadv_profiles}, but now for models with advection. 
The colored solid lines refer to models including collisional losses, while the corresponding dot-dot-dashed lines refer to models not accounting for collisional losses. 
Unlike the results shown in \autoref{fig:noadv_profiles}, now the CR pressure profile significantly changes with $\sigma_\parallel$. For $\sigma_\parallel = 10^{-29}$~cm$^{-2}$~s, the profile is relatively flat and does not show significant variations as a function of $z$, while for $\sigma_\parallel = 10^{-27}$~cm$^{-2}$~s, the CR pressure peaks in the mid-plane, where the gas velocity is relatively low and mainly oriented in $xy$-direction, and decreases at higher $z$. 
Comparing profiles in \autoref{fig:Gev_profiles} with \autoref{fig:noadv_profiles} for each $\sigma$, we see that the CR pressure in the disk decreases by about one order of magnitude when advection is included. As a consequence, the mid-plane CR pressure is comparable to the other relevant pressure for $10^{-29} \lesssim\sigma_\parallel \lesssim 10^{-28}$~cm$^{-2}$~s, while this is true only for a much lower scattering coefficient ($\sigma_\parallel = 10^{-30}$~cm$^{-2}$~s) in the absence of advection. The results of \autoref{fig:Gev_snaps} and \autoref{fig:Gev_profiles} demonstrate that accounting for advection of CRs by galactic winds is crucial in models of CR propagation, since CRs can easily escape from the galactic disk by flowing out along with the hot fast-moving gas. We further explore this point in \autoref{Streaming vs Diffusion vs Advection}.

Another important result of \autoref{fig:Gev_profiles} comes from the comparison of the vertical profiles of CR pressure obtained in the absence and in the presence of CR collisional losses. The change of CR pressure is almost negligible in models with $\sigma_\parallel \leqslant 10^{-28}$~cm$^{-2}$~s, while it is more significant in the model assuming $\sigma_\parallel = 10^{-27}$~cm$^{-2}$~s, especially in the mid-plane, where $P_\mathrm{c}$ decreases by $\sim 25 \,\%$ when CR losses are included. We note that the rate of CR energy losses is proportional to the gas density (see \autoref{Gammaloss2}). Therefore, CR losses are more effective for relatively high scattering coefficients, as this traps CRs in denser portions of the ISM for a longer time.

\begin{figure*}
\centering
\includegraphics[width=\textwidth]{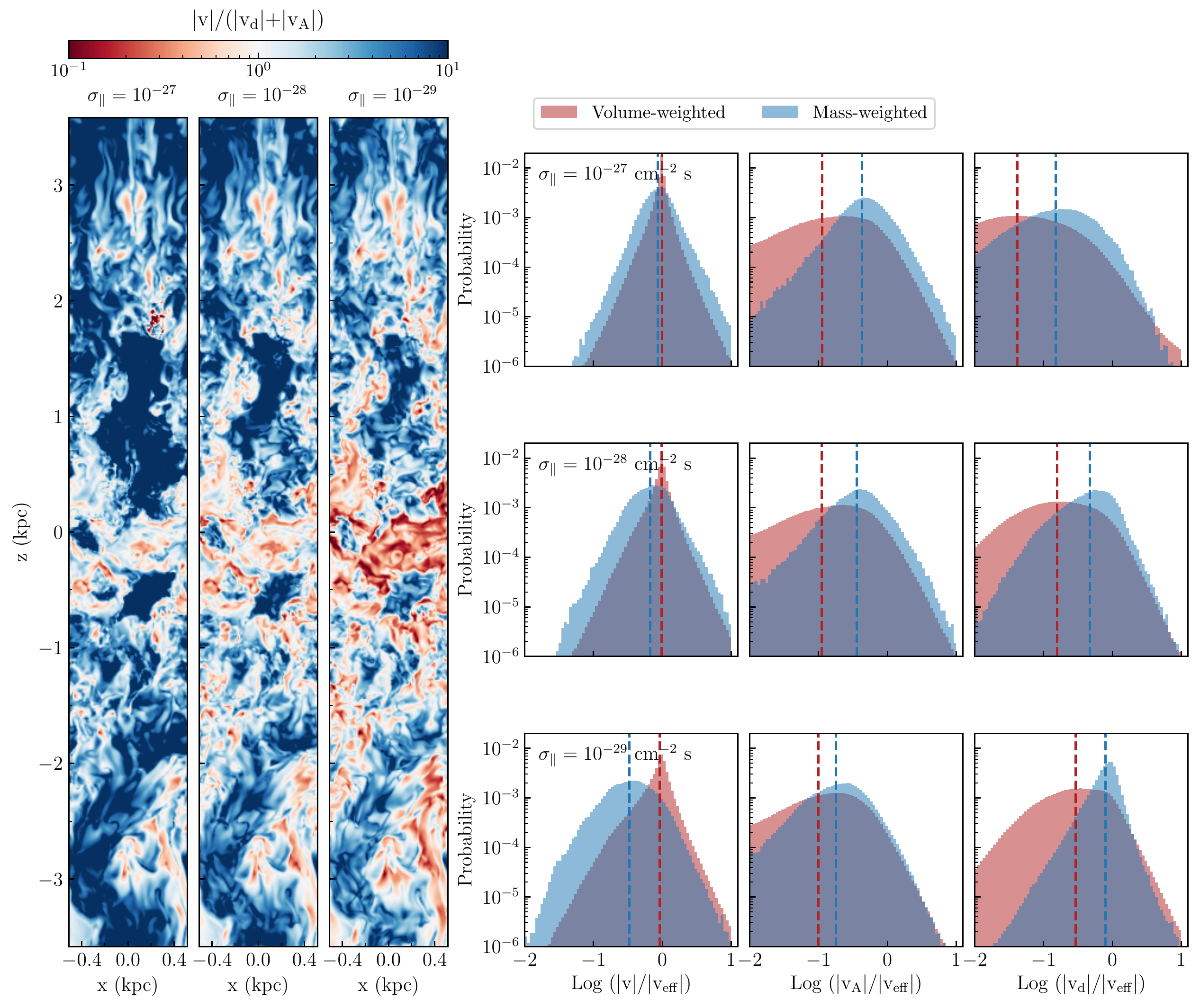}
\caption{Analysis of the relative contribution of streaming, diffusion, and advection to the propagation of CRs in models assuming spatially-constant scattering. \textit{Left side}: distribution on the grid of the ratio between the gas speed $v$ and the sum of the Alfv\'{e}n and diffusive speeds $\vert v_\mathrm{A} \vert + \vert v_\mathrm{d}\vert$ for models with $\sigma_\parallel = 10^{-27}$~cm$^{-2}$~s (\textit{left panel}), $\sigma_\parallel = 10^{-28}$~cm$^{-2}$~s (\textit{middle panel}) and $\sigma_\parallel = 10^{-29}$~cm$^{-2}$~s (\textit{right panel}). \textit{Right side}: each column shows the volume-weighted (red histograms) and mass-weighted (blue histograms) probability distribution of the ratio of $\vert v \vert$ (\textit{left column}), $\vert v_\mathrm{A} \vert$ (\textit{middle column}), $\vert v_\mathrm{d} \vert$ (\textit{right column}) relative to $\vert v_\mathrm{eff} \vert \equiv \vert 3/4\,\mathbf{F_\mathrm{c}}/e_\mathrm{c} \vert$ (an effective CR transport speed). Each row displays results for a model with given $\sigma_\parallel$, from $10^{-27}$~cm$^{-2}$~s (\textit{top}) to $10^{-29}$~cm$^{-2}$ (\textit{bottom}). The red and blue dashed lines indicate the median values of the volume-weighted and mass-weighted distributions, respectively. The analysis is for the $t = 286$~Myr TIGRESS snapshot.
}
\label{fig:FsvsFdvsFa}
\end{figure*}

\subsubsection{Importance of advective transport}
\label{Streaming vs Diffusion vs Advection}

We have seen that advection by fast-moving gas plays a key role in rapidly carrying CRs far from their injection sites. In \autoref{fig:FsvsFdvsFa}, we further quantify the relative contribution of advection compared to streaming and diffusion. The left side of \autoref{fig:FsvsFdvsFa} displays the distribution on grid of the ratio between the advection speed and the sum of the Alfv\'{e}n and diffusive speeds, $\vert v \vert$/$(\vert v_\mathrm{A} \vert + \vert v_\mathrm{d} \vert)$. We show results for models with three different  $\sigma_\parallel$ for the same  $t = 286$~Myr MHD snapshot. 
For all values of $\sigma_\parallel$, advection completely dominates in the hot gas, and is marginally more important than diffusion and streaming in much of the remaining volume. 
In higher-density gas, which fills much of the mid-plane and is present in clumps/filaments at high latitude, advection is subdominant.  
For the higher-density regions, the importance of advective transport decreases at lower $\sigma_\parallel$ as diffusion becomes more and more effective.  

The right panel of \autoref{fig:FsvsFdvsFa} shows the volume-weighted (red histograms) and mass-weighted (blue histograms) probability distributions of $\vert v \vert$/$\vert v_\mathrm{eff} \vert$, $\vert v_\mathrm{A} \vert$/$\vert v_\mathrm{eff} \vert$, and $\vert v_\mathrm{d} \vert$/$\vert v_\mathrm{eff} \vert$, with $\vert v_\mathrm{eff} \vert = \vert 3/4\,\mathbf{F_\mathrm{c}}/e_\mathrm{c} \vert$ the effective CR propagation speed, for the three choices of the scattering coefficient\footnote{ We note that the moduli of individual propagation-speed components, $v$, $v_\mathrm{A}$ and $v_\mathrm{d}$, can exceed the modulus of the effective propagation speed $v_\mathrm{eff}$. This is mostly due to vector cancellation in $v_\mathrm{eff}$, but also to the presence of zones out of steady-state equilibrium, for which \autoref{SteadyFlux} does not hold. In fact, even if the overall system is approximately in equilibrium, there are always a number of cells far from such condition. These cells are usually characterized by $\sigma_\mathrm{tot,\parallel} \approx 0$ (either because the magnetic field is nearly perpendicular to the CR pressure gradient,
or because of very low scattering coefficients). In this case, the RHS of \autoref{CRflux} approaches zero.}. When volume-weighted, transport of CRs is mostly through advection with the ambient gas, as on average the gas velocity dominates over the other relevant velocities, regardless of the value of $\sigma_\parallel$. However, when weighted by gas mass, the distribution shifts to lower values of $\vert v \vert$/$\vert v_\mathrm{t} \vert$. As previously noted, streaming and diffusion are more important in regions characterized by higher densities. In the model with $\sigma_\parallel = 10^{-29}$~cm$^{-2}$~s, the mass-weighted median of the diffusive speed distribution is higher than the medians of the advective and streaming speed distributions.  Thus, when the CR scattering coefficient is relatively low, diffusion is the main transport mechanism of CR propagation in higher-density regions.  For $\sigma_\parallel = 10^{-29}$~cm$^{-2}$~s, diffusion dominates over streaming even in terms of volume.  For the higher values  $\sigma_\parallel = 10^{-28}$~cm$^{-2}$~s and $10^{-27}$~cm$^{-2}$~s, however, both the mass-weighted median diffusion speed and streaming speed are lower than the advection speed.  

In \autoref{Streaming vs diffusion}, for models without advection, we have seen that a low scattering coefficient ($\sigma_\parallel < 10^{-29}$~cm$^{-2}$~s) is required for diffusion to be dominant over streaming. However, once advection is included, the median diffusion speed exceeds the median streaming speed even for $\sigma_\parallel \sim 10^{-28}$~cm$^{-2}$~s.  
It is striking that once advection is  once included, it becomes the main CR transport mechanism in many high-latitude regions that would otherwise be dominated by streaming (compare \autoref{fig:FsvsFdvsFa} with \autoref{fig:FsvsFd}) and where CRs would be trapped by tangled magnetic fields. For the same reason, the diffusive flux in the direction perpendicular to the magnetic field, which is crucial for the propagation of CRs 
when advection is neglected (see \autoref{Perpendicular Diffusion}), plays a minor role in the presence of advection. For example, in the model with $\sigma_\parallel = 10^{-28}$~cm$^{-2}$~s, if we suppress perpendicular diffusion entirely when advection is turned on, it leads to less than a factor 2 variation of CR pressure near the mid-plane.
In contrast, for the analogous case without advection, variations of $\sigma_\perp$ lead to much more significant variations of CR pressure (see \autoref{fig:diffpar-diffperp}).

\subsubsection{Time-averaged results}

\label{sec:time-averaged results}

\begin{figure*}
\centering
\includegraphics[width=\textwidth]{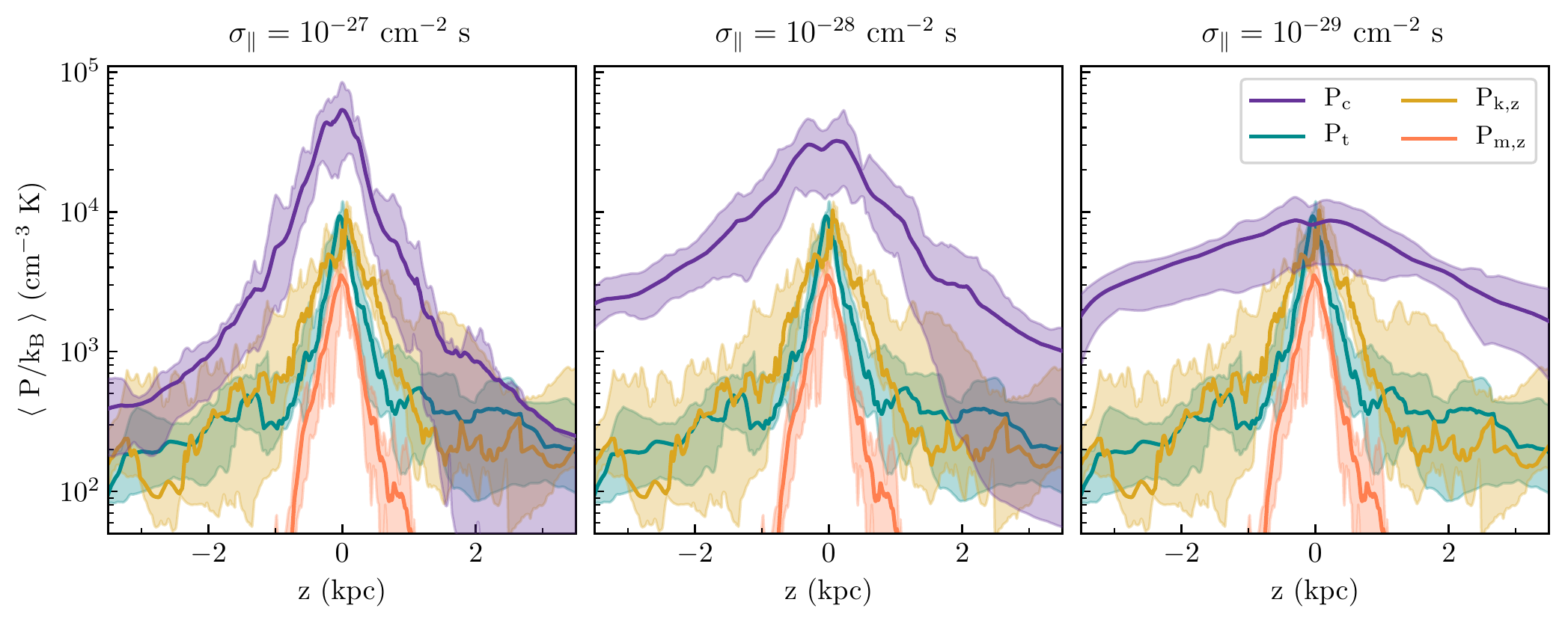}
\caption{Horizontally- and temporally-averaged vertical profiles of CR pressure $P_\mathrm{c}$ (purple), thermal pressure $P_\mathrm{t}$ (cyan), kinetic pressure $P_\mathrm{k,z}$ (yellow) and magnetic stress $P_\mathrm{m,z}$ (orange) in models including advection and assuming spatially-constant scattering. From left to right, panels show results from models with  $\sigma_\parallel = 10^{-27}$, $10^{-28}$, and $10^{-29}$~cm$^{-2}$~s. The shaded area covers the 16th and 84th percentiles from the temporal distribution.}
\label{fig:Gev_profiles_timeaver}
\end{figure*}

In \autoref{NoAdv} and above in \autoref{Adv}, we have analyzed results from a single TIGRESS snapshot. Here we use 10 post-processed TIGRESS snapshots  to investigate the temporally-averaged CR distribution in models including all the relevant mechanisms of CR transport, i.e. diffusion, streaming, and advection. As shown by \citet{Vijayan+20}, the gas properties are in a statistically steady state when averaged over several star-formation cycles. Therefore, averaging the CR pressure at different times (over $t=200-550$~Myr), we are able to study mean trends.

\autoref{fig:Gev_profiles_timeaver} shows the horizontally- and temporally-averaged profiles of CR pressure, thermal pressure, vertical kinetic pressure and vertical magnetic stress as a function of $z$ for models of CR propagation with different $\sigma_\parallel$. As highlighted in the discussion of \autoref{fig:Gev_profiles}, the CR pressure profile becomes flatter and smoother for low scattering rate since CRs escape from the mid-plane more easily and what would otherwise be inhomogeneities are erased by strong diffusion. 
In the mid-plane, the CR pressure is comparable to the thermal and vertical kinetic pressures for  $\sigma_\parallel \simeq 10^{-29}$~cm$^{-2}$~s. For higher scattering coefficient ($\sigma_{\parallel} \ge 10^{-29}$~cm$^{-2}$~s), the mid-plane CR pressure is above equipartition.  We note that the value of $\sigma_\parallel$ required to obtain pressure equipartition is slightly higher for the snapshot analyzed in \autoref{fig:Gev_profiles}. That snapshot is representative of an outflow-dominated period, when advection by fast-moving winds is particularly effective at removing CRs from the disk. 

We point out that for all cases, the CR scale height  ($\gtrsim \,1$~kpc) is larger that the scale height of thermal and kinetic pressure   \citep[$\sim 300-400$~pc,][]{Kim&Ostriker17,Vijayan+20}. This suggests that in conditions typical of our solar neighborhood,
the force exerted by CRs on the  gas ($\propto \partial P_\mathrm{c}/\partial z$) is less important to supporting the vertical weight of the galactic disk than the thermal and kinetic forces, especially if $\sigma_{\parallel} < 10^{-28}$~cm$^{-2}$~s. At high latitudes, however, the CR force dominates over the other forces, which suggests that CRs may be important in accelerating galactic winds from the extra-planar corona/fountain region.

Finally, for each transport model, we have calculated the time-averaged  individual sink/source energy terms.  These consist of integrals over the whole simulation domain of the terms on the RHS of \autoref{CRenergy} (${\bf v}_\mathrm{s}\cdot \nabla P_\mathrm{c}$ and ${\bf v}\cdot \nabla P_\mathrm{c}$ in steady state), as well as the integral of $\Lambda_\mathrm{coll}(E) n_H e_\mathrm{c}$. The average CR energy injected per unit time is the same for all propagation models, equal to $1.8 \times 10^{38}$~erg~s$^{-1}$. The rates of collisional and streaming energy losses and the rate of work exchange with the gas decrease in absolute value as $\sigma_\parallel$ decreases. In all cases, we find that the $\mathbf{v}\cdot \nabla P_\mathrm{c}$ energy exchange term is positive, i.e. on average the gas is doing work on the CR population.
Detailed examination of the simulations shows that the largest contributions to the work term come from the midplane region, at interfaces where hot gas (superbubbles) is expanding at high velocity into warm/cold gas where CR densities are high.  
Relative to the input, for $\sigma_\parallel = 10^{-27}$~cm$^{-2}$~s we find the collisional loss is 0.68, the streaming loss is 2.1, and the gain from the gas 
is 2.1.  For $\sigma_\parallel = 10^{-28}$~cm$^{-2}$~s, the relative collisional loss is  0.37, the streaming loss is 1.2, and the gain from the gas is 1.8.  For $\sigma_\parallel = 10^{-29}$~cm$^{-2}$~s, the relative collisional loss is  0.078, the streaming loss is 0.13, and the gain from the gas is 0.72.

Depending on the adopted value of $\sigma_\parallel$, different models have different CR grammage. The grammage gives a measure  of the column of gas traversed by CRs during their propagation, defined for an 
individual particle as $X = \int \rho v_\mathrm{p} dt = \int \rho v_\mathrm{p} dE/\dot{E} = \int \rho v_\mathrm{p} dE/(n_\mathrm{H} \Lambda_\mathrm{coll} (E) E) \approx (\mu_\mathrm{H} m_\mathrm{p} v_\mathrm{p} / \Lambda_\mathrm{coll}(E) ) \times \Delta E /E$, with $\mu_\mathrm{H} = 1.4$. Averaging over particles, $\Delta E/E$ is the mean fractional energy loss suffered by an individual particle from collisions, which is related to the collisional energy loss rate $\dot{E}_\mathrm{loss}$ and energy injection rate $\dot{E}_\mathrm{inj}$ over 
the whole domain by $\Delta E/E \approx \dot{E}_\mathrm{loss}/\dot{E}_\mathrm{inj}$.
The grammage can then be calculated as
\begin{equation}\label{eq:grammage}
X = \mu_\mathrm{H} m_p v_\mathrm{p}\frac{\int d^3 x\, n_H e_\mathrm{c}}{\dot{E}_\mathrm{inj},}
\end{equation}
where $\dot E_\mathrm{loss}$ is obtained by integrating $\Lambda_{\rm coll}(E) n_H e_c$ over the domain.
Clearly, the grammage increases if the CR energy density is concentrated near the mid-plane where the gas density is is high. We find $X \sim 103$~g~cm$^{-2}$ for $\sigma_\parallel = 10^{-27}$~cm$^{-2}$~s, $X \sim 60$~g~cm$^{-2}$ for $\sigma_\parallel = 10^{-28}$~cm$^{-2}$~s, and $X \sim 12$~g~cm$^{-2}$ for $\sigma_\parallel = 10^{-29}$~cm$^{-2}$~s. We note that the grammage obtained assuming $\sigma_\parallel = 10^{-29}$~cm$^{-2}$~s is in good agreement with the CR grammage measured at the Earth \citep[$\sim 10$~g~cm$^{-2}$, e.g.][]{Hanasz+21}.

\subsection{CR pressure vs. gas density in the absence and in the presence of advection}
\label{CR pressure-gas density}

\begin{figure*}
\centering
\includegraphics[width=0.95\textwidth]{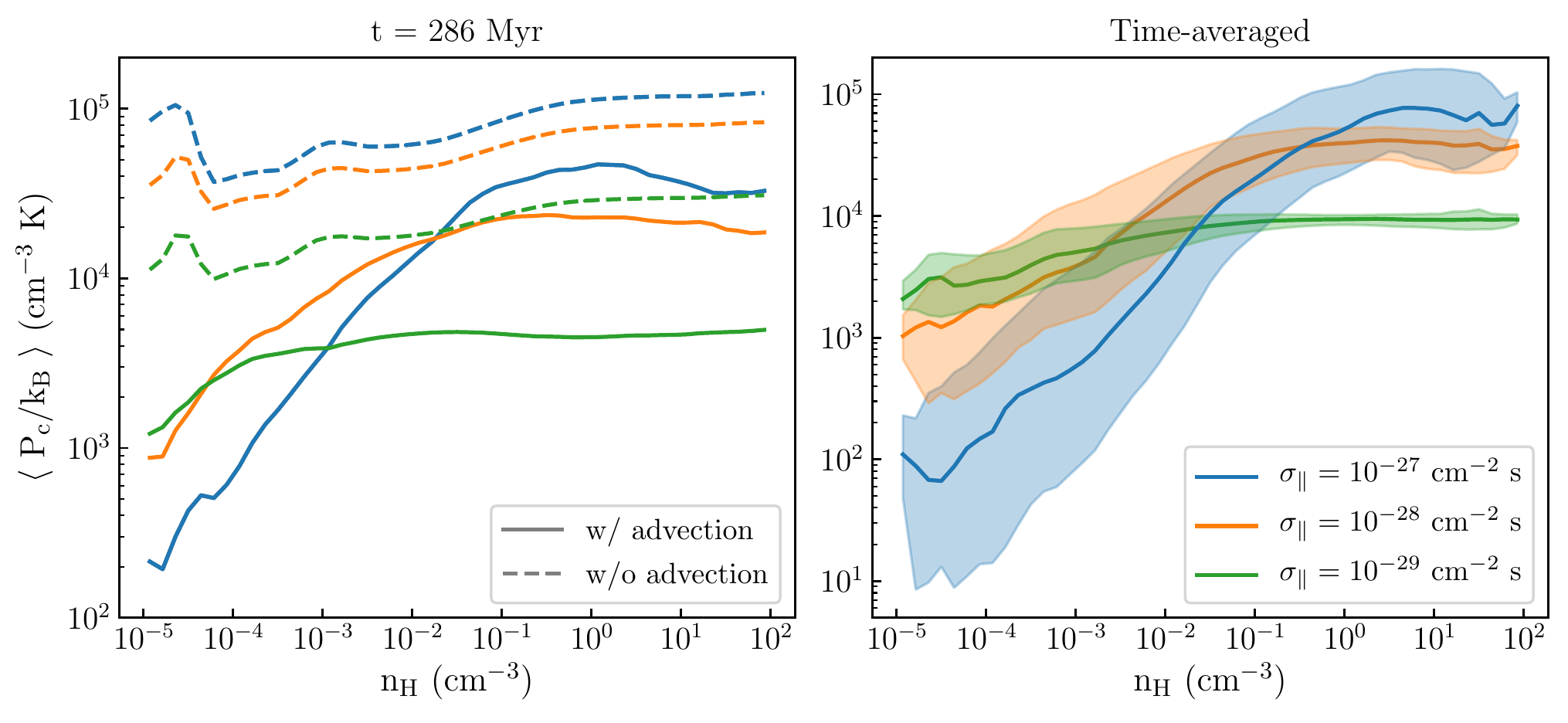}
\caption{Mean CR pressure $P_\mathrm{c}$ as a function of hydrogen density $n_\mathrm{H}$ in models with spatially-constant scattering coefficient: $\sigma_\parallel = 10^{-27}$~cm$^{-2}$~s (blue lines), $\sigma_\parallel = 10^{-28}$~cm$^{-2}$~s (orange lines) and $\sigma_\parallel = 10^{-29}$~cm$^{-2}$~s (green lines). \textit{Left panel:} Comparison between models neglecting advection (dashed lines) and models including advection (solid lines) for single snapshot at $t = 286$~Myr. \textit{Right panel:} Temporally-averaged results for models including advection. The shaded region covers the 16th and 84th percentiles from the temporal distribution.}
\label{fig:PcvsnH}
\end{figure*}

We conclude our study of constant-$\sigma$ models by analyzing how the CR pressure varies with the local gas density. In the left panel of \autoref{fig:PcvsnH}, we show the mean value of $P_\mathrm{c}$ as a function of $n_\mathrm{H}$ from models either including (solid lines) or neglecting (dashed lines) advection, based on the $t = 286$~Myr snapshot. We compare results obtained for  $\sigma_\parallel=10^{-27}$, $10^{-28}$, and $10^{-29}$~cm$^{-2}$~s. As highlighted in \autoref{Adv}, at given $\sigma_\parallel$ the mean CR pressure decreases when advection is included. 
Advection makes the most difference when the scattering coefficient is 
relatively high ($\sigma_\parallel \simeq 10^{-28}-10^{-27}$~cm$^{-2}$~s). 
In these cases, when advection is included the mean value of $P_\mathrm{c}$ rapidly decreases for $n_\mathrm{H} \lesssim 0.1$~cm$^{-3}$. This is because the low-density regions generally consist of gas heated and accelerated to high velocity by SN shocks, and the high-velocity flows remove CRs efficiently.

The right panel \autoref{fig:PcvsnH} shows the temporally-averaged mean of $P_\mathrm{c}$ as a function of $n_\mathrm{H}$. Only models including advection are considered here. In all models, the average value of $P_\mathrm{c}$ flattens at sufficiently high densities where diffusion of CRs dominates over advection (see \autoref{fig:FsvsFdvsFa}).  As noted above, the higher the scattering coefficient the stronger the correlation between CR pressure and gas density. In the model with $\sigma_\parallel = 10^{-27}$~cm$^{-2}$~s, the average value of $P_\mathrm{c}$ rapidly increases with $n_\mathrm{H}$ up to  $n_H \sim 1~\mathrm{cm}^{-3}$, since CRs are strongly confined within the midplane
and advection is increasingly ineffective in the high-density regions where velocities are relatively low. The correlation between $P_\mathrm{c}$ and $n_\mathrm{H}$ weakens at lower $\sigma_\parallel$ since diffusion is more and more effective in smoothing out CR inhomogeneities and allowing CRs to leave the midplane region where they are deposited. Moreover, the increasing effectiveness of diffusion results in a lower scatter of $P_\mathrm{c}$ around its mean value.  

\section{High-energy cosmic rays:\\ models with variable scattering coefficient}
\label{Gev-Variablesigma}

\begin{figure*}
\centering
\includegraphics[width=0.75\textwidth]{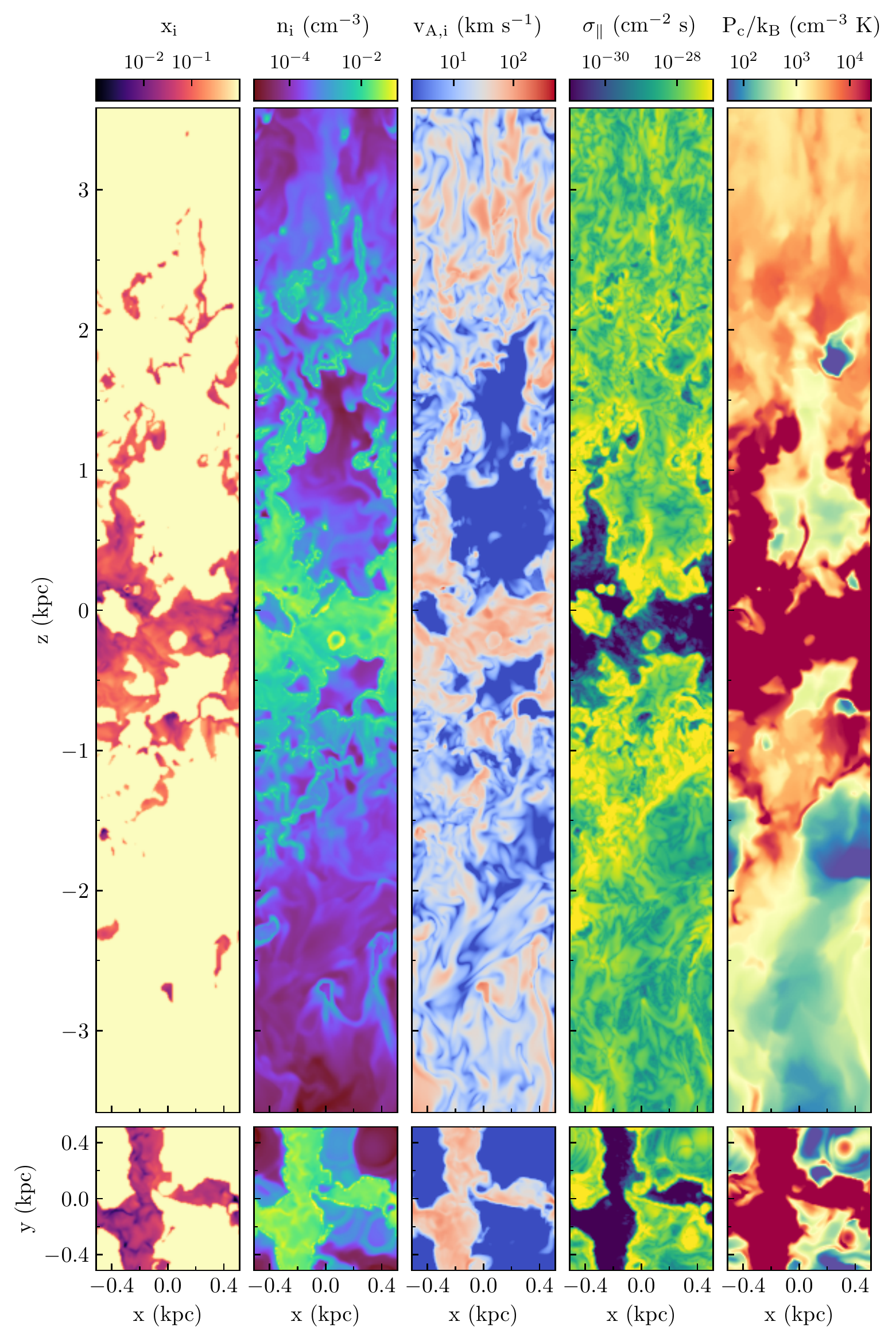}
\caption{Distribution on the grid of -- from left to right -- ionization fraction $x_\mathrm{i}$, ion density $n_\mathrm{i}$, ion Alfv\'{e}n speed $v_\mathrm{A,i}$, scattering coefficient parallel in the magnetic field direction $\sigma_\parallel$ and CR pressure $P_\mathrm{c}$ in the self-consistent CR propagation model assuming $\sigma_\perp = 10\, \sigma_\parallel$.  The snapshot is taken at $t = 286$~Myr and the slices are extracted at the center of the simulation box.}
\label{fig:varsigma_snaps}
\end{figure*}

\begin{figure*}
\centering
\includegraphics[width=0.95\textwidth]{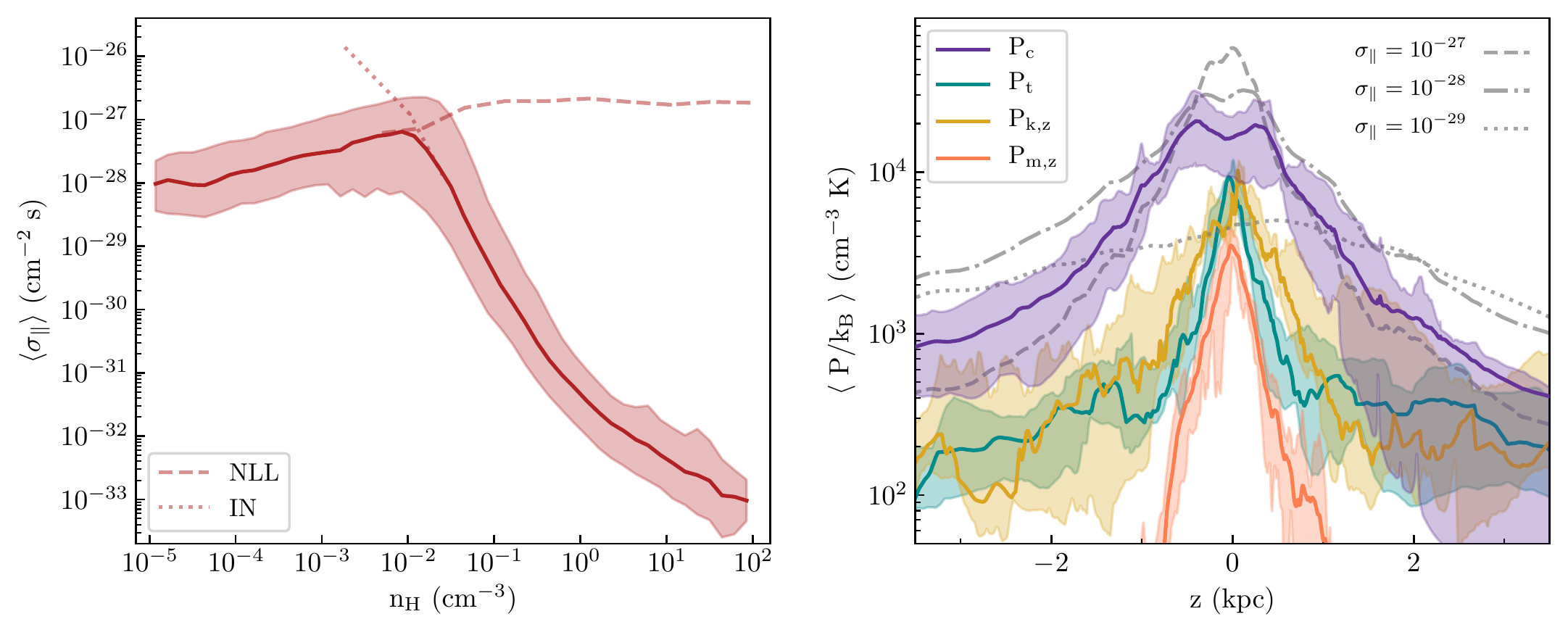}
\caption{Outcomes of the CR propagation model for the spatially-variable scattering treatment with  $\sigma_\perp = 10\, \sigma_\parallel$. \textit{Left panel}: temporally-averaged median of the scattering coefficient $\sigma_\parallel$ as a function of hydrogen density $n_\mathrm{H}$ (solid line). The shaded area covers the temporally-averaged variations, 
calculated as the difference between the 16th and 84th percentiles and the median of the distribution. The dashed and dotted lines show $\sigma_\mathrm{\parallel,NLL}$ (\autoref{NLL}) and  $\sigma_\mathrm{\parallel,IN}$ (\autoref{IN}), demonstrating that nonlinear Landau damping and ion-neutral damping is more important at low and high density, respectively. 
\textit{Right panel}: horizontally- and temporally-averaged vertical profiles of CR pressure $P_\mathrm{c}$ (purple), thermal pressure $P_\mathrm{t}$ (cyan), vertical kinetic pressure $P_\mathrm{k,z}$ (yellow) and vertical magnetic stress $P_\mathrm{m,z}$ (orange). The shaded areas cover the 16th and 84th percentiles of temporal fluctuations. The dashed, dot-dashed and dotted lines indicate the horizontally- and temporally-averaged vertical profiles of  $P_\mathrm{c}$ from the models with constant $\sigma_\parallel=10^{-27}$~cm$^{-2}$~s, $10^{-28}$~cm$^{-2}$~s, and $10^{-29}$~cm$^{-2}$~s, respectively. }
\label{fig:Gev_profiles_varsigma}
\end{figure*}

\begin{figure*}
\centering
\includegraphics[width=0.95\textwidth]{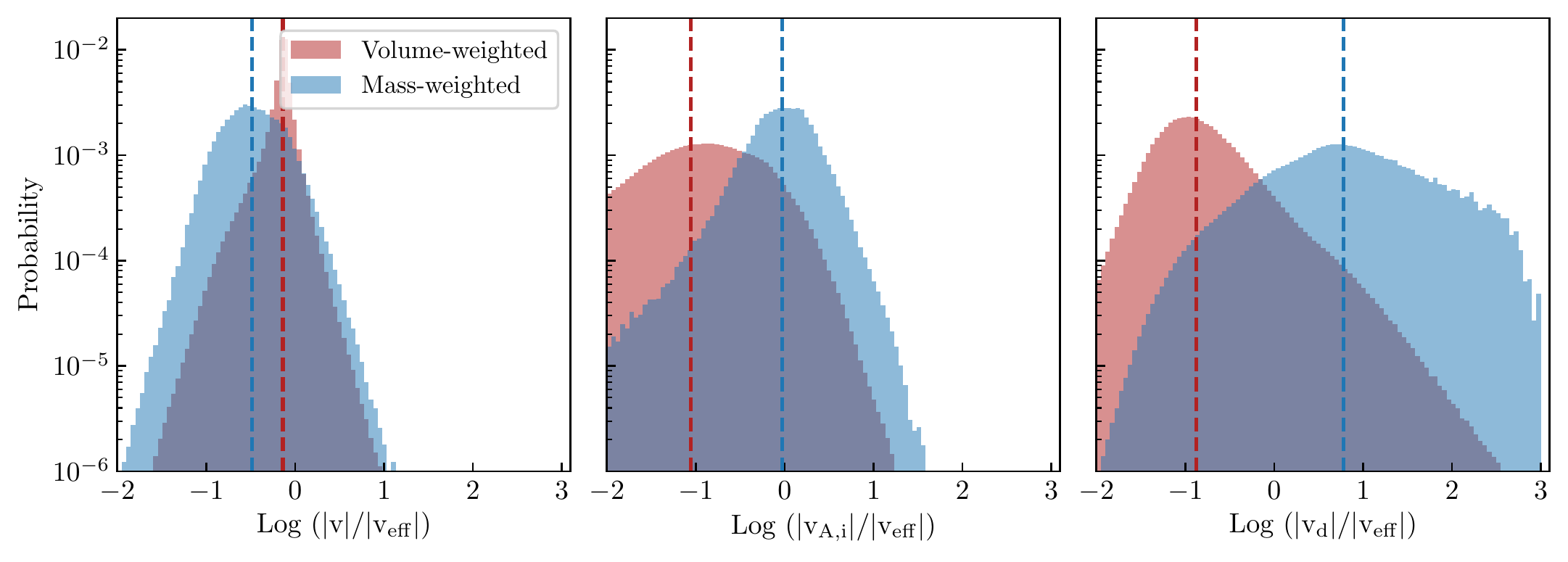}
\caption{Relative contribution to the total CR flux from advection, streaming, and diffusive terms, for the self-consistent model with  $\sigma_\perp = 10\, \sigma_\parallel$.
Volume-weighted (red histograms) and mass-weighted (blue histograms) show probability distributions of the ratio between advection speed $v$ (\textit{left panel}), ion Alfv\'{e}n speed $v_\mathrm{A,i}$ (\textit{middle panel}), diffusive speed $v_\mathrm{d}$ (\textit{right panel}), and the effective total CR propagation speed defined as $v_\mathrm{eff}\equiv 3/4\,\vert \mathbf{F}_\mathrm{c}\vert/e_\mathrm{c} $,  The red and blue dashed lines indicate the median values of the volume-weighted and mass-weighted distributions, respectively. The analysis is performed on the snapshots at $t = 286$~Myr.
}
\label{fig:FsvsFdvsFa_varsigma}
\end{figure*}

In this section, we investigate the distribution of CRs when we adopt a 
spatially-varying 
$\sigma$, computed under the assumption that CRs are scattered by streaming-driven Alfv\'{e}n waves (the self-confinement scenario), as described in \autoref{sigma}. The value of $\sigma_\parallel$ varies across the simulation box depending on the local properties of CRs and thermal gas, and we use the ion Alfv\'{e}n speed $v_\mathrm{A,i}$ (rather than $v_\mathrm{A}$) in the CR energy and momentum equations (\autoref{CRenergy} and \autoref{CRflux}) and the computation of the scattering rates (\autoref{IN} and \autoref{NLL}). As we shall show, in the higher-density gas where gas velocities are low and advection is ineffective, $v_\mathrm{A,i}/v_\mathrm{A} = x_i^{-1/2}$ can exceed 10 since $x_i\lesssim 10^{-2}$. An accurate estimate of the ionization fraction (which depends on the low-energy CRs)  is therefore important for proper computation of CR transport in the neutral gas, which comprises most of the mass in the ISM.  

To enable comparison with the models adopting constant scattering coefficient (\autoref{Adv}), we first discuss the results of the self-consistent model assuming anisotropic diffusion and $\sigma_\perp = 10\, \sigma_\parallel$. For the same snapshot shown in 
\autoref{MHDsim},  
\autoref{fig:varsigma_snaps} shows the distribution on the grid of some MHD quantities relevant for the self-consistent calculation: ion fraction $x_\mathrm{i}$, ion density $n_\mathrm{i}$, and ion Alfv\'{e}n speed $v_\mathrm{A,i}$, as well as the computed scattering coefficient $\sigma_\parallel$ and $P_\mathrm{c}$. The ion fraction (see \autoref{IonFrac} and \autoref{EqIonFrac}) is $x_i\simeq 1$ in regions with densities $n_\mathrm{H} \lesssim 10^{-2}$~cm$^{-3}$, meaning that gas is mostly ionized in those regions. In the mid-plane and in a few high-density filaments/clouds at higher latitudes, $x_i < 0.1$. The ion density is given by the product of the ion fraction and the hydrogen density. Therefore, $n_\mathrm{i}\approx n_\mathrm{H}$ ($n_\mathrm{i}\ll n_\mathrm{H}$) for $n_\mathrm{H}\lesssim 10^{-2}$~cm$^{-3}$ ($n_\mathrm{H}\gtrsim 0.1$~cm$^{-3}$). 
The scattering coefficient distribution closely follows the distribution of these three MHD quantities, since it is inversely proportional to the ion Alfv\'{e}n speed and to the ion density (see \autoref{IN} and \autoref{NLL}). In particular, $\sigma_\parallel$ is relatively high ($\simeq 10^{-28}$~cm$^{-2}$~s) in low-density regions ($n_\mathrm{H} < 10^{-2}$~cm$^{-3}$) and quite low ($\ll 10^{-29}$~cm$^{-2}$~s) in higher-density regions  ($n_\mathrm{H} > 10^{-1}$~cm$^{-3}$). Intermediate-density regions at the interface between neutral and ionized gas are characterized by the highest values of $\sigma_\parallel$ ($\gtrsim 10^{-28}$~cm$^{-2}$~s).

\subsection{Scattering rate coefficient and vertical profiles}\label{sec:GeVprof}

We now turn to results based on ten post-processed snapshots.  The left panel of \autoref{fig:Gev_profiles_varsigma} quantitatively analyzes the variation of $\sigma_\parallel$ with density $n_\mathrm{H}$, showing its temporally-averaged median value. The dashed and dotted lines respectively show $\sigma_\parallel$ assuming  only  nonlinear Landau damping (\autoref{NLL}),   
and only ion-neutral damping (\autoref{IN}). 
Nonlinear Landau damping dominates at low density, where the gas is well ionized. The resulting scattering coefficient has a weak explicit dependence on the hydrogen density, $\sigma_\mathrm{\parallel}= \sigma_\mathrm{\parallel,NLL} \propto (v_\mathrm{A,i} n_\mathrm{i})^{-1/2} \propto {n_\mathrm{H}}^{-1/4}$. Rather than decreasing with $n_\mathrm{H}$, however, $\sigma_\mathrm{\parallel,NLL}$ in 
\autoref{fig:Gev_profiles_varsigma}
slowly increases,  which we attribute to the increase of the CR pressure gradient in higher-density gas with $\sigma_\mathrm{\parallel,NLL}\propto (\vert \hat B \cdot\nabla  P_\mathrm{c}\vert   )^{1/2}$.
Indeed, as pointed out in \autoref{Adv}, advection of CRs is particularly effective in the fast-moving low-density gas, thus selectively reducing the CR pressure in these regions. 
Above $n_\mathrm{H}\sim 10^{-2}$~cm$^{-3}$, gas becomes mostly neutral and ion-neutral damping becomes stronger than nonlinear Landau damping, so that $\sigma_\parallel = \sigma_\mathrm{\parallel,IN}$.  
In this case, 
the scattering coefficient decreases with increasing the gas density, 
$\sigma_\mathrm{\parallel,IN} \propto ({v_\mathrm{A,i} n_\mathrm{i} n_\mathrm{n}})^{-1} 
\propto n_\mathrm{H}^{-5/4}$.
Putting the different regimes together, $\sigma_\parallel$ slowly increases from $\simeq 10^{-28}$~cm$^{-2}$~s at $n_\mathrm{H} \simeq 10^{-5}$~cm$^{-3}$ to $\simeq 10^{-27}$~cm$^{-2}$~s at $n_\mathrm{H} \simeq 10^{-2}$~cm$^{-3}$ and rapidly decreases at higher densities, reaching a value of $\simeq 10^{-33}$~cm$^{-2}$~s at $n_\mathrm{H} \simeq 10^{2}$~cm$^{-3}$. At $n_\mathrm{H} \simeq 10^{-1}$~cm$^{-3}$, the average scattering coefficient is a few times $10^{-30}$~cm$^{-2}$~s.

The above results for the dependence of scattering rate on density are useful for interpreting 
the CR pressure distribution displayed in the far right panel of \autoref{fig:varsigma_snaps}. The overall CR distribution follows the gas density distribution, as for the models with uniform $\sigma_\parallel = 10^{-27}-10^{-28}$~cm$^{-2}$~s (see \autoref{fig:Gev_snaps}). Much of the simulation volume is occupied by gas at low density, characterized by $\sigma_\parallel \gtrsim 10^{-28}$~cm$^{-2}$~s. Therefore, it is not surprising that the overall CR distribution resembles that of models with high scattering coefficients and ineffective diffusion. The difference with respect to those models arises in regions at higher density $n_\mathrm{H}\gtrsim 0.1$~cm$^{-3}$, where the gas is mostly neutral. The very low scattering coefficient in this regime ($\sigma_\parallel \lesssim 10^{-29}$~cm$^{-2}$~s) makes diffusion particularly effective in smoothing out CR inhomogeneities within the dense gas.  

The right panel of \autoref{fig:Gev_profiles_varsigma} shows the horizontally- and temporally-averaged vertical profiles of CR pressure, thermal pressure, vertical kinetic pressure and magnetic stress. The profiles of CR pressure obtained in simulations with uniform $\sigma_\parallel$ are also displayed for comparison. In the mid-plane, the average CR pressure is higher than the average thermal and kinetic pressures by about a factor of 2. The variable-$\sigma$ model has central $P_\mathrm{c}$ slightly lower than the $\sigma_\parallel = 10^{-28}$~cm$^{-2}$~s, and high-altitude wings similar to the  $\sigma_\parallel = 10^{-27}$~cm$^{-2}$~s model.  
Interestingly, even though most of the mass in the disk is at $n_\mathrm{H} \simeq 0.1-1$~cm$^{-3}$, where  $\sigma_\parallel \lesssim 10^{-30}$~cm$^{-2}$~s, the mid-plane CR pressure is higher than that obtained with constant $\sigma_\parallel = 10^{-29}$~cm$^{-2}$~s everywhere at $|z| \lesssim 1$~kpc. Even though both diffusion and streaming are highly effective in the high-density regions of the disk (see also \autoref{Streaming vs Diffusion vs Advection_Variable Sigma}), the propagation of CRs out of the dense gas depends on the properties of the surrounding hotter and lower density gas which has much higher scattering rates.  As a result, CRs are effectively trapped in the midplane region.  We conclude that the overall distribution of CRs depends on their propagation in the low-density, hot gas that sandwiches the disk at high-altitude.  
  
\subsection{Role of streaming, diffusive and advective transport}
\label{Streaming vs Diffusion vs Advection_Variable Sigma}

In this section, we evaluate the relative contributions of diffusion, streaming, and advection to the overall CR transport when $\sigma_\parallel$ is self-consistently calculated from a balance of the growth and damping rate of resonant Alfv\'en waves.  
\autoref{fig:FsvsFdvsFa_varsigma} shows the volume-weighted (red histograms) and 
mass-weighted (blue histograms) probability distributions of $\vert v \vert$/$\vert v_\mathrm{eff} \vert$, $\vert v_\mathrm{A,i} \vert$/$\vert v_\mathrm{eff} \vert$ and $\vert v_\mathrm{d} \vert$/$\vert v_\mathrm{eff} \vert$, which are the contributions to the total flux from advection, streaming, and diffusion\footnote{In \autoref{Streaming vs Diffusion vs Advection}, we have explained that the moduli of $v$, $v_\mathrm{A}$ and $v_\mathrm{d}$ can exceed the modulus of $v_\mathrm{eff}$ in zones out of steady-state equilibrium characterized by $\sigma_\mathrm{tot, \parallel} \approx 0$. In models with variable $\sigma$, deviations from equilibrium happen mainly in higher-density regions characterized by $\sigma_\parallel \ll 10^{-30}$~cm$^{-2}$~s$~\approx 0$. This explains why the mass-weighted distribution of $v_\mathrm{d}/v_\mathrm{eff}$, dominated by CRs in higher-density regions, extends orders of magnitude above unity.}. For the volume-weighted distributions, the overall profiles and median values are similar to those obtained adopting $\sigma_\parallel  = 10^{-28}$~cm$^{-2}$~s. Indeed, most of the simulation volume is occupied by low-density gas ($n_\mathrm{H} < 10^{-2}$~cm$^{-3}$, see \autoref{MHDsim}), where the average scattering coefficient is $\gtrsim 10^{-28}$~cm$^{-2}$~s (see \autoref{fig:Gev_profiles_varsigma}). As for the models with constant $\sigma_\parallel$, advection with the gas contributes the most to the CR propagation when weighted by volume. 

In contrast, if we consider the mass-weighted distributions, both diffusion and streaming transport dominate over advection. In higher-density regions containing most of the gas mass, the scattering coefficient decreases to very low values (see the left panel of \autoref{fig:Gev_profiles_varsigma}) due to ion-neutral damping, and CR  diffusion becomes quite strong. At the same time, the CR streaming velocity  at the ion Alfv\'{e}n speed $v_\mathrm{A,i}$ is significantly higher than the ideal Alfv\'{e}n speed $v_\mathrm{A}$ adopted in models with constant $\sigma_\parallel$.  
For gas at densities above 1~cm$^{-3}$, the mean value of $v_\mathrm{A,i}$ is $\simeq 60\, \kms$, and the mean ratio $v_\mathrm{A,i}/v_\mathrm{A} = x_\mathrm{i}^{-1/2}$ is $\simeq 7$. 

\begin{figure*}
\centering
\includegraphics[width=0.95\textwidth]{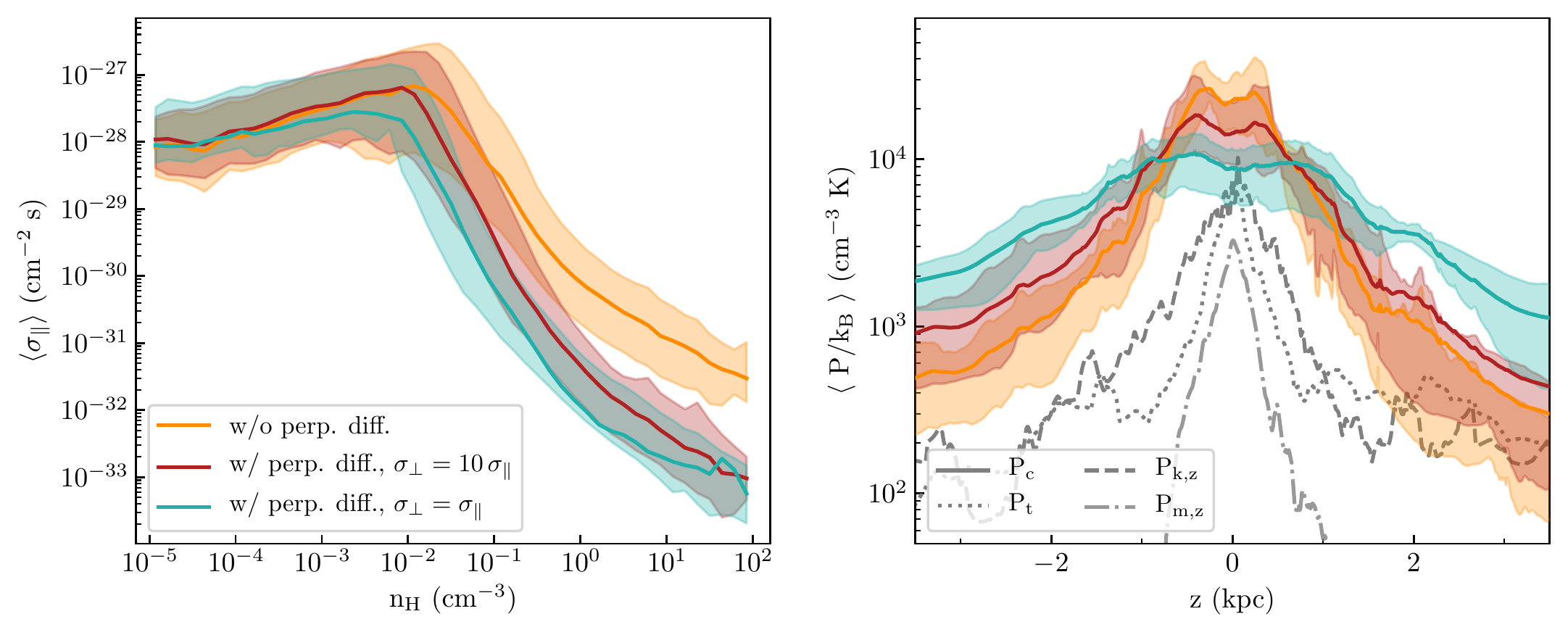}
\caption{Outcomes of self-consistent CR transport models with different treatments of diffusion perpendicular to the magnetic field. We show cases without perpendicular diffusion (orange line), with $\sigma_\perp= 10\, \sigma_\parallel$ (red line) and with isotropic diffusion $\sigma_\perp= \sigma_\parallel$ (turquoise line). \textit{Left panel}: temporally-averaged median of the scattering coefficient $\sigma_\parallel$ as a function of hydrogen density $n_\mathrm{H}$. 
\textit{Right panel}: horizontally- and temporally-averaged vertical profiles of CR pressure $P_\mathrm{c}$. For both panels, the shaded areas cover the 16th and 84th percentiles of temporal fluctuations. The gray lines show the horizontally- and temporally-averaged profiles of thermal pressure $P_\mathrm{t}$ (dotted line), vertical kinetic pressure $P_\mathrm{k,z}$ (dashed line) and vertical magnetic stress $P_\mathrm{m,z}$ (dot-dashed line).}
\label{fig:PerpDiff}
\end{figure*}

In the self-consistent model, we assume that the low-energy slope of the CR spectrum  is $\delta = -0.35$ (see \autoref{spectrum}). 
This enters in the calculation of both $\sigma_\parallel$ and $v_\mathrm{A,i}$ through $n_\mathrm{i}$, since  $x_i=n_i/n_\mathrm{H}$ depends on the low-energy CR ionization rate in high density/low temperature regions (see \autoref{IonFrac}).\footnote{The slope of the low-energy CR spectrum $\delta$ also enters in the calculation of $\sigma_\parallel$ through $n_1$  (see \autoref{spectrum}). However, for the CRs with energies of $\sim 1$~GeV that we are considering in this section, the value of $n_1$ is almost independent on the value of $\delta$.} Since $v_\mathrm{d} = \vert \hat \mathbf{B}\cdot \nabla P_\mathrm{c} \vert /(4 P_\mathrm{c} \sigma_\parallel)$, the ratio between  $v_\mathrm{A,i} \propto   1/\sqrt{n_\mathrm{i}}$ and 
$v_\mathrm{d} \propto  \sqrt{n_\mathrm{i}}$ scales linearly in $1/x_i$. An increase/decrease of the adopted value of $\delta$ would lead to a lower/higher CR ionization rate $\zeta_\mathrm{H}$ (see \autoref{CRionrate}), with $x_i \propto \zeta_\mathrm{H}^{1/2}$ in the warm neutral gas (see \autoref{EqIonFrac}). Thus, streaming would be relatively more important compared to diffusion if the relative abundance of low-energy CRs  is reduced (a flatter distribution -- i.e. higher $\delta$).  Even though the relative contribution of diffusion and streaming transport varies with $\delta$, the CR distribution is only  weakly affected. We show results of models assuming different values of $\delta$ in \autoref{AppendixA3}.

Finally, we point out that the outcomes of \autoref{fig:FsvsFdvsFa_varsigma} refer to the self-consistent model assuming $\sigma_\perp = 10\, \sigma_\parallel$. Clearly, the relative importance of CR diffusion increases/decreases with increasing/decreasing $\sigma_\perp$, as we show in the next section.

\subsection{Models with different perpendicular diffusion}
\label{Diffusion perpendicular to the magnetic field}

For a uniform background magnetic field, CRs diffuse along the magnetic field direction only, due to pitch-angle scattering. However, should magnetic field turbulence be present, CRs can also diffuse perpendicular to the mean magnetic field \citep[e.g.][]{Zweibel13,Shalchi2020} There are several different regimes \citep[see][]{Shalchi2019} with perpendicular diffusion coefficient $\kappa_\perp \sim \kappa_\parallel (\delta B/B)^2$ for $(\delta B/B)$ the fractional magnetic field perturbation and $\kappa_\parallel \equiv 1/\sigma_\parallel$ a parallel transport coefficient. 
In the case of diffusive parallel transport, the corresponding  perpendicular scattering coefficient would be $\sigma_\perp \sim \sigma_\parallel \, (B/ \delta B)^2 $. 
At the gyroradius scale, $\sim 10^{-6}$~pc, $\delta B/B \ll 1$ and diffusion perpendicular to the mean magnetic field would be  negligible compared to parallel diffusion.  While in our simulations we directly follow the transport along the magnetic field, we cannot resolve this all the way down to the kinetic scales, and there would be an effective perpendicular diffusion corresponding to magnetic field perturbations (perpendicular ``wandering'') that are unresolved by our grid. If we extrapolate the large-scale power down to the grid scale ($8$~pc) in our simulations, $(\delta B/B)^2$ is non-negligible, of order $\sim 0.1$ if perturbations are order-unity at the scale height of the disk ($\sim 300$~pc).  The implied  effective perpendicular diffusion would then be an order of magnitude below the resolved parallel diffusion. While to some extent motivating the default choice $\sigma_\perp = 10\, \sigma_\parallel$, this is by no means rigorous.   
To address the issue of uncertain perpendicular diffusion, in this section we explore the effect of different choices of $\sigma_\perp$ on the distribution of CR pressure. In addition to the default model ($\sigma_\parallel = 10\, \sigma_\perp$) discussed above, we post-process the TIGRESS snapshots with a model ignoring perpendicular diffusion ($\sigma_\perp \gg 1$) and with a model assuming isotropic diffusion ($\sigma_\perp = \sigma_\parallel$). 

The comparison of results for the three perpendicular diffusion choices is shown in \autoref{fig:PerpDiff}. The left panel shows the temporally-averaged median value of $\sigma_\parallel$ as a function of $n_\mathrm{H}$. The profiles of $\sigma_\parallel$ produced by the default model and by the model without perpendicular diffusion are nearly identical up to $n_\mathrm{H} \simeq 10^{-2}$~cm$^{-3}$. At higher density, $\sigma_\parallel$ decreases faster in the presence of perpendicular diffusion. 
The model with isotropic diffusion has marginally slower growth in the low-density regime compared to the other two models, while decreasing at high density similar to the default case. These differences can be attributed to the different CR pressure gradients in the three models. The value of  $\sigma_\parallel$ is proportional to $(\vert \hat{\mathbf{B}} \cdot  \nabla {P_\mathrm{c}} \vert)^{1/2}$  at low densities and to ${\vert \hat{\mathbf{B}} \cdot  \nabla  {P_\mathrm{c}} \vert}$ at high densities. CR pressure gradients are more easily smoothed out when the overall diffusion is more effective. This explain why $\sigma_\parallel$ is lower in case of isotropic diffusion. 

The right panel of \autoref{fig:PerpDiff} shows the horizontally- and temporally-averaged vertical profiles of CR pressure for the three models of perpendicular diffusion. As expected, the CR pressure profile becomes flatter and flatter with increasing  perpendicular diffusion. Interestingly, the profile of the default model is closer to that of the model ignoring perpendicular diffusion than to that of the model assuming isotropic diffusion. 
As discussed above, the overall CR propagation efficiency mostly depends on transport in the low/intermediate-density gas ($n_\mathrm{H} < 0.1$~cm$^{-3}$). In the low-density regime ($n_\mathrm{H} < 10^{-2}$~cm$^{-3}$) advection is by far the dominant transport mechanism  (see \autoref{Streaming vs Diffusion vs Advection_Variable Sigma}) and the presence or absence of perpendicular diffusion plays only a marginal role, as confirmed by the almost-identical profiles of $\sigma_\parallel$ in the models with pure parallel diffusion and $\sigma_\parallel = 10\, \sigma_\perp$, and the very similar model with $\sigma_\parallel = \sigma_\perp$. Therefore, the transport of CRs 
is expected to proceed in the same way at low densities. The slightly-different vertical profiles of $P_\mathrm{c}$ for the
pure parallel diffusion and $\sigma_\parallel = 10\, \sigma_\perp$
models are due to the different diffusivity in the intermediate-density gas ($n_\mathrm{H} \simeq 10^{-2}-10^{-1}$~cm$^{-3}$), generally located at the interface between cold/warm and hot gas where advective transport plays a smaller role. In these regions, perpendicular diffusion effectively contributes to transporting CRs perpendicular to the local magnetic field. The lower CR pressure gradients cause the parallel scattering coefficient to decrease ($\sigma_\parallel$ decreases by an order of magnitude when $\sigma_\perp = 10\, \sigma_\parallel$, at $n_\mathrm{H} \sim 10^{-1}$~cm$^{-3}$) and enhancing the overall diffusion. At density  $n_\mathrm{H} \gtrsim 1$~cm$^{-3}$) there is appreciably higher  $\sigma_\parallel$ for the pure parallel diffusion model than the models with nonzero $\sigma_\perp$, but this has negligible effect on the overall CR distribution  because $\sigma$ is still extremely low ($<10^{-30}$~cm$^{-2}$~s; see \autoref{CR pressure-gas density-variablesigma} and \autoref{fig:PcvsnH-variablesigma}).

In the presence of isotropic diffusion the CR pressure profile is much smoother than in the case with $\sigma_\perp = 10\, \sigma_\parallel$, even though the parallel scattering coefficients are quite similar at very low-density and differ by a factor of a few at  higher densities. However, the factor-of-a-few reduction in $\sigma_\parallel$ corresponds to reduction of a few tens in $\sigma_\perp$ compared to the $\sigma_\perp = 10\, \sigma_\parallel$ case.  The effect of reduced $\sigma_\perp$ may be amplified at the interfaces between the midplane layer of warm/cold gas and the surrounding mostly-hot corona because the magnetic field is preferentially horizontal near the midplane region.  The result of isotropic diffusion is significantly more effective CR diffusion overall, and a lower central CR pressure.

Finally, we summarize the relative importance of individual energy sink/source terms for the three transport models analyzed here. We calculate the time-averaged integral of the sink/source terms in the RHS of \autoref{CRenergy} over the whole simulation domain (see \autoref{sec:time-averaged results} for a comparison with models assuming constant scattering). The average CR energy injected per unit time is $1.8 \times 10^{38}$~erg~s$^{-1}$ for all models. For the model  without perpendicular diffusion, we find that relative to the input, the collisional loss is 0.34, the streaming loss is 1.9, and the energy gain from the gas  is 1.6.  For $\sigma_\perp = 10 \, \sigma_\parallel$, the relative collisional loss is  0.23, the streaming loss is 1.4, and the gain from the gas  is 1.2.  With isotropic diffusion, the relative collisional loss is  0.12, the streaming loss is 0.83, and the gain from the  gas is 1.05.
As noted for the constant-diffusivity models, the total rates of energy transfer decrease with
increasing diffusivity. In each  model, the absolute value of the streaming loss rate and of the adiabatic gain rate are comparable, while the collisional loss rate is only 10-20\% of the other terms in absolute value. 

From the total rate of collisonal losses and the total rate of energy injection, we calculate that the grammage (\autoref{eq:grammage}).  This is $\sim 53$~g~cm$^2$ in the absence of perpendicular diffusion, $\sim 34$~g~cm$^2$ when $\sigma_\perp = 10\, \sigma_\parallel$, and $\sim 20$~g~cm$^2$ for isotropic diffusion. These values are from 5 to 2 times larger than the grammage measured at the Earth 
\citep[$\sim 10$~g~cm$^{-2}$, e.g.][]{Hanasz+21}. We point out that the grammage is proportional to the rate of fractional losses, which depends on the CR energy density (see \autoref{eq:grammage}). The higher grammage measured in our physically-motivated models reflects the fact that the predicted CR energy density is slightly larger that the observed one. We refer to \autoref{CR pressure in the Galactic disk} for a discussion about possible reasons for this mismatch.

\subsection{Relation between CR pressure and gas density}
\label{CR pressure-gas density-variablesigma}

\begin{figure}
\centering
\includegraphics[width=0.48\textwidth]{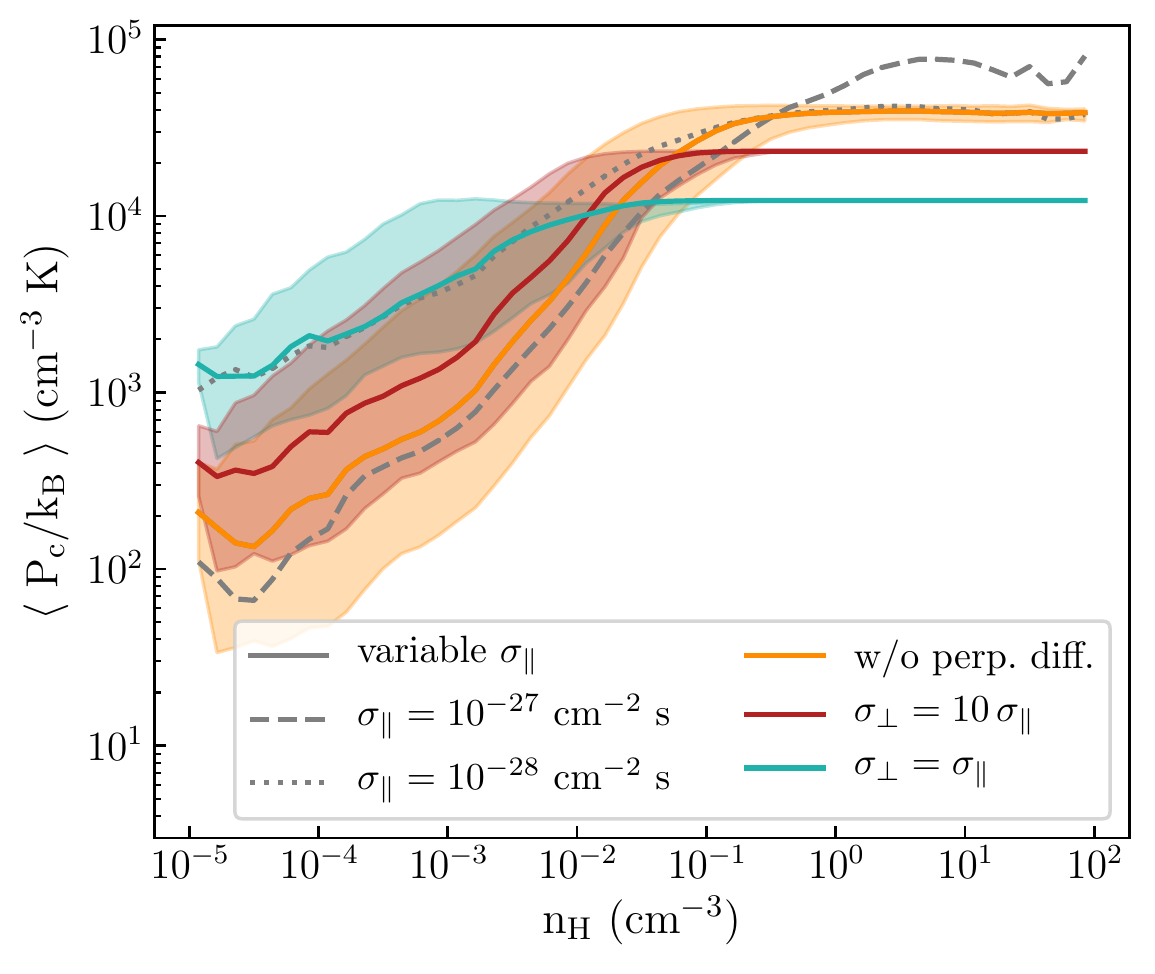}
\caption{Temporally-averaged mean CR pressure $P_\mathrm{c}$ as a function of hydrogen density $n_\mathrm{H}$ 
from self-consistent CR transport models with different treatments of diffusion perpendicular to the magnetic field. We show cases without perpendicular diffusion (orange line), with $\sigma_\perp= 10\, \sigma_\parallel$ (red line) and with isotropic diffusion $\sigma_\perp= \sigma_\parallel$ (turquoise line). The shaded areas cover the 16th to 84th percentiles of the temporally-averaged variations around the mean. For comparison, the results from models with $\sigma_\parallel = 10^{-27}$~cm$^{-2}$~s (dashed gray line) and with $\sigma_\parallel = 10^{-28}$~cm$^{-2}$~s (dotted gray line) are also shown.  
}
\label{fig:PcvsnH-variablesigma}
\end{figure}

We now analyze the relation between CR pressure and gas density in models with variable scattering coefficient. In \autoref{fig:PcvsnH-variablesigma}, we show the temporally-averaged mean of
$P_\mathrm{c}$ as a function  of $n_\mathrm{H}$ from the self-consistent models without perpendicular diffusion (orange),  with  $\sigma_\perp =10\, \sigma_\parallel$ (red), and with isotropic diffusion (turquoise). For comparison, we also plot the mean values of $P_\mathrm{c}$ from the models with $\sigma_\parallel = 10^{-27}$~cm$^{-2}$~s and with $\sigma_\parallel = 10^{-28}$~cm$^{-2}$~s. In all three self-consistent models, the mean value of $P_\mathrm{c}$ increases with $n_\mathrm{H}$ at low density while having a constant value in the high-density regime. The slope of $\log P_\mathrm{c}$ vs.~$\log n_\mathrm{H}$ in the low/intermediate-density regime and the  value of $P_\mathrm{c}$  in the high-density plateau both decrease as the efficiency of perpendicular diffusion increases. 
As expected, the correlation between $P_\mathrm{c}$ and $n_\mathrm{H}$ weakens with isotropic diffusion, since CR inhomogeneities caused by non-uniform advection are more easily erased, while correlations strengthen in the absence of perpendicular diffusion. As explained for \autoref{fig:PerpDiff}, the value of $P_\mathrm{c}$ in the high-density gas is mainly determined by the propagation efficiency at the interface between cold/warm and hot gas. CRs are trapped in the dense neutral gas for a longer time when diffusion is less effective, thus increasing their pressure.  
The $P_\mathrm{c}$-$n_\mathrm{H}$ relation in the case without perpendicular diffusion 
resembles that of the  $\sigma_\parallel = 10^{-27}$~cm$^{-2}$~s model in the low-density regime, while it coincides with that of the $\sigma_\parallel = 10^{-28}$~cm$^{-2}$~s model at high densities. This evidence suggests that the effective scattering coefficient for the most realistic model of CR propagation assuming pure parallel diffusion is between $10^{-27}$~cm$^{-2}$~s and $10^{-28}$~cm$^{-2}$~s.   

It is interesting to note that, in addition to the extremely constant value of $P_\mathrm{c}$ at densities above $n_\mathrm{H}\gtrsim 0.01-0.1$ (i.e. in the warm/cold gas), the three self-consistent models predict negligible scatter around the mean value of $P_\mathrm{c}$, unlike the models with constant scattering coefficient in the range $\sigma_\parallel > 10^{-29}$~cm$^{-2}$~s (see \autoref{fig:PcvsnH}). This is because ion-neutral damping reduces $\sigma_\parallel$ below  $10^{-29}$~cm$^{-2}$~s in the high-density gas for all cases (see left panel of  \autoref{fig:PerpDiff}), and this is low enough to make the CR pressure extremely uniform. This result is particularly important to understanding the dynamical effects of CRs. The absence of CR pressure gradients in the denser regions of the galactic disk implies that CRs do not apply forces to the gas there. This, together with the comparison between the CR and other pressure profiles in the right panel of  \autoref{fig:PerpDiff}, suggests that CRs are not important to vertical support of the ISM disk in the midplane region, while at the same time having potentially great importance to galactic wind launching/fountain dynamics which takes place at high altitudes ($|z|\gtrsim 0.5$~kpc).

\section{Low-energy cosmic rays}
\label{Low-EnergyCRs}

\begin{figure*}
\centering
\includegraphics[width=\textwidth]{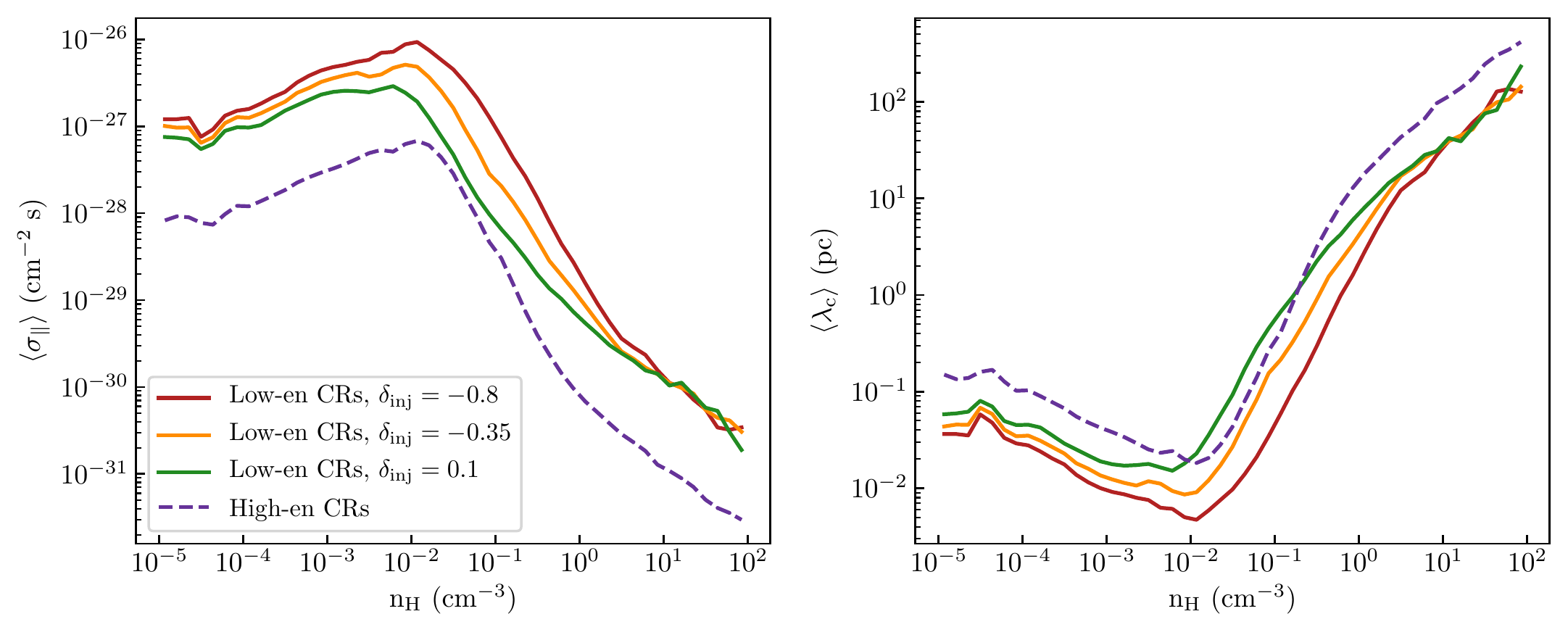}
\caption{Temporally-averaged median of the scattering coefficient $\sigma_\parallel$ (\textit{left panel}) and mean free path $\lambda_\mathrm{c}$ (\textit{right panel}) of high-energy (dashed lines) and low-energy (solid lines) CRs.   For low-energy CRs, three different values of the low-energy slope of the source spectrum have been explored: -0.8 (red lines), -0.35 (orange lines) and 0.1 (green lines). All cases assume no diffusion in the direction perpendicular to the magnetic field lines.}
\label{fig:GevvsMeV_sigma}
\end{figure*}

In this section, we investigate the propagation and distribution of low-energy ($\sim 30$~MeV) CRs using models with variable scattering coefficient (see \autoref{Losses} and \autoref{spectrum} for the treatment of low-energy CRs) ignoring the presence of diffusion perpendicular to the magnetic field direction. Different models are characterized by different assumptions for the fraction of low-energy CRs injected per supernova event. The differing injection fractions correspond to different assumptions for the low-energy slope of the CR injection spectrum $\delta_\mathrm{inj}$, which is observationally quite uncertain. In this work, we explore three values of $\delta_\mathrm{inj}$: $-0.8$, $-0.35$, $0.1$.

\subsection{Scattering rate coefficient and mean free path}\label{sec:MeVsigma}

In the left panel of \autoref{fig:GevvsMeV_sigma}, we compare the temporally-averaged value of the scattering coefficient of high- and low-energy CRs as a function of hydrogen density. Even though the overall profiles are similar, i.e. $\sigma_\parallel$ slightly increases with $n_\mathrm{H}$ up to $n_\mathrm{H} \simeq 10^{-2}$~cm$^{-3}$ and rapidly decreases at higher densities, the scattering coefficient of low-energy CRs at a given density increases with decreasing $\delta_\mathrm{inj}$ and, regardless of the value of $\delta_\mathrm{inj}$, is always higher than the scattering coefficient of high-energy CRs. We note that $\sigma_\parallel$ depends on $n_\mathrm{1}$ (\autoref{GrowthRate2}), which in turns depends both on the shape of the CR energy spectrum and on the CR kinetic energy. In particular, in the low-density regime, where $\Gamma_\mathrm{nll}> \Gamma_\mathrm{in}$, $\sigma_\parallel \propto \sqrt{n_\mathrm{1}}$, while in the high-density regime, where $\Gamma_\mathrm{nll} < \Gamma_\mathrm{in}$, $\sigma_\parallel \propto {n_\mathrm{1}}$  (see \autoref{NLL} and \autoref{IN}). In \autoref{AppendixA1}, we show that the value of $n_\mathrm{1}$ at $E_\mathrm{k} \sim 30$~MeV increases by a factor of $\sim 7$ when the low-energy slope of the spectrum decreases from $0.1$ to $-0.8$. In \autoref{fig:GevvsMeV_sigma}, we can in fact see that the average ratio between the value of $\sigma_\parallel$ predicted by the model assuming $\delta_\mathrm{inj}=-0.8$ and the value of $\sigma_\parallel$ predicted by the model assuming $\delta_\mathrm{inj}=0.1$ is $\approx 2-3$ at $n_\mathrm{H} \lesssim 10^{-2}$~cm$^{-3}$, where non-linear Landau damping dominates, and slightly less than one order of magnitude at $10^{-2} < n_\mathrm{H} < 1$~cm$^{-3}$, where ion-neutral damping dominates. At higher densities, the distributions of $\sigma_\parallel$ predicted by the three different models for low-energy CRs nearly overlap. In this density regime, the scattering coefficient decreases with increasing the CR ionization rate (because $\sigma_\parallel \propto n_i^{-1/2} \propto \zeta_\mathrm{c}^{-1/4}$; see \autoref{AppendixA3}), which, in turn, increases with decreasing $\delta_\mathrm{inj}$. Thus, the tendency for $\sigma_\parallel$ to increase with $n_1$ at lower $\delta_\mathrm{inj}$ is counterbalanced by the decrease of $n_i^{-1/2}$. 

In \autoref{AppendixA1}, we also show that, regardless of the value of $\delta$, $n_\mathrm{1}$ is always higher at $E_\mathrm{k} = 30$~MeV than at $E_\mathrm{k} = 1$~GeV for a given spectrum normalization. In particular, the value of $n_1$ at $E_\mathrm{k} = 30$~MeV is a factor of $\sim 3$ higher than the value of $n_1$ at $E_\mathrm{k} = 1$~GeV if $\delta = 0.1$ and more than one order of magnitude if $\delta = -0.8$. In \autoref{fig:GevvsMeV_sigma}, we can however observe that $\sigma_\parallel$ decreases by more than a factor $\sqrt{n_1}$ going from low-energy to high-energy CRs at $n_\mathrm{H} \lesssim 10^{-2}$~cm$^{-3}$.  The reason is that, in the low-density regime, the scattering rate is inversely proportional to the particle speed $v_\mathrm{p}$ (see \autoref{NLL}), which is higher for CRs with $E_\mathrm{k} = 1$~GeV ($v_\mathrm{p} \simeq 2.6 \times 10^{10}$~cm~s$^{-1}$) than for CRs with $E_\mathrm{k} = 30$~MeV ($v_\mathrm{p} \simeq 7.4 \times 10^9$~cm~s$^{-1}$). Moreover, as diffusion becomes more important for high-energy CRs, the scale heights of their distribution decrease ($\sigma_\parallel = \sigma_{\parallel,\mathrm{NLL}} \propto ({\mid \hat{B} \cdot \nabla P_\mathrm{c}\mid}/P_\mathrm{c})^{-1/2}$). 

The right panel of \autoref{fig:GevvsMeV_sigma} shows the temporally-averaged mean free path of high- and low-energy CRs as a function of hydrogen density. The mean free path $\lambda_\mathrm{c}$ is calculated as $(v_\mathrm{p} \sigma_\parallel)^{-1}$, where $v_\mathrm{p}$ is the CR velocity (\autoref{eq:proton_velocity}). Since the speed of CRs with $E_\mathrm{k} = 1$~GeV is higher than the speed of CRs with $E_\mathrm{k} = 30$~MeV, the mean free path of high-energy CRs is only slightly larger the mean free path predicted by the propagation models for low-energy CRs assuming $-0.35 < \delta_\mathrm{inj} < 0.1$, even though the scattering coefficient is lower. For high-energy (low-energy) CRs, the average mean free path decreases from $\lambda_\mathrm{c} \simeq 0.1-0.2$~pc ($\lambda_\mathrm{c} \simeq 0.03-0.06$~pc) at $n_\mathrm{H} = 10^{-5}$~cm$^{-3}$ to $\lambda_\mathrm{c} \simeq 0.01-0.03$~pc ($\lambda_\mathrm{c} \simeq 0.05-0.1$~pc) at $n_\mathrm{H} \simeq 10^{-2}$~cm$^{-3}$, where scattering is highly effective. At higher densities, the mean free path quickly increases as the scattering coefficient decreases. At $n_\mathrm{H} \simeq 10-10^{2}$~cm$^{-3}$ -- the characteristic density of cold atomic and diffuse molecular clouds -- $\lambda_\mathrm{c} \sim 30-300$~pc for low-energy CRs and slightly higher for high-energy CRs. With a mean free path in the cold dense gas comparable to the size of individual clouds, CRs freely stream across them (subject, however, to the increased collisional losses at higher density). 

\begin{figure*}
\centering
\includegraphics[width=\textwidth]{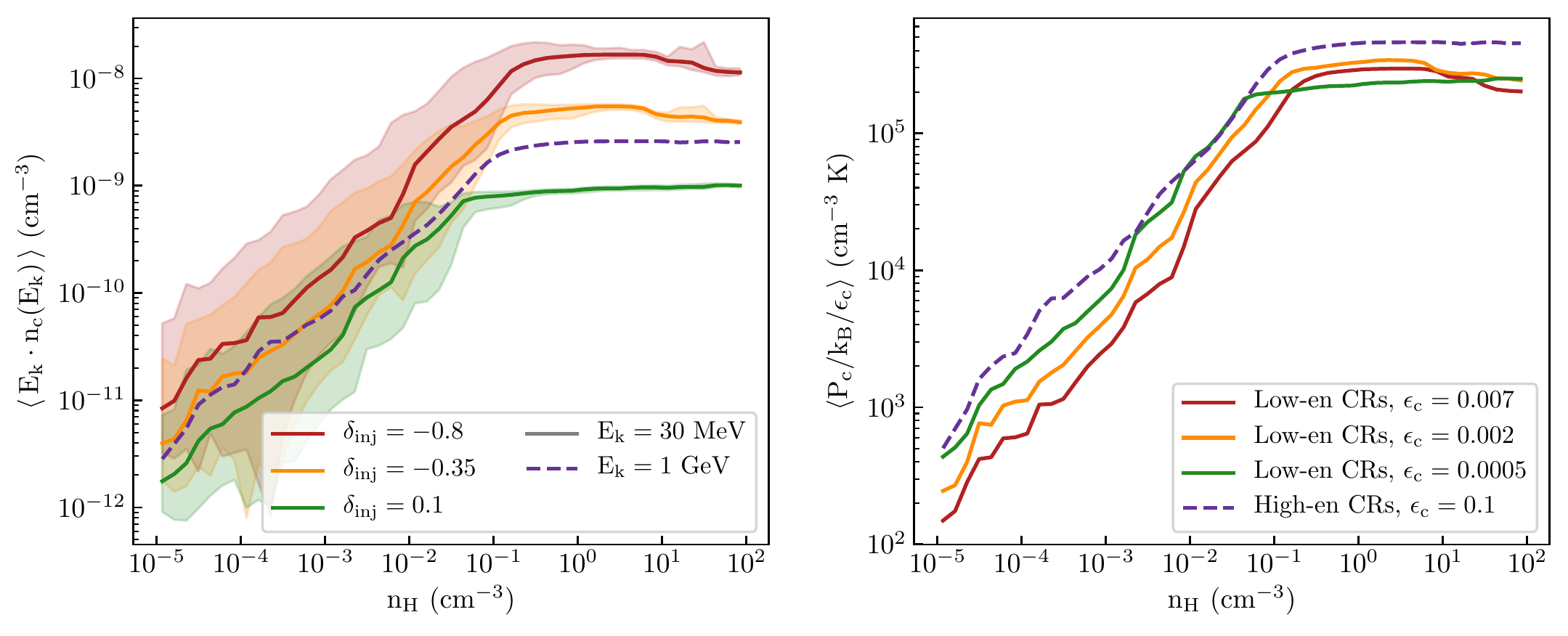}
\caption{Comparison of the self-consistent transport model for high-energy and low-energy CRs, for models without perpendicular diffusion. For low-energy CRs, three different values of the low-energy slope of the injected spectrum have been explored: $\delta_\mathrm{inj}=$ -0.8 (red lines), -0.35 (orange lines) and 0.1 (green lines). \textit{Left panel}: temporally-averaged median density of CRs in a bin of width $E_\mathrm{k}$ as a function of hydrogen density $n_\mathrm{H}$, for  $E_\mathrm{k} = 30$~MeV (solid lines) and $E_\mathrm{k} = 1$~GeV (dashed line). The shaded areas cover the 16th and 84th percentiles of the distribution.
\textit{Right panel}: temporally-averaged median pressure $P_\mathrm{c}$ of high-energy (dashed line) and low-energy (solid lines) CRs as a function of hydrogen density $n_\mathrm{H}$. The value of $P_\mathrm{c}$ is divided by $\epsilon_\mathrm{c}$, the fraction of supernova energy converted into CRs with a given kinetic energy. The fraction  $\epsilon_\mathrm{c}=0.1$ for high-energy CRs, while it depends on the assumption made for $\delta_\mathrm{inj}$ for low-energy CRs.}
\label{fig:GevvsMeV_P}
\end{figure*}

\subsection{Density dependence and CR losses}\label{sec:MeVdens}

The results of the propagation models for low-energy CRs and a comparison with results for high-energy CRs are displayed in \autoref{fig:GevvsMeV_P}. 
The left panel shows the temporally-averaged median density, $E_\mathrm{k} n_\mathrm{c}(E_\mathrm{k})$, of CRs with kinetic energy $E_\mathrm{k} \simeq 30$~MeV and $E_\mathrm{k} \simeq 1$~GeV in a bin of width $E_\mathrm{k}$ as a function of hydrogen density $n_\mathrm{H}$. For low-energy CRs, $E_\mathrm{k} n_\mathrm{c}(E_\mathrm{k} = 30\, \mathrm{MeV}) = e_\mathrm{c} (\mathrm{MeV})/\Delta E_\mathrm{k} \cdot E_\mathrm{k}/E(E_\mathrm{k})$, with $\Delta E_\mathrm{k}=1$~MeV (see \autoref{spectrum}). For high-energy CRs, the normalization of $n_\mathrm{c}(E_\mathrm{k})$ is calculated from the energy density $e_\mathrm{c}(\mathrm{GeV})$ using \autoref{SpecNorm}, while the low-energy slope is assumed to be $-0.35$ (default model). At a given $n_\mathrm{H}$, the average CR density increases for lower $\delta_\mathrm{inj}$ for the $E_\mathrm{k} = 30$~MeV CRs, and for $\delta_\mathrm{inj}=-0.35$ and $-0.8$ the number density is also higher than for the GeV CRs, consistent with the injection spectrum.
Despite the shift in normalization, the distributions of CR density predicted by the four models are roughly similar, with $n_\mathrm{c}$ increasing up to $n_\mathrm{H} \sim 0.01-0.1$~cm$^{-3}$ and flattening at higher densities. However, unlike the model assuming $\delta_\mathrm{inj} = 0.1$, the models with $\delta_\mathrm{inj} = -0.35$ and $-0.8$ predict a slight decrease of CR density at $n_\mathrm{H} \gtrsim 10$~cm$^{-3}$. For these models, the higher scattering rates in the intermediate/low density gas (see \autoref{fig:GevvsMeV_sigma}) trap CRs more effectively near the midplane, and this provides more time for CRs in the dense gas to lose energy. Since the rate of energy losses increases with $n_\mathrm{H}$ (\autoref{Gammaloss2}), the CR density decreases with $n_\mathrm{H}$.

For a more direct comparison between the energy-density distributions of high-energy and low-energy CRs, in the right panel of \autoref{fig:GevvsMeV_P} we show the average distributions of $P_\mathrm{c} (\mathrm{GeV})/\epsilon_\mathrm{c}(\mathrm{GeV})$ and $P_\mathrm{c} (\mathrm{MeV})/\epsilon_\mathrm{c}(\mathrm{MeV})$, where $\epsilon_\mathrm{c}(\mathrm{GeV})$ and $\epsilon_\mathrm{c}(\mathrm{MeV})$ are the fractions of supernova energy converted into GeV and MeV CRs respectively, as a function of $n_\mathrm{H}$. $\epsilon_\mathrm{c}$ is set to  $0.1$ for high-energy CRs, while it depends on the assumption made for $\delta_\mathrm{inj}$ for low-energy CRs, as shown in the legend of \autoref{fig:GevvsMeV_P}. As explained in \autoref{CR pressure-gas density} and \autoref{CR pressure-gas density-variablesigma}, the effect of increasing scattering is to prevent the propagation of CRs from high-density to low-density regions. As a consequence, for higher $\sigma$ the CR pressure tends to decrease in low-density regions and increase in higher-density regions (see also \autoref{fig:PcvsnH} and \autoref{fig:PcvsnH-variablesigma}). In \autoref{fig:GevvsMeV_P}, at low gas densities $P_\mathrm{c}/\epsilon_\mathrm{c}$ indeed decreases going from high-energy to low-energy CRs and going from the model with $\delta_\mathrm{inj}=0.1$ to the model $\delta_\mathrm{inj}=-0.8$. However, in the high density regime, $P_\mathrm{c}/\epsilon_\mathrm{c}$ is always lower for low-energy than for high-energy CRs, even though the higher scattering of the MeV CRs traps them more effectively. The reason is that low-energy CRs undergo more significant collisional energy losses, which are particularly effective in the dense gas. 

We find that the average fraction of injected energy lost via collisions with the ambient gas is $\sim 0.59, \,\sim 0.52$ and $\sim 0.41$ for low-energy CRs models adopting $\delta_\mathrm{inj} = -0.8, \,- 0.35$ and $ 0.1$, respectively. These losses exceed the fractional loss $f_\mathrm{coll} \sim 0.34$ for high-energy CRs.  Based on \autoref{eq:grammage},  
the time-averaged grammage of low-energy CRs is $\sim 12$, $\sim 10$ and $\sim 8$~g~cm$^{-2}$ for the case $\delta_\mathrm{inj} = -0.8, \,- 0.35$ and $ 0.1$, respectively, 
lower than the time-averaged grammage of high-energy CRs $\sim 53$~g~cm$^{-2}$. We note that even though the fractional loss is larger for low-energy CRs than for high-energy CRs, the latter are characterized by a larger velocity, which explains why their grammage exceeds the grammage of low-energy CRs.

\subsection{Role of streaming, diffusive, and advective transport}

\begin{figure*}
\centering
\includegraphics[width=\textwidth]{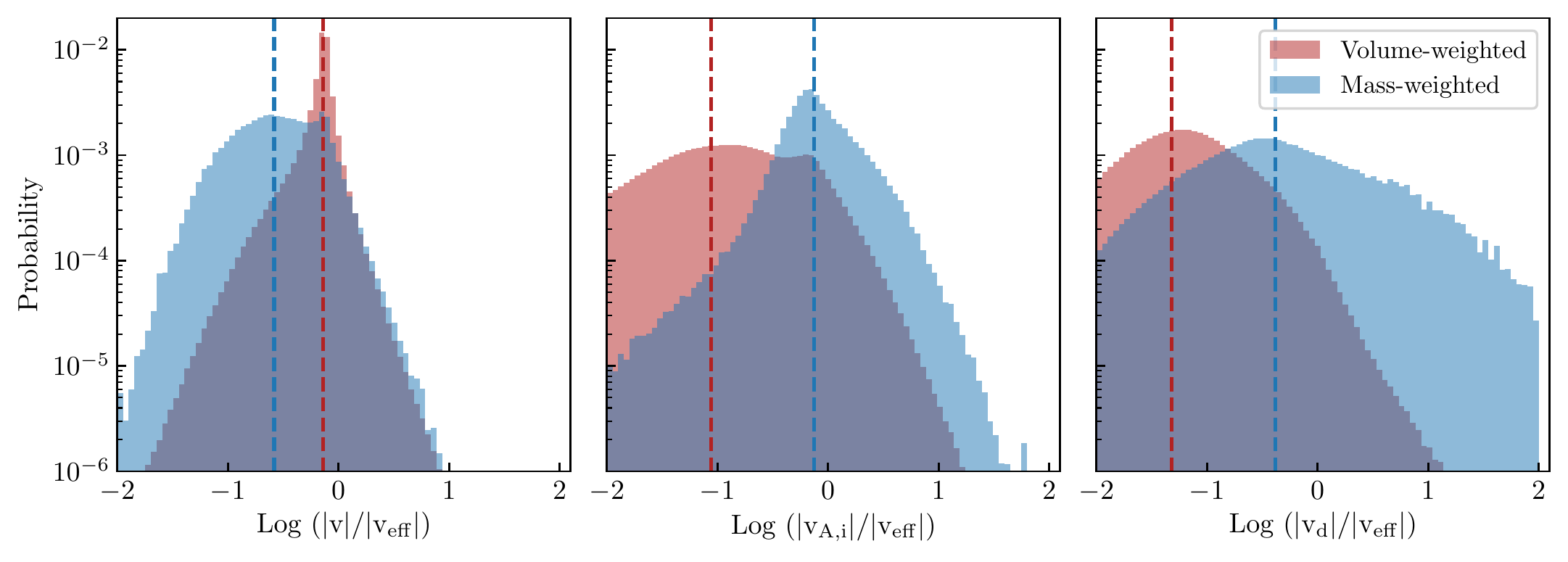}
\caption{Relative contribution to the total flux of low-energy CRs from advection, streaming, and diffusive terms, for the self-consistent model without perpendicular diffusion.
Volume-weighted (red histograms) and mass-weighted (blue histograms) show probability distributions of the ratio between advection speed $v$ (\textit{left panel}), ion Alfv\'{e}n speed $v_\mathrm{A,i}$ (\textit{middle panel}), diffusive speed $v_\mathrm{d}$ (\textit{right panel}), and the effective total CR propagation speed defined as $v_\mathrm{eff}$. The red and blue dashed lines indicate the median values of the volume-weighted and mass-weighted distributions, respectively. The analysis is performed on the snapshot at $t = 286$~Myr adopting the model with $\delta_\mathrm{inj} = -0.35$.}
\label{fig:Hist_MeV}
\end{figure*}

In this section, we evaluate the relative contribution of streaming, diffusion and advection to the overall propagation of low-energy CRs. \autoref{fig:Hist_MeV} shows the volume-weighted (red histograms) and mass-weighted (blue histograms) probability distributions of $\vert v \vert$/$\vert v_\mathrm{eff} \vert$, $\vert v_\mathrm{A,i} \vert$/$\vert v_\mathrm{eff} \vert$ and $\vert v_\mathrm{d} \vert$/$\vert v_\mathrm{eff} \vert$ for the model adopting $\delta_\mathrm{inj} = -0.35$. As for high-energy CRs, advection contributes the most to the transport of CRs when weighted by volume, while streaming and diffusion dominates over advection when weighted by gas mass (see \autoref{Streaming vs Diffusion vs Advection_Variable Sigma}). However, unlike high-energy CRs, the streaming velocity of low-energy CRs is on average larger than their diffusion velocity. While the streaming velocity distribution is the same for low-energy and high-energy CRs\footnote{Strictly speaking this could differ, but the spectrum normalization used to calculate the CR ionization rate (\autoref{CRionrate}), relevant for the calculation of the ion Alfv\`en speed, is the same for low-energy and high-energy CRs}, the diffusion velocity distribution is different. In \autoref{sec:MeVsigma}, we have indeed seen that in higher-density regions, containing the bulk of the gas mass, the scattering coefficient of low-energy CRs is almost one order of magnitude larger than the scattering coefficient of high-energy CRs (see \autoref{fig:GevvsMeV_sigma}), which results in a lower diffusion velocity for low-energy CRs compared to high-energy CRs. We therefore conclude that streaming is the primary transport mechanism for low-energy CRs in the midplane regions containing most of the ISM mass. We find that diffusion dominates over streaming in highly-dense regions ($n_\mathrm{H} > 10$~cm$^{-3}$) only.  A caveat, however, is that if the overall level of CRs were decreased (e.g. with altered magnetic field topology -- see \autoref{CR pressure in the Galactic disk}), the scattering rate would drop and this would tend to enhance diffusion.  

\subsection{CR spectrum and ionization rate}
\label{Low-energy slope}

\begin{figure*}
\centering
\includegraphics[width=\textwidth]{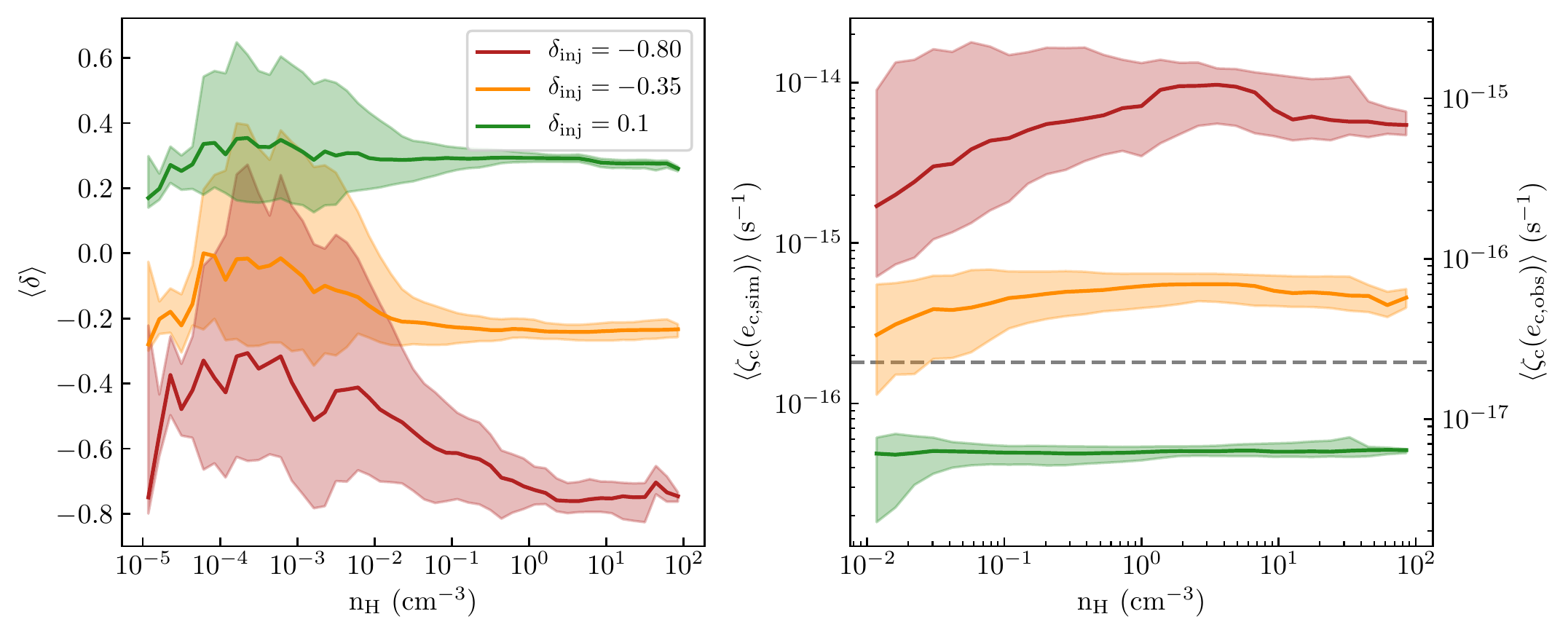}
\caption{Temporally-averaged median of the low-energy slope $\delta$ (\textit{left panel}) and primary CR ionization rate of atomic hydrogen $\zeta_\mathrm{c}$ (\textit{right panel}) as a function of hydrogen density $n_\mathrm{H}$ in models with $\delta_\mathrm{inj} = -0.8$ (red lines), $\delta_\mathrm{inj} = -0.35$ (orange lines) and $\delta_\mathrm{inj} = 0.1$ (green lines). In the right panel, the left-hand-side $y-$axis denotes $\zeta_\mathrm{c}(e_\mathrm{c,sim})$, the primary CR ionization rate calculated adopting the spectrum normalization $C$ (\autoref{SpecNorm}) predicted by the self-consistent GeV propagation model without perpendicular diffusion analysed in \autoref{Diffusion perpendicular to the magnetic field}, while the right-hand-side $y-$axis denotes $\zeta_\mathrm{c}(e_\mathrm{c,obs})$, the primary CR ionization rate calculated assuming that the mean energy density of high-energy CRs near the mid-plane is 1~eV~cm$^{-3}$, as observed in the solar neighborhood.  
In both plots, the shaded areas cover the 16th and 84th percentiles of the distribution at a given $n_\mathrm{H}$.}
\label{fig:CRionrate}
\end{figure*}

We now focus on the effect of different choices of $\delta_\mathrm{inj}$ on the average CR ionization rate in the mid-plane region of the galactic disk ($\vert z \vert \, \leqslant 250$~pc). \autoref{fig:CRionrate} shows the temporally-averaged low-energy slope of the CR spectrum (\autoref{delta_spectrum}) and primary CR ionization rate of atomic hydrogen (\autoref{CRionrate}) as a function of hydrogen density obtained under the three different assumptions of $\delta_\mathrm{inj}$. 
As the propagation of high-energy and low-energy CRs differ, with the latter scattering more and having more significant collisional energy losses, the local slope of the spectrum differs from the slope of the injected spectrum. The local slope $\delta$ is equal to $\delta_\mathrm{inj}$ at very low densities ($n_\mathrm{H} \sim 10^{-5}$~cm$^{-3}$) only, i.e. at the typical densities of supernova remnants where CRs are injected, and increases with the gas density up to $n_\mathrm{H} \simeq 10^{-3}$~cm$^{-3}$. Since diffusion 
is less effective for low-energy CRs, the ratio between $e_\mathrm{c}(\mathrm{GeV})$ and $e_\mathrm{c}(\mathrm{MeV})$ is higher than their injection ratio in the low-density regime away from injection sites (see previous section and \autoref{fig:GevvsMeV_P}), thus making the CR spectrum flatter at $n\sim 10^{-3}$~cm$^{-3}$. At higher densities, $\delta$ approaches a constant value for $n_\mathrm{H} \gtrsim 0.1-1$~cm$^{-3}$ which is slightly larger than  $\delta_\mathrm{inj}$. This reflects the near-constant level of  $e_\mathrm{c}$ for both high-energy and low-energy CRs at high gas densities. We note that the scatter in the distribution of $\delta$ increases at a given $n_\mathrm{H}$ with decreasing $\delta_\mathrm{inj}$ as diffusion becomes less effective.

In the right panel of \autoref{fig:CRionrate}, the trend of the CR ionization rate $\zeta_\mathrm{c}$ is analyzed for $n_\mathrm{H} \geqslant 10^{-2}$~cm$^{-3}$ only, since hydrogen is fully ionized at lower densities. The overall distribution of $\zeta_\mathrm{c}$ reflects that of $e_\mathrm{c}$ for low-energy CRs (\autoref{fig:GevvsMeV_P}), i.e. $\zeta_\mathrm{c}$ becomes more uniform with increasing $\delta_\mathrm{inj}$. In the local Milky Way, the primary CR ionization rate of atomic hydrogen measured in local diffuse molecular clouds ($n_\mathrm{H} \simeq 100$~cm$^{-3}$) is $\zeta_\mathrm{c} = 1.8^{+1.3}_{-1.1} \times 10^{-16}$~s$^{-1}$ (e.g.~\citealt{Indriolo&McCall12, Indriolo+15, Bacalla+19}, see also review by \citealt{Padovani+20} and references therein). These numbers lie between the value of $\zeta_\mathrm{c} \simeq 6 \times 10^{-17}$~s$^{-1}$ predicted by the model assuming $\delta_\mathrm{inj} = 0.1$ and the value of $\zeta_\mathrm{c} \simeq 5 \times 10^{-16}$~s$^{-1}$ predicted by the model assuming $\delta_\mathrm{inj} = -0.35$ at $n_\mathrm{H} \simeq 100$~cm$^{-3}$. However, as we shall discuss in \autoref{CR pressure in the Galactic disk}, the average energy density of high-energy CRs predicted by the self-consistent model
is in fact larger than the average energy density measured in the local ISM, and
the normalization of the ionization rate is set by the GeV energy density (in this case, using the model without perpendicular diffusion).  
To match the observed midplane GeV energy density of  $1$~eV~cm$^{-3}$, we can reduce the ionization rates predicted by our models, $\zeta_\mathrm{c}(e_\mathrm{sim})$, by a factor of 8; the resulting  $\zeta_\mathrm{c}(e_\mathrm{obs})$ is  indicated on the right-hand-side $y-$axis of \autoref{fig:CRionrate}. Accounting for this reduction, the observed CR ionization rate lies between the value of $\zeta_\mathrm{c}\simeq 7 \times 10^{-17}$~s$^{-1}$ predicted by the model adopting $\delta_\mathrm{inj} = -0.35$ and the value of $\zeta_\mathrm{c}\simeq 7 \times 10^{-16}$~s$^{-1}$ predicted by the model adopting $\delta_\mathrm{inj} = -0.8$. We note that the values of $\zeta_\mathrm{c}$ displayed in \autoref{fig:CRionrate} have been obtained adopting $E_\mathrm{k,min} = 10^5$~eV in \autoref{CRionrate}. Using $E_\mathrm{k,min} = 10^6$~eV, consistent with the minimum CR energy probed by Voyager~1, the value of $\zeta_\mathrm{c}(e_\mathrm{obs})$ for the $\delta_\mathrm{inj} = -0.8$ model would be in better agreement with the observed value (see \autoref{AppendixA2}).

We conclude that $-0.35 < \delta < -0.8$ might be a good approximation for the low-energy slope of the injected CR energy spectrum and that the average value of $\delta$ at the average ISM densities ($n_\mathrm{H} \simeq 0.1-1$~cm$^{-3}$) is likely to lie between $-0.7$ and $-0.25$. Moreover, we note that the slight anti-correlation between CR ionization rate and hydrogen density predicted by the model adopting $\delta_\mathrm{inj} = -0.8$ at $n_\mathrm{H}>1$~cm$^{-3}$ is generally in agreement with  observations of diffuse molecular clouds in the solar neighborhood \citep[e.g.,][but note that the observed anti-correlation is between the CR ionization rate and column density, rather than volume density]{Neufeld&Wolfire17}. We further discuss these results in \autoref{CR ionization rate}.

\section{Discussion}
\label{Discussion}

\subsection{CR pressure in the Galactic disk}
\label{CR pressure in the Galactic disk}

Except for the 
case with isotropic diffusion, the self-consistent propagation models presented in \autoref{Gev-Variablesigma} predict that near the mid-plane the average pressure of CRs with kinetic energies of $\gtrsim$ 1 GeV is $P_\mathrm{c}/k_B = 2 - 3 \times 10^4 \,{\rm cm^{-3}\ K} $, under the assumption that the energy input rate is $\epsilon_\mathrm{c}=10$\% of the SN rate.  This is a few times larger than the midplane thermal, kinetic, and magnetic field pressures, each of which is $\approx 10^4 \,{\rm cm^{-3}\ K}$ in the warm/cold atomic gas which comprises most of the ISM's mass (see \autoref{MHDsim} and  \autoref{fig:PerpDiff}, and note that the magnetic pressure is lower in the hot gas).  While still close to
equipartition, the CR pressure here exceeds that of the other ISM components.  This can be compared with the local  Milky Way, where the estimated cosmic ray, thermal, kinetic, and magnetic pressures are individually in the range $\sim  3000-10000\,{\rm cm^{-3}\ K}$, i.e. somewhat closer to 
equipartition with each other (as well as slightly smaller than in the simulation).  

When star formation feedback dominates other energy inputs to the ISM, approximate equipartition can be understood based on input rates (mostly from radiation and supernovae) and the response of the ISM to the various forms of input.  The individual pressures are set by balancing far-UV (photoelectric)  heating and cooling for thermal pressure \citep{Ostriker2010}, balancing momentum flux injection from supernovae with kinetic turbulent pressure  \citep{Ostriker2011}, and applying turbulent driving in combination with shear to maintain the pressure in the magnetic field \citep{KimOstriker2015}.  The ratios between midplane pressure components and the star formation rate per unit area $\Sigma_\mathrm{SFR}$ are the feedback ``yield'' components \citep[][Ostriker \& Kim 2021, in prep.]{Kim2013,Kim+20,WT_Kim2020}, and  the ratios among the individual pressure components simply reflects the relative feedback yields.  

Just as for the other pressure components that derive from star formation feedback, there must be a relationship between the midplane CR pressure $P_\mathrm{c}(0)$ and $\Sigma_\mathrm{SFR}$. 
In the case of negligible losses (collisional or via work on the gas), the average vertical flux of CR energy 
above the SN input layer would be $F_\mathrm{c,z}=(1/2)\epsilon_\mathrm{c}  E_\mathrm{SN}  \Sigma_\mathrm{SFR}  / m_\star$, where $m_\star$ is the total mass of new stars per supernova \citep[$95.5 M_\odot$  in][from a Kroupa IMF]{Kim&Ostriker17}; this ``no-losses'' CR flux is $2 \times 10^{45}\, \mathrm{erg\, yr^{-1} \, kpc^{-2}}$. We can also relate the flux and pressure by
$P_\mathrm{c}(0)=\sigma_\mathrm{eff} H_\mathrm{c,eff} F_\mathrm{c,z}$ for  $H_\mathrm{c,eff}= \langle |d\ln P_\mathrm{c}/dz| \rangle^{-1}$ an effective CR scale height and  $\sigma_\mathrm{eff}^{-1}$ an effective diffusion coefficient. With $H_\mathrm{c,eff}\sim 1.3$~kpc from our self-consistent simulations with $\sigma_\perp = 10\, \sigma_\parallel$, the midplane pressure-flux relation would be satisfied for $\sigma_\mathrm{eff}=6 \times 10^{-29}$~cm$^{-2}$~s, while $H_\mathrm{c,eff}\sim 0.7$~kpc and $\sigma_\mathrm{eff}=2 \times 10^{-28}$~cm$^{-2}$~s for the model without perpendicular diffusion.  Note that these values of $\sigma_\mathrm{eff}$ use the actual midplane CR pressure and vertical CR flux at $|z|=1$~kpc  ($2.4 \times 10^{45}\, \mathrm{erg\, yr^{-1} \, kpc^{-2}}$ or $1.9 \times 10^{45}\, \mathrm{erg\, yr^{-1} \, kpc^{-2}}$, respectively), which differ slightly from the ``no-losses'' vertical CR flux. 
By comparison, the $P_\mathrm{c}$ vs.~$n_\mathrm{H}$ relation in our self-consistent models
is best matched by the constant $\sigma$ models 
when $\sigma_\parallel= 10^{-27}-10^{-28}$~cm$^{-2}$~s (depending on the density range; see \autoref{fig:PcvsnH-variablesigma}), consistent with the measured peaks in $\sigma_\parallel$ in \autoref{fig:PerpDiff}. The lower $\sigma_\mathrm{eff}$ can be understood since (1) in fact advection dominates in the lowest-density gas, and (2) Alfv\`enic streaming and diffusion are comparable when realistic ionization is taken into account (\autoref{fig:FsvsFdvsFa_varsigma}). Both of these effects increase the rate of transport, contributing to a reduction in $\sigma_\mathrm{eff}$ compared to the actual scattering rate.     
The corresponding CR feedback ``yield'' 
\begin{equation}
    \Upsilon_\mathrm{c} \equiv \frac{P_\mathrm{c}(0)}{\Sigma_\mathrm{SFR}} =  \frac{1}{2}\epsilon_\mathrm{c} \sigma_\mathrm{eff} H_\mathrm{c,eff} \frac{E_\mathrm{SNe}}{m_*}  
    \label{eq:UpsilonCR}
\end{equation}
 would then be $\Upsilon_\mathrm{c}\sim 700 \,\kms$ or $\sim 1000\,\kms$, respectively, respectively, for the $\sigma_\perp = 10\, \sigma_\parallel$ model or the model without perpendicular diffusion.

In the TIGRESS simulations (and presumably for the real ISM as well), the disk as a whole is in vertical dynamical equilibrium, with the ISM weight ${\cal W}$ balanced by the difference 
$\Delta P$ between midplane pressure and pressure at the top of the atomic/molecular layer.  In the current TIGRESS simulations, the weight is balanced by the 
sum of the midplane thermal, kinetic, and magnetic pressure \citep[][Ostriker \& Kim 2021, in prep.]{Vijayan+20,WT_Kim2020}.   
If $P_\mathrm{tot}(0) \approx \Delta P \approx \mathcal{W}$, we then have 
$\Sigma_\mathrm{SFR}={\cal W}/\Upsilon$ using the total feedback yield $\Upsilon$.    
In the pressure-regulated, feedback-modulated theory, 
$\Upsilon$
then controls the star formation rate, 
with thermal+turbulent+magnetic terms yielding  $\Upsilon \sim 10^3 \,{\rm km/s}$ for the solar neighborhood model \citep{Kim&Ostriker17}.

In principle, CR pressure could also contribute to the vertical support of the  disk, and in doing so participate in regulating the star formation rate. However, it is important to note that only the \emph{difference} $\Delta P$ between midplane and high-altitude pressure 
contributes to vertical support against gravity in the ISM. 
For the thermal, kinetic, 
and magnetic pressure, the high-altitude 
($\sim 0.5$~kpc) values are very small compared to the midplane values, so that 
$\Delta P$ is essentially the same as the 
midplane value.  For the CRs, in contrast, the pressure is nearly uniform within the neutral gas layer (see \autoref{fig:varsigma_snaps}, \autoref{fig:PerpDiff}, \autoref{fig:PcvsnH-variablesigma}) because ion-neutral collisions damp resonant Alfv\'en waves, keeping the scattering rate quite small. As a consequence, $\Delta P_\mathrm{c}\ll P_\mathrm{c}(0)$, and the contribution of CRs to supporting the ISM weight 
is expected to be small.  As a consequence, CRs would also not contribute to the control of star formation on $\sim$~kpc scales;  $\Upsilon_\mathrm{c}$ would not be included in the total $\Upsilon$ that is used to predict 
the SFR via $\Sigma_\mathrm{SFR} = \mathcal{W}/\Upsilon$.

The fact that the midplane CR pressure is a factor of $\sim 5-8$ larger than observed Milky Way values is in part because all pressures are slightly enhanced in this particular TIGRESS simulation compared to the solar neighborhood. Fine-tuning of the adopted galactic model, together with inclusion of ionizing radiation feedback to create \ion{H}{2} regions, could reduce this.  However, the CR pressure is \emph{more} enhanced than other pressures.  One possible reason for this is that the TIGRESS MHD simulation does not self-consistently include CRs.  For the reasons explained above, we do not expect that inclusion of CRs in the MHD simulation would appreciably reduce the SFR.  However, there could potentially be a significant difference to the magnetic structure at high altitude.  \autoref{fig:Gev_snaps} shows that there is generally a very large CR pressure gradient between the mostly-neutral midplane gas and the surrounding corona,  \autoref{MHDsim} shows that the magnetic field is preferentially horizontal in the midplane gas, and \autoref{fig:diffpar-diffperp} and \autoref{fig:PerpDiff} show that magnetic geometry and low perpendicular diffusion can significantly limit  CR transport.  It is likely that if the back-reaction of the CR pressure on the gas were included,  the  strain would cause the magnetic field lines at high altitude to open up in the direction perpendicular to disk \citep{Parker69}. This rearrangement of magnetic field topology would enable CRs that would  otherwise be trapped in the ISM to stream and diffuse out of the disk along the magnetic field lines, leading to a significant decrease in the CR pressure near the mid-plane.

With fully time dependent simulations, we will be able to determine whether the CR pressure is reduced to be closer to the other pressures.  If this is not the case, it would instead suggest that modification of the scattering rate coefficients (see \autoref{sigma}) is required.  
Considering the case with $\sigma_\perp = 10\,\sigma_\parallel$, we find that an increase of the damping rates (\autoref{DampingRateIN} and \autoref{DampingRateNLL}) or a reduction of the Alfv\'{e}n-wave growth rate (\autoref{GrowthRate2}) by a factor of $\sim 10$ would be required for the CR pressure to be consistent with the other pressures near the mid-plane.
In fact, recent MHD-PIC simulations of nonlinear streaming instability and quasilinear diffusion with local damping suggest an effective scattering rate about half of the traditional theoretical value $(\pi/8)(\delta B/B)^2  \Omega$  \citep{Bambic2021}, already alleviating some of the tension.

Finally, it is worth pointing out that the observed CR pressure in the Milky Way is effectively a single evolutionary snapshot of the solar neighborhood, while the CR pressures shown in \autoref{fig:Gev_profiles_varsigma} and \autoref{fig:PerpDiff} are the result of temporal averaging. Hence, the comparison with observations should be taken with a grain of a salt. For example, if we consider the individual snapshot at $t=536$~Myr for the model with $\sigma_\perp = 10 \sigma_\parallel$, the midplane CR energy density is 1.5 eV~cm$^{-3}$, in good agreement with the observed value of $\sim 1$~eV~cm$^{-3}$.

\subsection{CR pressure vs. magnetic pressure}

In this section, we investigate the relation between CR pressure and magnetic pressure.
Pressure equality (or energy-density equipartition) between CRs and magnetic fields is commonly assumed in order to infer the magnetic field strength from synchrotron observations of star-forming galaxies \citep[e.g.][]{Longair94, Beck&Krause05}. An argument used to justify the equipartition assumption is that CRs and magnetic fields have a common source of energy. While the former are accelerated in supernova shocks, the latter are amplified by ISM turbulence, that, in turn, is driven by supernova feedback. Another argument for equipartition derives from the fact that CRs are confined by magnetic fields; a CR pressure exceeding the magnetic pressure would not be able to maintain this confinement.  However, while the above and related arguments suggest a relationship should exist between magnetic and CR pressure, there is no robust physical reason to support the assumption of equipartition, especially on local scales.  

In practice, observational evidence shows that in Milky Way-like galaxies, CR and magnetic pressures in midplane gas are similar on scales $ \gtrsim 1$~kpc. However, there are some observational signatures, supported by recent MHD simulations \citep[][]{Seta&Rainer19}, showing that pressure equality is unlikely on spatial scales of the order of 100 pc \citep[][]{Stepanov+14}. The discrepancy between CR pressure and magnetic pressure is even stronger in starburst galaxies, where the magnetic energy density is significantly larger than the CR energy density \citep{Yoast-Hull+16}.

\begin{figure}
\centering
\includegraphics[width=0.48\textwidth]{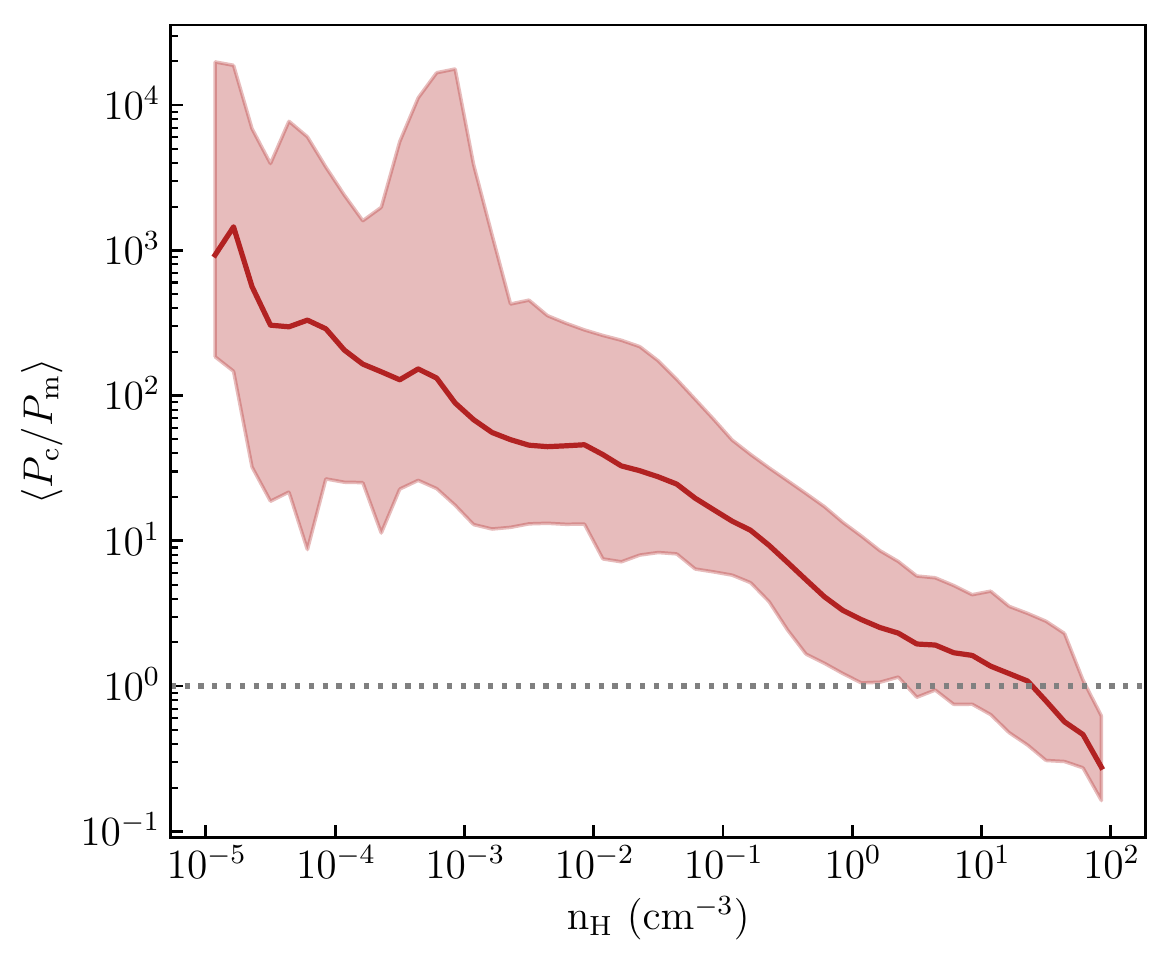}
\caption{Temporally-averaged median of the ratio between CR pressure $P_\mathrm{c}$ and magnetic pressure $P_\mathrm{m}$ as a function of hydrogen density $n_\mathrm{H}$ for the model with variable $\sigma_\parallel$ and no  diffusion perpendicular to the magnetic field direction. The shaded area covers the temporally-averaged 16th and 84th percentile variations around the median profile. 
The dotted line indicates equal pressure.}
\label{fig:PcPm}
\end{figure}

Here, we use the outcomes of the self-consistent model for high-energy CRs to evaluate the median ratio of CR pressure $P_\mathrm{c}$ and magnetic pressure\footnote{The magnetic pressure $P_\mathrm{m}$ is calculated as $(B_\mathrm{x}^2 + B_\mathrm{y}^2 + B_\mathrm{z}^2)/8 \pi$. Note that the magnetic pressure is always higher than the vertical magnetic stress ($P_\mathrm{m,z} = (B_\mathrm{x}^2 + B_\mathrm{y}^2 - B_\mathrm{z}^2)/8 \pi$) discussed in the rest of this paper.  For an isotropic magnetic field, the ratio would be a factor of three.} $P_\mathrm{m}$ as a function of hydrogen density $n_\mathrm{H}$ (see \autoref{fig:PcPm}). Clearly, the assumption of pressure equality is not valid for most of the ISM. The median value of  $P_\mathrm{c}/P_\mathrm{m}$ decreases with increasing the density: it is orders of magnitude above unity at low densities ($n_\mathrm{H} \lesssim 10^{-2}$~cm$^{-3}$) and decreases below unity at $n_\mathrm{H} > 1$~cm$^{-3}$. 

At the average midplane-density of the ISM ($n_\mathrm{H} \approx 0.1-1$~cm$^{-3}$), the median ratio is $\sim 3-30$. This explains why the average CR pressure and magnetic pressure become comparable near $z\simeq0$ in our model (see right panel of \autoref{fig:Gev_profiles_varsigma}) and to some extent justifies traditional premises for interpreting synchrotron emission when 
averaged on kpc-scales. At higher densities, the magnetic pressure is higher than the CR pressure. Indeed, while the latter is completely uniform for $n_\mathrm{H} > 10^{-1}$~cm$^{-3}$ (see \autoref{fig:PcvsnH}), the former increases in the densest parts of the ISM undergoing gravitational collapse. The median value of $P_\mathrm{c}/P_\mathrm{m}$ is $\simeq \, 0.1$ in regions with $n_\mathrm{H} \simeq 10^{2}$~cm$^{-3}$, whose typical size scale is $\lesssim 100$~pc (see \autoref{MHDsim}). This result is in agreement with what found by \citet {Stepanov+14} in the Milky Way, M31 and the LMC, i.e. that, when measured on spatial scales of the order of 100~pc, the magnetic energy density is larger than what expected from the assumption of pressure equality. Also, the variation of $P_\mathrm{c}/P_\mathrm{m}$ by almost one order of magnitude around the median value indicates the lack of a strict correlation between $P_\mathrm{c} $ and $P_\mathrm{m}$ at every density. Observed synchrotron emission is of course produced by relativistic electrons rather than the CR protons studied here, but our results serve as a caution in assuming pressure equality (or correlation) to infer local properties of the magnetic field from synchrotron observations. 

\subsection{Comparison with other works}
\label{Comparison with other works}

As mentioned in \autoref{Introduction}, many numerical studies have aimed to constrain the propagation of CRs on galactic scales with direct measurements of CR energy density in our Galaxy. These works generally assume temporally- and spatially-constant isotropic diffusion and often ignore the presence of CR advection \citep[e.g.][]{Trotta+11, Cummings+16, Johannesson+16}. They find that the isotropically-averaged scattering coefficient required to match the observed CR spectrum is of the order of $10^{-28}-10^{-29}$~cm$^{-2}$~s. At face value, these numbers appear to agree with our finding that, under the assumption of spatially-constant scattering, a value of $\sigma_\parallel\sim 10^{-29}$~cm$^{-2}$~s is required for the CR pressure to be comparable with the other relevant pressures near the galactic plane (see \autoref{fig:Gev_profiles_timeaver}) -- as observed in the solar neighborhood.
However, we have also seen that advection by fast-moving magnetized gas is crucial for transporting CRs out of the disk (\autoref{Streaming vs Diffusion vs Advection}). 
In propagation models ignoring advection, the primarily-horizontal magnetic field near the midplane makes the value of the perpendicular diffusion coefficient more important than the parallel coefficient;   \autoref{fig:noadv_profiles} and  \autoref{fig:diffpar-diffperp} show that $\sigma_\perp\sim 10^{-29}$~cm$^{-2}$ is required for the mid-plane CR pressure to be comparable to other pressures. 
While this essentially agrees with the results of traditional models that assume isotropic diffusion, it also points out their serious physical flaw: first, the value $\sigma_\perp\sim 10^{-29}$~cm$^{-2}$ is unrealistically low, and second, in our simulations advection is actually playing the role imputed to perpendicular diffusion.  

An important difference between our models \emph{without} CR advection and those mentioned above is that, while the latter make use of simplified analytic prescriptions to model the gas distribution within the Milky Way, in our model the background gas distribution is that predicted by the TIGRESS simulation of our solar neighborhood. The high resolution and the sophisticated physics included in TIGRESS allows for an accurate reproduction of the multiphase star-forming ISM. A realistic distribution of the background thermal gas and magnetic field is crucial for a detailed modeling of the CR propagation. For example, the analysis of our models without advection has shown that CRs can be easily trapped in regions with either  highly-tangled magnetic fields  with relatively high Alfv\'{e}n speed (at high altitude) or with primarily-horizontal magnetic fields (near the mid-plane).  
This explains why (unrealistically) low perpendicular scattering coefficients are required for the CR pressure to decrease to the observed values. If we neglected the real structure of the magnetic field and assumed open magnetic field lines, as usually done in analytic models of CR propagation, CRs would 
easily stream outward, and the needed parallel scattering coefficients would be higher. 

Another shortcoming of most
models is that they ignore the dependence of the scattering coefficient on the properties of the background gas. In \autoref{Gev-Variablesigma}, we have seen that in realistic models with non-uniform scattering, $\sigma_\parallel$ is relatively high in the ionized gas, while it rapidly decreases with the increase of density in the neutral gas. In the ionized gas that dominates the volume outside the midplane, the average value of $\sigma_\parallel$ (a few $\times \, 10^{-28}$~cm$^{-2}$~s) is higher than that predicted by simple diffusive models of CR propagation in the Milky Way. In contrast, in the neutral gas the scattering rates are far lower than in simple constant-diffusion models, and this low scattering leads to  
a very smooth CR  distribution in the mid-plane.
Furthermore, the actual value of the CR pressure in the neutral gas is not set by local transport, but the efficiency of CR propagation in the surrounding ionized gas. 
One can think at the gaseous disk as composed of a thinner layer of warm/cold neutral gas surrounded by a thicker layer of warm and hot ionised gas. If the transport of CRs is slow in the ionized gas, CRs remain trapped in the neutral gas, even though the local diffusivity is extremely large. 

It is also interesting to compare our results to those of \citet{Hopkins2021} based on cosmological zoom-in FIRE simulations \citep{Hopkins+18} with CRs.  The authors explore a variety of models, from some assuming constant scattering to more realistic models based either on the self-confinement or on the extrinsic-turbulence picture. 
Their results are generally similar to ours, while differing in some details.  In common with our conclusions, they find that there is no single diffusivity that characterizes transport, with the ISM phase structure leading to orders of magnitude variation.  They also find, as we have emphasized based on our models, that rapid transport of CRs in the neutral gas is not the main limitation on CR residence times (and CR pressure) in dense gas.  Rather, the main confinement of CRs is provided by the  surrounding ionized gas.  
Also, from the analysis of their simulations of Milky Way-like galaxies assuming constant scattering, they find that $\sigma_\parallel 
\sim 10^{-29}$~cm$^{-2}$~s is required to match the CR energy density $e_\mathrm{c}\sim 1$~eV~cm$^{-3}$ 
measured in the solar neighborhood, 
similar to the results shown in the right panel of our 
 \autoref{fig:Gev_profiles_timeaver}.

To our knowledge,  \citet{Hopkins2021} is the only work to date that has tested transport with a variable-scattering model based on self-confinement, and our self-consistent model without perpendicular diffusion is most similar to their default self-confinement model.    However, it should be noted that there are some non-negligible differences between our and \citeauthor{Hopkins2021}'s model. One difference regards the damping processes taken into account to compute $\sigma_\parallel$ (see \autoref{sigma}). While we consider ion-neutral and nonlinear Laundau damping only, \citet{Hopkins2021} also include turbulent and linear Landau damping in their calculations. Depending on the conditions of the background thermal gas, the addition of damping mechanisms may reduce the growth of Alfv\`{e}n waves, and, as a consequence, the CR scattering rate. Another difference is in the procedure to calculate $n_1$ in \autoref{IN} and \autoref{NLL}. \citet{Hopkins2021} approximate $n_1$ with $e_\mathrm{c}/E$, where $E$ is equal to 1~GeV. However, deriving $n_1$ from a realistic CR spectrum, we find that its actual value is almost one order of magnitude lower than $e_\mathrm{c}/E$. This results in a lower normalization of the scattering coefficient (a factor of $\sqrt{10}$ when $\Gamma_\mathrm{nll} > \Gamma_\mathrm{in}$ and a factor of 10 when $\Gamma_\mathrm{nll} < \Gamma_\mathrm{in}$) in our work. Also, to derive the ion number density in \autoref{IN} and \autoref{NLL}, we calculate the ionization fraction in the primarily-neutral gas based on the low-energy cosmic ray ionization rate, which is not implemented by \citet{Hopkins2021}. Since \citeauthor{Hopkins2021} report CR energy density averaged over radial shells but not the corresponding thermal, turbulent, or magnetic pressures (or values of $\Sigma_\mathrm{SFR}$), it is difficult to make detailed comparisons. However, we can note that 
the mid-plane CR pressure in our work is less than a factor of 2 lower than that found in their self-confinement model at $R=8$~kpc.

\subsection{CR spectrum and ionization rate in the ISM}
\label{CR ionization rate}

As discussed in \autoref{spectrum}, the low-energy slope of the CR spectrum in the solar neighborhood is highly uncertain. A simple extrapolation of the Voyager~1 data down to energies of 1 MeV predicts $\delta\approx 0.1$. However, this value fails to reproduce the rate of CR ionization measured in nearby diffuse molecular clouds ($n\approx100$~cm$^{-3}$, $T\approx 100$~K). \citet{Padovani+18} found that $\delta$ must be $\simeq -0.8$ at the edges of the clouds in order to match the inferred $\zeta_\mathrm{c}$ based on abundances of molecular ions \citep{Neufeld&Wolfire17}. As low-energy CRs penetrate the dense clouds, they lose a significant portion of their energy due to collisional interactions with the surrounding gas. Therefore, the low-energy slope of the spectrum tends to increase from the initial value (becoming flatter). Following \citeauthor{Padovani+18}, \citet{Silsbee+19} tried to constrain the value of $\delta$ using alternative models of CR propagation. They confirmed the value of $-0.8$ under the assumption of free streaming, as in \citet{Padovani+18}. However, they found that the low-energy slope decreases to $\delta = -1.0$ for the model where CRs freely stream along magnetic field lines above a given column density and are scattered by MHD waves below such threshold, and to $\delta = -1.2$ under the assumption of pure scattering. In \autoref{Low-EnergyCRs}, we have found that low-energy CRs freely stream along the magnetic field lines at the typical densities of diffuse molecular clouds (see
\autoref{fig:Hist_MeV}),
suggesting that the transport model proposed by \citet{Padovani+18} is more representative of the actual propagation of CRs at high gas densities. 

In \autoref{Low-energy slope}, we have used the predictions of our self-consistent models for the propagation of high-energy and low-energy CRs to constrain the low-energy slope of the injected CR spectrum, $\delta_\mathrm{inj}$, which is an input for our models.  
Our choices for $\delta_\mathrm{inj}$ correspond an assumption for the fraction of supernova energy going into production of CRs with energies of about $30$~MeV. We have explored three different values of $\delta_\mathrm{inj}$: $\delta_\mathrm{inj} = -0.8$, in agreement with \citet{Padovani+18}, $\delta_\mathrm{inj} = 0.1$, in agreement with the low-energy slope extrapolated from the Voyager spectrum, and $\delta_\mathrm{inj} = -0.35$, an intermediate value between $-0.8$ and $0.1$. We have then computed the local value of $\delta$ from the local energy density of both low-energy and high-energy CRs (\autoref{delta_spectrum}) and inferred the corresponding CR ionization rate.  
Comparing the CR ionization rates predicted by the three models for low-energy CRs with the CR ionization rate measured in diffuse molecular clouds, we find that a value of $-0.8 < \delta_\mathrm{inj} < -0.35$ is required to reproduce the observations. A significantly higher/lower value would result in a CR ionization rate below/above the observed range of values.  

We caution that our simulations lack the resolution to capture structures with densities above $\sim 100$~cm$^{-3}$. As the rate of collisional losses increases with the gas density, we expect the attenuation of CRs energy-density and flux to be stronger should the internal structure of individual clouds be resolved. As a consequence, the CR ionization rate might be lower than that shown in \autoref{fig:CRionrate} for a given choice of $\delta_\mathrm{inj}$, likely making the model assuming $\delta_\mathrm{inj} = -0.8$ more realistic than the model assuming $\delta_\mathrm{inj} = -0.35$.

\section{Final Summary}
\label{Conclusions}

This work investigates the propagation of CRs in a galactic environment with conditions similar to the solar neighborhood, taking into account a realistic spatial distribution of multiphase gas density, velocity, and magnetic field. For this purpose, we extract a set of snapshots from a TIGRESS MHD simulation \citep{Kim&Ostriker17, Kim+20} with spatial resolution $\Delta x = 8$~pc and post-process them using the algorithm for CR transport implemented in \textit{Athena}++ by \citet{Jiang&Oh18}. By comparing to post-processed TIGRESS simulations with the same conditions at both higher and lower resolution, we demonstrate that a resolution $ \Delta x \leq 16$~pc is required to achieve convergence of the CR properties analyzed in this paper (see \autoref{fig:Resolution}). 

We consider a wide range of CR transport models, from simple models including either diffusion or streaming only, to models including both diffusion and streaming but neglecting advection, to models including advection.  We first consider models in which the diffusivity is spatially constant, and 
analyze the effect of different choices of the scattering coefficient.  We then explore the 
physically-motivated case in which the scattering coefficient varies spatially.  The properties of the background gas and spatial distribution of CRs enter together in determining the scattering rate coefficient,  under the assumption that CRs are scattered by streaming-driven Alfv\'{e}n waves and that the wave amplitude is set by the balance of growth and damping (considering  both  ion-neutral damping and non-linear Landau damping). We separately evaluate transport of CRs with kinetic energies of  $\sim 1$ GeV (high-energy CRs) and $\sim 30$ MeV (low-energy CRs), respectively important for the dynamics and for the ionization of the ISM. 

Our main conclusions are as follows:
\begin{itemize}[leftmargin=*]
\item \textit{Advection by fast-moving, hot gas plays a key role in removing CRs from the disk}. Streaming and diffusion parallel to the magnetic field are relatively ineffective in transporting CRs from the mid-plane to the coronal region, since magnetic field lines are mainly oriented in the $x-y$ direction near the midplane in the warm gas, while the Alfv\'en speed is low in hot superbubbles (\autoref{fig:FsvsFdvsFa}). In transport models neglecting advection, diffusion perpendicular to the magnetic field direction becomes crucial for the propagation of CRs (\autoref{fig:diffpar-diffperp}). 
In the absence of advection and for the case of spatially-constant diffusivity, we find that scattering coefficients $\sigma_\parallel \lesssim \sigma_\perp \sim 10^{-29}$~cm$^{-2}$~s are required for the CR pressure to be in equipartition with thermal, kinetic and magnetic pressure near the Galactic plane (\autoref{fig:noadv_profiles}). In contrast, a value of $\sigma_\parallel \sim 10^{-29}$~cm$^{-2} \ll \sigma_\perp$~s is sufficient to reach pressure equipartition in the presence of advection (\autoref{fig:Gev_profiles} and \autoref{fig:Gev_profiles_timeaver}).

\item \textit{There is no single diffusivity.} For our variable-diffusion model,  the scattering coefficient varies over more than four
orders of magnitude depending on properties of the ambient gas (left panel of \autoref{fig:Gev_profiles_varsigma} and forth panel from left of \autoref{fig:varsigma_snaps}). 
Clearly, realistic spatial and thermal distributions of the background gas, as well as an accurate calculation of the ionization state, is crucial for a proper computation of $\sigma_\parallel$. For high-energy CRs, we find that $\sigma_\parallel$
is roughly constant and relatively high ($\simeq 10^{-28}$~cm$^{-2}$~s) in low-density regions ($n_\mathrm{H} < 10^{-2}$~cm$^{-3}$) where nonlinear Landau damping dominates. The scattering rate coefficient decreases to very low values ($\ll 10^{-29}$~cm$^{-2}$~s) in higher-density regions ($n_\mathrm{H} > 10^{-1}$~cm$^{-3}$) of primarily-neutral  gas where ion-neutral damping dominates.  The maximum value of $\sigma_\parallel$ ($\simeq 10^{-27}$~cm$^{-2}$~s) is reached at intermediate gas densities ($n_\mathrm{H} \sim 10^{-2}$~cm$^{-3}$), at the interface between neutral and fully-ionized gas.

\item \textit {Diffusion and streaming regulates the propagation of CRs within most of the ISM}. Our physically-motivated model accounting for variable $\sigma_\parallel$ predicts that diffusion largely dominates over advection in the higher-density, lower-temperature gas that comprises most of the mass of the ISM (\autoref{fig:FsvsFdvsFa_varsigma}). Gas velocities are high in the hot gas but much lower in the warm/cold gas, 
while at the same time ion-neutral damping in these phases keeps the scattering coefficient low. The higher density regions are also characterized by the highest values of streaming velocity, since the relatively  high value of the magnetic field and low ion density leads to a high ion Alfv\'{e}n speed. With ionization fraction $x_i \sim 0.01-0.1$ determined by the 
CR ionization rate, the ion Alfv\'{e}n speed of $50-100\,\kms$ exceeds the advection speed of $\sim 10\,\kms$ (see third panel from left of \autoref{fig:varsigma_snaps}).  Still, the mass-weighted diffusion speed exceeds the mass-weighted streaming speed for GeV CRs.  

\item \textit{The overall distribution of CRs depends on how effective their propagation is in the low-density gas}. Even though the scattering rate is very low within most of the ISM's mass near the mid-plane, CRs are strongly confined within this region.  CR transport out of the midplane
is limited by the high scattering rate in
surrounding lower-density, hotter, higher-ionization gas.
As a consequence, the overall CR distribution is in better agreement  with uniform diffusivity models (including advection)
that have relatively high $\sigma_\parallel \sim 10^{-27}-10^{-28}$~cm$^{-2}$~s, rather than models with values of the scattering coefficient similar to local values in the neutral gas (right panel of \autoref{fig:Gev_profiles_varsigma}). 
In realistic models, the CR pressure is strongly phase dependent and CRs are extremely  uniform at densities above $n_\mathrm{H} \sim 0.1$~cm$^{-3}$, while the constant-diffusion models with advection have a range of CR pressure at high density (\autoref{fig:PcvsnH-variablesigma}). In contrast, constant-diffusion 
models \emph{without advection} have smooth CR distributions across all phases, and the correlation of CR pressure with 
gas density is a side-effect of preferential CR deposition near the midplane where star formation and supernovae  are concentrated  (\autoref{fig:PcvsnH}). 

\item \textit{Low-energy CRs have less effective diffusion and more significant collisional losses compared to high-energy CRs}. Even though the scattering-coefficient distribution as a function of gas density is qualitatively similar for high- and low-energy CRs, the value of $\sigma_\parallel$ at a given density increases with decreasing kinetic energy. Also, for low-energy CRs, the scattering coefficient depends somewhat on
the low-energy slope of the injected energy-spectrum of CRs (left panel of \autoref{fig:GevvsMeV_sigma}).  We find that $\sigma_\parallel$ increases with decreasing slope (steepening spectrum). In addition to different diffusion, high-energy and low-energy CRs undergo different energy losses via interaction with the ambient gas since the rate of collisional losses is more than a factor of 2 larger for low-energy CRs. For low-energy CRs, the fraction of energy losses increases for steeper spectral slope since CRs are trapped in the dense gas for a longer time when the scattering rate is higher, thus losing more energy.

\item \textit{Streaming exceeds diffusion for low-energy CRs} (\autoref{fig:Hist_MeV}).
Since diffusion is less effective for low-energy CRs, their average streaming velocity is higher than their average diffusion velocity at $n_\mathrm{H} \sim 0.1-1$~cm$^{-3}$. Diffusion becomes dominant for $n_\mathrm{H} > 10$~cm$^{-3}$. At $n_\mathrm{H} \sim 10-100$~cm$^{-3}$, the mean free path of both high- and low-energy CRs is comparable to the size of diffuse cold atomic/molecular clouds (right panel of \autoref{fig:GevvsMeV_sigma}), meaning that CRs freely stream across them, subject to collisional energy losses only.
\end{itemize}

Although this paper has not directly studied dynamical effects of CRs in galaxies, our results have implications that are highly relevant to ISM dynamics.
First,  considering that the CR distribution 
in our physically-motivated model is extremely uniform in primarily-neutral gas,
our results predict that CR pressure gradient forces are negligible compared to the other forces associated with thermal pressure and Reynolds and Maxwell stresses. We therefore expect CRs not to contribute significantly to the dynamical equilibrium of of the ISM gas, or to immediate regulation of star formation rates. Nevertheless, CR pressure gradients are very large at the interface between the mostly-neutral disk and the surrounding lower-density corona, suggesting that CRs may significantly contribute to the gas dynamics of this region, including driving galactic winds (which regulate star formation over long timescales).  Clearly, fully self-consistent simulations with MHD and CRs are required to corroborate our expectations.

Because low-energy CRs suffer greater collisional losses than high-energy CRs, the ``evolved'' spectrum tends to be flatter than the injection spectrum.  This is partly  offset by the higher scattering rates for low-energy CRs, which trap them more effectively than high-energy  CRs. Even so, the local low-energy spectrum is always flatter than the injection spectrum.  
Although our models do  show the expected trend of decreasing low-energy CR density in dense gas (here, at $n_\mathrm{H} \gtrsim 3$\,cm$^{-3}$), our limited resolution does 
not allow us to constrain the spectrum based on differential CR ionization with density. 
Nevertheless, our models do suggest that a slope of the low-energy CR spectrum similar to $\delta \sim -0.5$  would be compatible with observed ionization rates.  A steeper slope would produce excess ionization, while a flatter slope would produce insufficient ionization.  Simulations at higher resolution, with self-consistent dynamics, will be needed to test and refine these conclusions.

\section*{Acknowledgements}
We thank the anonymous referee for valuable comments and
suggestions. We are grateful to Chang-Goo Kim and Munan Gong for sharing their expertise and technical tools. This  work was supported in part by grant 510940 from  the Simons Foundation to E.~C. Ostriker, and in part by Max-Planck/Princeton Center for Plasma Physics (NSF grant PHY-1804048).   Computational resources were
provided by the Princeton Institute for Computational Science and Engineering
(PICSciE) and the Office of Information Technology's High Performance Computing
Center at Princeton University. The Center for Computational Astrophysics at the Flatiron Institute is supported by the Simons Foundation.

\bibliography{bib}{}
\bibliographystyle{aasjournal}

\appendix

\section{Cosmic-ray spectrum dependence}
\label{AppendixA}

\subsection{Relation between $n_\mathrm{1}$ and $n_\mathrm{c}$ and dependence on low-energy spectral slope $\delta$}
\label{AppendixA1}

\begin{figure}
\centering
\includegraphics[width=0.55\textwidth]{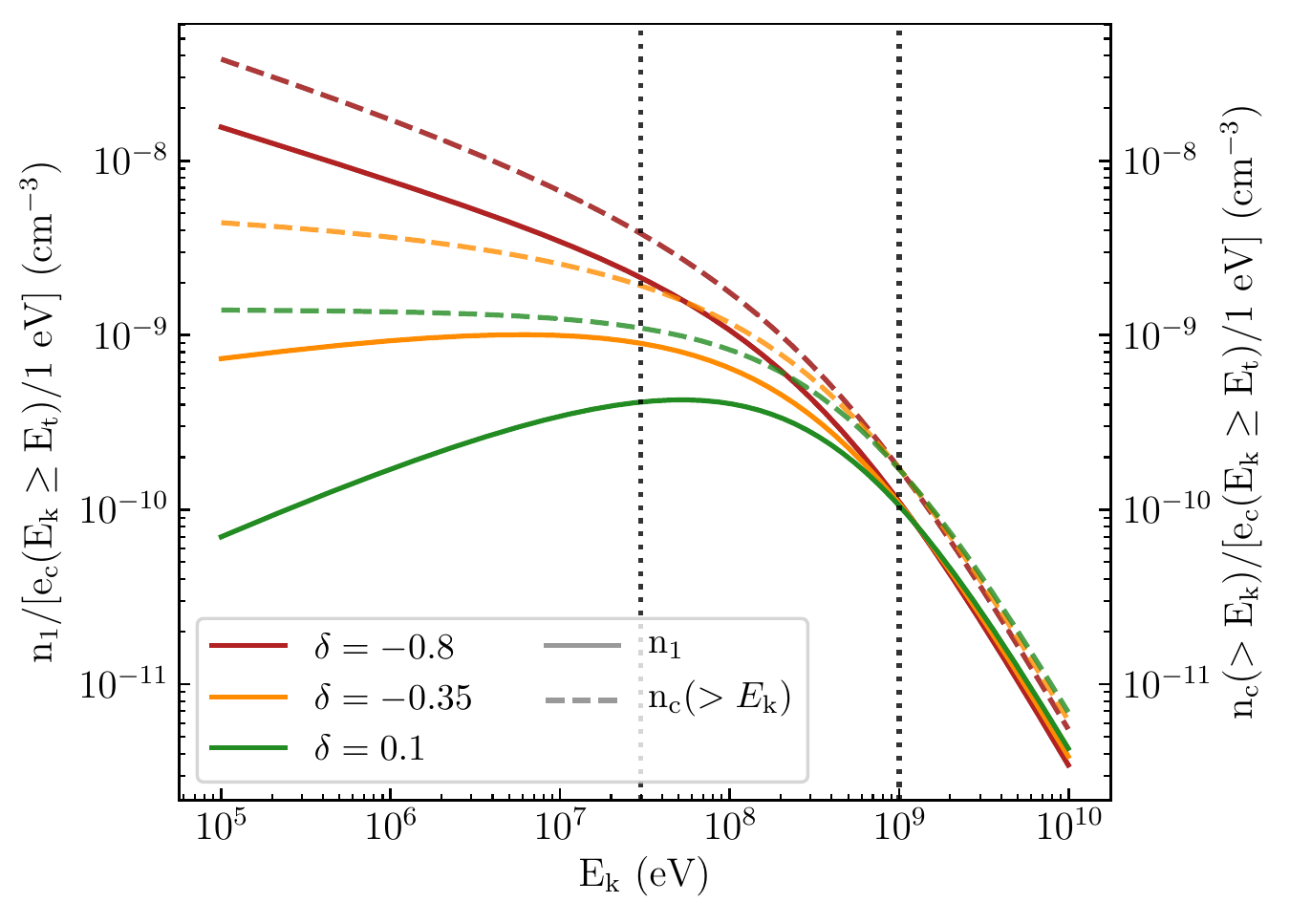}
\caption{Trend of $n_1$ (solid lines) and $n_\mathrm{c}(>E_\mathrm{k})$ (dashed lines) as a function of $E_\mathrm{k}=[(cp_1)^2 + (m c^2)^2]^{1/2} - m c^2$ for different low-energy slopes in the adopted CR spectral form (\autoref{CRspectrum}): $\delta = 0.1$ (green lines), $\delta = -0.35$ (orange lines), $\delta=-0.8$ (red lines). To normalize we divide  $n_\mathrm{1}$ and $n_\mathrm{c}$ by $e_\mathrm{c}(E_\mathrm{k} \geq E_\mathrm{t})/(1 \, \rm{eV})$, where this is the total energy density of CRs with kinetic energies above $E_\mathrm{t}$ = 650 MeV. The dotted vertical lines indicate where $E_\mathrm{k} = 30$~MeV and $E_\mathrm{k} = 1$~GeV.}
\label{fig:n1}
\end{figure}

\autoref{fig:n1} shows the trend of $n_\mathrm{1}$ (\autoref{eq:n1}) as a function of $E_\mathrm{k}=(c^2 p_1^2 + (m c^2)^2)^{1/2} - m c^2$ for three different values of $\delta$ for the low-energy slope of the CR energy-flux spectrum (\autoref{CRspectrum}). To normalize, $n_\mathrm{1}$ is divided by $e_\mathrm{c}(E_\mathrm{k} \geq E_\mathrm{t})/(1 \, \rm{eV})$, with $e_\mathrm{c}(E_\mathrm{k} \geq E_\mathrm{t})$ the total energy density of CRs with kinetic energies above the break of the spectrum $E_\mathrm{t}$.
Clearly, different choices of $\delta$ affect the trend of $n_\mathrm{1}$ only at low kinetic energy, where the three curves diverge at low energy.
The value of $n_1$ at a given energy increases with decreasing (more negative) $\delta$. For example, at $E_\mathrm{k} = 30$~MeV, the value of $n_\mathrm{1}$ at $\delta=-0.8$ is nearly one order of magnitude larger than the value of $n_\mathrm{1}$ at $\delta=0.1$. At kinetic energies above $E_\mathrm{t}$, the value of $n_\mathrm{1}$ is almost independent of $\delta$. 

To calculate the value of $\sigma_\parallel$ (\autoref{IN} and \autoref{NLL}) for high-energy CRs, we adopt the value of $n_\mathrm{1}$ at $E_\mathrm{k} = 1$~GeV, that is $\sim 10^{-10} \,e_\mathrm{c}(E_k \geq E_t)/1 \, \rm{eV})$~cm$^{-3}$ (which is extremely insensitive to $\delta$.) We note that the value of $n_\mathrm{1}$ at $E_\mathrm{k} = 1$~GeV is a factor of $\sim 3$ lower than the value of $n_\mathrm{1}$ at $E_\mathrm{k} = 30$~MeV for $\delta = 0.1$; this becomes more than a factor of 10 for $\delta = -0.8$. As a result, the normalization for the scattering coefficient is lower for high-energy CRs compared to low-energy CRs in all the transport models analysed in this paper.

For reference, the dashed lines in \autoref{fig:n1} show the CR number density $n_\mathrm{c} (>E_\mathrm{k}) = 4\pi\int_{p_1} ^\infty F(p) p ^2 dp$ as a function of $E_\mathrm{k}$ for the same three choices of $\delta$. The trend of $n_\mathrm{c}$ is the same as $n_\mathrm{1}$ at high kinetic energy, where the spectrum follows a power-law distribution ($j(E_\mathrm{k}) \sim C E_\mathrm{k}^{-2.7}$, $F(p)\sim C p^{-4.7}$). Here, $n_\mathrm{1} = n_\mathrm{c} (>E_\mathrm{k}) (3+r)/(2+r)$, where $r = - 4.7$ is the high-energy slope of $F(p)$. At low kinetic energy, the trends of $n_\mathrm{c}$ diverge with decreasing energy. Of course, unlike $n_1$, $n_c$ monotonically increases towards lower energy. 

\subsection{Dependence of $\zeta_\mathrm{c}$ on $\delta$ and $E_\mathrm{k,min}$}
\label{AppendixA2}

\begin{figure}
\centering
\includegraphics[width=0.5\textwidth]{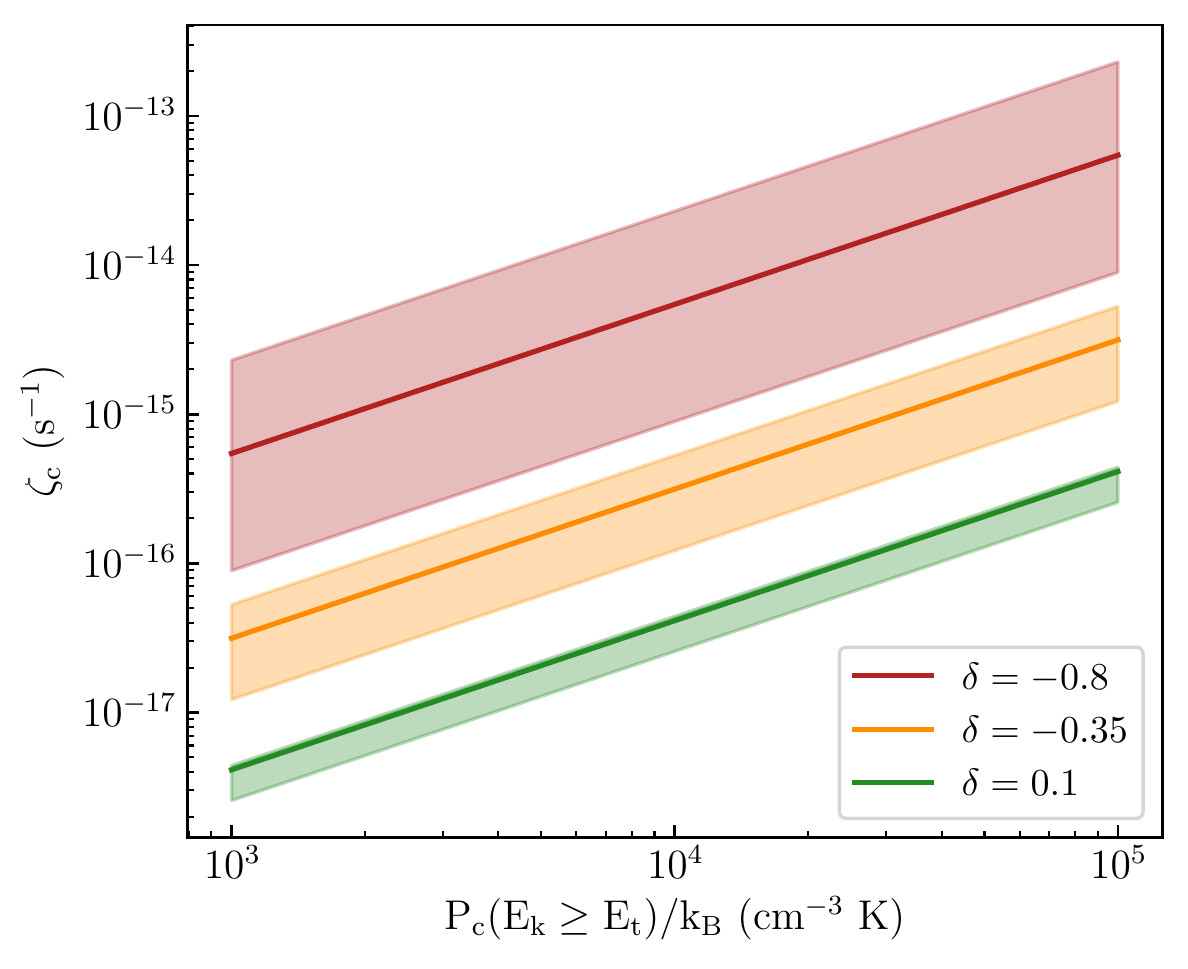}
\caption{Trend of the primary CR ionization rate per hydrogen atom $\zeta_\mathrm{c}$ as a function of the total pressure $P_\mathrm{c}/k_\mathrm{B}$ of high-energy CRs ($E_\mathrm{k} \geq E_\mathrm{t} = 650$~MeV) for $\delta = 0.1$ (green line), $\delta = -0.35$ (orange line), $\delta=-0.8$ (red lines). The solid line is obtained adopting $E_\mathrm{k,min} = 10^5$~eV in \autoref{CRionrate}. The lower and upper boundaries of the shaded area indicate the value of $\zeta_\mathrm{c}$ obtained adopting $E_\mathrm{k,min} = 10^6$~eV and $E_\mathrm{k,min} = 10^4$~eV, respectively.}
\label{fig:zeta}
\end{figure}

The dependence of the primary CR ionization rate per hydrogen atom (\autoref{CRionrate}) on the low-energy slope of the CR spectrum is displayed in \autoref{fig:zeta}, showing the value of $\zeta_\mathrm{c}$ as a function of $P_\mathrm{c} (E_\mathrm{k} \geq E_\mathrm{t})/k_\mathrm{B}$ for three different choices of $\delta$. $\zeta_\mathrm{c}$ linearly increases with $P_\mathrm{c} (E_\mathrm{k} \geq E_\mathrm{t}) = e_\mathrm{c} (E_\mathrm{k} \geq E_\mathrm{t}) /3$, which enters in the calculation of $\zeta_\mathrm{c}$ through the spectrum normalization $C$ (\autoref{SpecNorm}). At a given CR pressure, the value of $\zeta_\mathrm{c}$ increases with decreasing $\delta$ since this corresponds to an increase in the number density of low-energy CRs which ionize the ambient gas (see dashed lines in \autoref{fig:n1}). 
The solid lines denote the value of $\zeta_\mathrm{c}$ obtained adopting $E_\mathrm{k,min} = 10^5$~eV in \autoref{CRionrate} (our default assumption), while the lower and upper boundaries of the shaded area indicate the value of $\zeta_\mathrm{c}$ obtained adopting $E_\mathrm{k,min} = 10^6$~eV and $E_\mathrm{k,min} = 10^4$~eV, respectively. 

The variation of $\zeta_\mathrm{c}$ with $E_\mathrm{k,min}$ significantly depends on $\delta$. At a given CR pressure, the value of $\zeta_\mathrm{c}$ varies by less than a factor of 2 when $\delta = 0.1$, but by more than one order of magnitude when $\delta = -0.8$. This can be understood based on the dashed lines in \autoref{fig:n1}. For $\delta = 0.1$, there is negligible increase in the CR number density towards lower $E_\mathrm{k}$ below $E_\mathrm{k} \sim 10^6$~eV, 
whereas for $\delta = - 0.8$ there is very large increase. This means the additional ionization from CRs with energies below $10^6$~eV is negligible for a CR distribution with $\delta = 0.1$, while it is significant for a CR distribution with $\delta = -0.8$. A value of $\zeta_\mathrm{c}$ comparable to the observed CR ionization rate \citep[$\zeta_\mathrm{c} \simeq 1.8 \times 10^{-16}$~s$^{-1}$, e.g.][]{Padovani+20} 
can be recovered for $P_\mathrm{c}/k_\mathrm{B}$ in the range  $\sim 4 - 10 \times 10^3$~cm$^{-3}$~K (similar or slightly larger than solar neighborhood estimates) by the model with $\delta = -0.35$ for $E_\mathrm{k,min} \sim 10^4-10^5$~eV or by the model with $\delta = -0.8$ for $E_\mathrm{k,min} \gtrsim 10^6$~eV.

\subsection{Propagation models for high-energy CRs assuming different $\delta$}
\label{AppendixA3}

\begin{figure*}
\centering
\includegraphics[width=\textwidth]{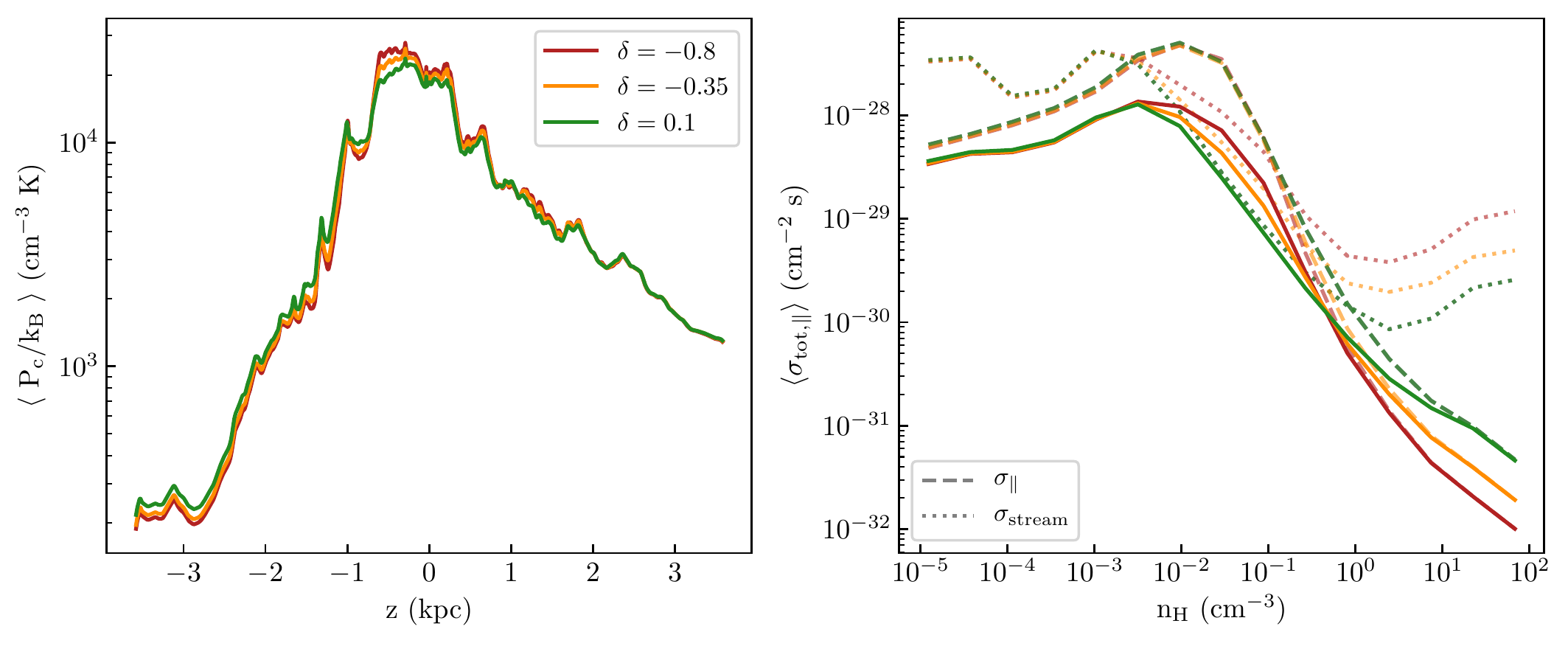}
\caption{Comparison of self-consistent propagation models for high-energy CRs assuming different low-energy slope of the CR spectrum, $\delta = -0.8$ (red lines), $\delta = -0.35$ (orange lines) and  $\delta = 0.1$ (green lines). \textit{Left panel}: horizontally-averaged vertical profiles of CR pressure. \textit{Right panel}: average particle-wave coefficients in the direction parallel to the mean magnetic field, $\sigma_\mathrm{tot, \parallel}$, as a function of hydrogen density $n_\mathrm{H}$. The dashed and dotted lines indicate the average scattering coefficients $\sigma_\mathrm{parallel}$ and streaming coefficients $\sigma_\mathrm{stream}$ as a function of $n_\mathrm{H}$, respectively. $\sigma_\mathrm{stream}$ is calculated as the inverse of the the second term on the RHS of \autoref{eq:sigmapar} ($1/\sigma_\mathrm{tot, \parallel} =1/ \sigma_\parallel + 1/\sigma_\mathrm{stream}$). The analysis is performed on the snapshot at $t = 286$~Myr.}
\label{fig:DifferentDelta}
\end{figure*}

The self-consistent model shown in \autoref{Gev-Variablesigma} for high-energy CRs is based on the assumption that the low-energy slope of the CR energy spectrum is $\delta = -0.35$. Here, we compare the results of the default model with those obtained for different choices of $\delta$, i.e. $\delta = -0.8$ and $\delta = 0.1$. 
The left panel of \autoref{fig:DifferentDelta} shows that the average vertical profiles of CR pressure are almost independent of the value of $\delta$. The three profiles are nearly identical except for a slight tendency to become steeper with decreasing $\delta$. We can note, for example, that at low latitudes ($\vert z \vert \lesssim 1$~kpc), the CR pressure slightly increases with decreasing $\delta$. 

To understood this behavior, we must consider how the roles of diffusion and streaming change with $\delta$. We recollect that both the scattering coefficient -- relevant for the calculation of the diffusive flux -- and the ion Alfv\'{e}n speed -- relevant for the calculation of the streaming flux -- depend on the ion number density.  In particular, $\sigma_\parallel \propto n_\mathrm{i}^{-0.5}$ when $\Gamma_\mathrm{in} > \Gamma_\mathrm{nll}$, and $\sigma_\parallel \propto n_\mathrm{i}^{-0.25}$ when $\Gamma_\mathrm{in} < \Gamma_\mathrm{nll}$, while $v_\mathrm{A,i} \propto n_\mathrm{i}^{-0.5}$. In turn, the ion number density depends on the rate of CR ionization, which is function of $\delta$ (see \autoref{AppendixA2}). A decrease of the low-energy slope of the CR spectrum entails an increase of the number of low-energy CRs, and, as a consequence, an increase of the ionization rate. In the low-density regime, both  $\sigma_\parallel$ and $v_\mathrm{A,i}$ must be similar for the three models since most of the gas is already ionized ($n_\mathrm{i} \simeq n_\mathrm{H}$) and the effect of different CR ionization rates is negligible. In the intermediate/high-density regime, where gas is partially or mostly neutral, we expect that the diffusive flux to increase with decreasing $\delta$ ($F_\mathrm{d,\parallel} \propto 1/\sigma_\parallel$), and the streaming flux to decrease with decreasing $\delta$. 

In the right panel of \autoref{fig:DifferentDelta}, the solid lines indicate the average particle-wave interaction coefficient along the magnetic field direction, $\sigma_\mathrm{tot, \parallel}$, as a function of hydrogen density for the three different choices of $\delta$. In the same plot, we use the dashed and dotted lines to indicate the average trend of the scattering coefficient $\sigma_\parallel$ and of the streaming coefficient $\sigma_\mathrm{stream}$. The latter is defined as the inverse of the second term on the RHS of \autoref{eq:sigmapar} ($1/\sigma_\mathrm{tot, \parallel} =1/ \sigma_\parallel + 1/\sigma_\mathrm{stream}$), $1/\sigma_\mathrm{stream}\sim v_\mathrm{A,i} H $.  
As anticipated above, both $\sigma_\parallel$ and $\sigma_\mathrm{stream}$ are independent of the choice of $\delta$ for $n_\mathrm{H} \lesssim 10^{-3}$~cm$^{-3}$. At higher density,
$\sigma_\mathrm{stream}$ increases with decreasing $\delta$, while $\sigma_\parallel$ remains independent of $\delta$ up to $n_\mathrm{H} \lesssim 10^{-1}$~cm$^{-3}$ and then decreases with decreasing $\delta$. 

The transition from the streaming-dominated to the diffusion-dominated regime slightly varies with $\delta$: it happens at $n_\mathrm{H} \sim 0.1$~cm$^{-3}$ for $\delta = -0.8$ and at $n_\mathrm{H} \sim 1$~cm$^{-3}$ for $\delta = 0.1$. In \autoref{Gev-Variablesigma}, we have seen that the overall distribution of CR pressure is regulated by the efficiency of CR propagation in the low-to-intermediate density gas: the less effective the CR propagation in these regions the more CRs are trapped in higher-density regions. In the intermediate-density regime ($n_\mathrm{H} \simeq 0.01-0.1$~cm$^{-3}$) $\sigma_\mathrm{stream} > \sigma_\parallel$, meaning that streaming dominates over diffusion ($F_\mathrm{s} = 
\vert \hat{\mathbf {B}} \cdot \nabla P_\mathrm{c} \vert / \sigma_\mathrm{stream} = (4/3)e_\mathrm{c} v_\mathrm{A,i} > F_\mathrm{d,\parallel} = 
\vert \hat{\mathbf {B}} \cdot \nabla P_\mathrm{c} \vert / \sigma_\parallel $). The higher CR pressure near the mid-plane and the steeper profiles predicted by models with lower $\delta$ can therefore be explained by the less effective CR streaming at intermediate densities. However, we note that while the variation in $\sigma_\mathrm{stream}$ increases with $n_\mathrm{H}$, the variation in $\sigma_\mathrm{tot,\parallel}$ -- which determines the total flux in the wave frame -- decreases as diffusion becomes more important.  In the intermediate-density regime, the average variation in $\sigma_\mathrm{tot, \parallel}$ is much lower than that in $\sigma_\mathrm{stream}$ and this explains why the vertical CR pressure profile is only weakly dependent on $\delta$.

\section{Sensitivity to numerical resolution}
\label{AppendixB}

\begin{figure*}
\centering
\includegraphics[width=0.9\textwidth]{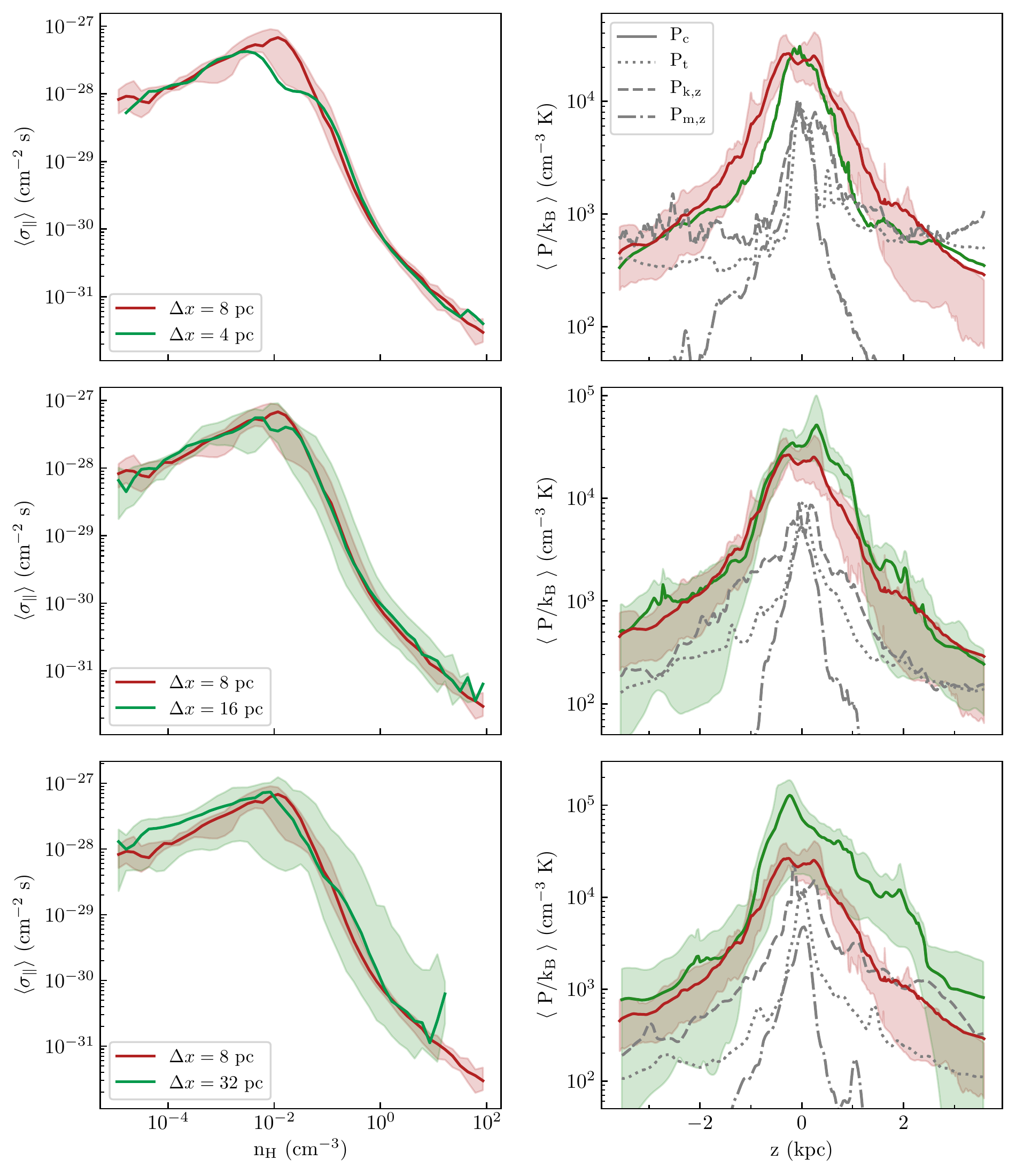}
\caption{Resolution comparison for self-consistent CR propagation model without perpendicular diffusion. From top to bottom, each row shows the comparison between the results obtained by post-processing the TIGRESS simulation at standard resolution ($\Delta x = 8$~pc -- red lines and shaded areas) and the results obtained from simulations with 
resolution $\Delta x = 4$~pc, 16~pc and 32~pc (green lines and shaded areas), respectively. While the results at resolutions $\Delta x \geq 8$~pc are averaged in time over multiple snapshots, the results at resolution $\Delta x = 4$~pc are for a single snapshot. 
\textit{Left panel}: median scattering coefficient $\sigma_\parallel$ as a function of hydrogen density $n_\mathrm{H}$. 
\textit{Right panel}: average vertical profiles of CR pressure $P_\mathrm{c}$. For the run at resolutions $\Delta x \geq 8$~pc, the shaded areas cover the  16th to 84th percentiles of temporal fluctuations. The gray lines show the horizontally-averaged profiles of thermal pressure $P_\mathrm{t}$ (dotted line), vertical kinetic pressure $P_\mathrm{k,z}$ (dashed line) and magnetic stress $P_\mathrm{m,z}$ (dot-dashed line) extracted from the high-resolution snapshot (\textit{first row}) and time-averaged over multiple snapshots at resolution $\Delta x = 16$~pc (\textit{second row}) and 32~pc (\textit{third row}).
}
\label{fig:Resolution}
\end{figure*}

In this section, we conduct studies to verify the robustness of our results to the numerical resolution of the MHD simulation. We apply the self-consistent models describing the propagation of high-energy CRs in the absence of perpendicular diffusion to a 
single snapshot extracted from the TIGRESS simulation 
at twice the standard resolution ($\Delta x = 4$~pc) and to multiple snapshots at lower resolutions ($\Delta x = 16$~pc and $\Delta x = 32$~pc).  The results shown for standard resolution are based on post-processing multiple snapshots and averaging over time (as in \autoref{fig:PerpDiff}).
The initial conditions of the simulations at different resolutions are identical to those of the standard simulation ($\Delta x = 8$~pc ) analyzed in this paper (see \autoref{MHD simulation}). 

In the first row of \autoref{fig:Resolution}, we compare the results from the standard run with higher spatial resolution results. The left panel shows the average scattering coefficient as a function of hydrogen density. 
Evidently, there is no systematic variation with resolution of the distribution of scattering coefficient. The deviation between curves for different resolution near  $n_\mathrm{H} \simeq 10^{-2}$~cm$^{-3}$ is most likely because 
the high-resolution result is from a single snapshot that has
local excursions from a  statistically-steady state,
rather than due to an intrinsic dependence on the spatial resolution. The right panel shows the average vertical profile of CR pressure for both resolutions. The average vertical profiles of thermal pressure, vertical kinetic pressure, and vertical magnetic stress from the high-resolution snapshot are also shown. The shaded areas around the temporally-averaged standard-resolution profiles (red lines) cover the 16th to 84th percentiles of temporal fluctuations. The high-resolution pressure profile is consistent with the low-resolution one and, most importantly, lies within the shaded area indicative of temporal fluctuations around the mean profile. As found for the standard resolution simulation, the CR pressure is a factor $\sim 3$ higher than the other relevant pressures in the mid-plane (see \autoref{Gev-Variablesigma} and \autoref{fig:Gev_profiles_varsigma}).

The second and third rows of \autoref{fig:Resolution} show the comparisons between the time-averaged results of the standard simulation and the time-averaged results of the simulations with lower resolutions, respectively $\Delta x = 16$~pc and $\Delta x = 32$~pc. For the run at 16~pc, the average scattering coefficient distribution is in very good agreement with the standard distribution, even though characterized by larger temporal fluctuations. The average CR pressure profile is consistent with the standard profile at high latitudes ($z \gtrsim 1$~kpc), while it lies slightly above that near the mid-plane. The agreement considerably worsens when we further halve the resolution. Even though the average scattering coefficient distributions obtained from the runs at 8~pc and 32~pc are roughly consistent, the latter presents temporal fluctuations over more than one order of magnitude, meaning that the distribution of $\sigma$ significantly change from one snapshot to another. The low-resolution average profile of CR pressure is always above the standard profile and, as for $\sigma_\parallel$, characterized by large temporal fluctuations. Near the mid-plane, the average CR pressure increases by almost one order of magnitude going from $\Delta x = 8~$pc to $\Delta x = 32~$pc. At the same time, the run at 32~pc presents higher thermal and kinetic pressures in the disk compared to the runs at $\Delta x \le 16$~pc, as well as a very large range of SFR \citep[see][]{Kim&Ostriker17}. We conclude that models with $\Delta x \le 16$~pc are converged, thus confirming that the spatial resolution of 8 pc is sufficient to achieve robust convergence of the CR properties analyzed in this paper. 

We note that moving-mesh simulations typically have resolution much lower than fixed-grid simulations in the low-density gas.  For example, cosmological zoom simulations with mass resolution of $10^4 M_\odot$ correspond to spatial resolution $\Delta x = 66$~pc $(n_\mathrm{H}/1\, \mathrm{cm}^{-3}$. The warm-cold ISM at $n_\mathrm{H}\sim 0.1-100\, \mathrm{cm}^{-3})^{-1/3}$ would have $\Delta x = 140-14$~pc, while the hot ISM at $n<0.01\, \mathrm{cm}^{-3})^{-1/3}$ would have $\Delta x > 300$~pc.  The  above analysis regarding resolution dependence suggests that while the scattering rate coefficients of these simulations may be in agreement with higher resolution simulations, the 
CR distribution itself may not be converged. 

\end{CJK*}
\end{document}